\documentclass[]{pasj02} 
\usepackage[switch,mathlines]{lineno} 

\usepackage{natbib}
\usepackage{amsfonts}
\usepackage{amsmath}

\usepackage{ulem} 
\usepackage{graphicx}
\usepackage{physics}
\usepackage[switch,mathlines]{lineno}
\usepackage{multirow}
\usepackage{threeparttable}

\definecolor{ao_en}{rgb}{0.0, 0.5, 0.0}

\newcounter{ncount}
\stepcounter{ncount}
\edef\TPHtwoform{\arabic{ncount}}
\stepcounter{ncount}
\edef\TPHphion{\arabic{ncount}}
\stepcounter{ncount}
\edef\TPHdisso{\arabic{ncount}}
\stepcounter{ncount}
\edef\TPPhotoele{\arabic{ncount}}
\stepcounter{ncount}
\edef\TPPHion{\arabic{ncount}}
\stepcounter{ncount}
\edef\TPHtwodiso{\arabic{ncount}}
\stepcounter{ncount}
\edef\TPHrecom{\arabic{ncount}}
\stepcounter{ncount}
\edef\TPHminus{\arabic{ncount}}
\stepcounter{ncount}
\edef\TPHexcit{\arabic{ncount}}
\stepcounter{ncount}
\edef\TPHeplus{\arabic{ncount}}
\stepcounter{ncount}
\edef\TPFreeFree{\arabic{ncount}}
\stepcounter{ncount}
\edef\TPCOMPT{\arabic{ncount}}
\stepcounter{ncount}
\edef\TPLINE{\arabic{ncount}}
\stepcounter{ncount}
\edef\TPCOOL{\arabic{ncount}}

\stepcounter{ncount}
\edef\TPPHM{\arabic{ncount}}
\stepcounter{ncount}
\edef\OMKI{\arabic{ncount}}
\stepcounter{ncount}
\edef\TPHFHYONE{\arabic{ncount}}
\stepcounter{ncount}
\edef\TPPBakesTielens{\arabic{ncount}}
\stepcounter{ncount}
\edef\TPPFKSM{\arabic{ncount}}
\stepcounter{ncount}
\edef\TPFERLAND{\arabic{ncount}}
\stepcounter{ncount}
\edef\TPPSUGI{\arabic{ncount}}
\stepcounter{ncount}
\edef\TPPHOSOKW{\arabic{ncount}}
\stepcounter{ncount}
\edef\TPPCEN{\arabic{ncount}}
\stepcounter{ncount}
\edef\TPPSKang{\arabic{ncount}}

\jyear{2024}
\Received{}
\Accepted{}

\graphicspath{{./}{figure/}} 

\begin{document}

\title{\textsc{SFUMATO\#}: A GPU-accelerated code for self-gravitational radiation hydrodynamics simulation with adaptive mesh refinement}

\author{
 Hajime \textsc{Fukushima}\altaffilmark{1}\altemailmark\orcid{0000-0000-0000-0000} \email{fukushima@ccs.tsukuba.ac.jp} 
 Tomoaki \textsc{Matsumoto}\altaffilmark{2}\orcid{0000-0000-0000-0000}
}
\altaffiltext{1}{Center for Computational Sciences, University of Tsukuba, Ten-nodai, 1-1-1 Tsukuba, Ibaraki 305-8577, Japan}
\altaffiltext{2}{Faculty of Sustainability Studies, Hosei University, Fujimi, Chiyoda, Tokyo 102-8160, Japan}




\maketitle

\begin{abstract}
We present a new implementation of the \textsc{SFUMATO} code, called \textsc{SFUMATO\#}, for solving self-gravitational radiation hydrodynamics problems using adaptive mesh refinement (AMR) with the CUDA/HIP programming frameworks.
The code incorporates a multigrid solver for self-gravity, radiation transfer with M1 closure and reduced speed of light approximation, non-equilibrium chemistry, thermal evolution, and sink particle schemes.
We develop new non-equilibrium chemistry and thermal solvers based on a linearized implicit method, whose accuracy is validated through a series of test problems by comparison with solutions obtained using the Newton-Raphson method.
By incorporating the heat capacity of dust grains, the dust temperature can be evolved without iterative energy-balance calculations.
From the perspective of computational cost, we demonstrate that adopting an increased pseudo dust heat capacity accelerates the chemistry solver while preserving accuracy, even when the value is increased by up to three orders of magnitude relative to the realistic value.
In addition, we perform a suite of test problems to confirm the validity of the other components of our implementation.
The code supports multi-GPU execution via MPI-based parallelization.
We measure the strong-scaling performance of the hydrodynamics and self-gravity solvers on both uniform and AMR grids, as well as the overall code performance using a giant molecular cloud simulation.
We find that the computational cost of the self-gravity solver increases with the number of MPI processes, indicating that efficient parallel performance is achieved only when the number of devices is chosen such that the cost of the self-gravity solver remains comparable to that of the other components.
\end{abstract}


\section{INTRODUCTION} \label{introduction}

In recent years, graphical processing units (GPUs) have been integrated into state-of-the-art high-performance computing systems.
GPUs provide substantially higher computational throughput compared to central processing units (CPUs), and their utilization has therefore become an important consideration in astrophysical simulations.
Executing simulations on a GPU requires the use of specialized programming frameworks, such as CUDA, HIP, or SYCL.
CUDA is restricted to NVIDIA GPUs, whereas HIP and SYCL are open-source frameworks. 
HIP supports GPUs from NVIDIA and AMD, and SYCL supports GPUs from NVIDIA, AMD, and Intel.
Generally, CPU and GPU memories are separate and do not have direct access to each other.
We need to use explicit instructions to transfer data between the CPU and GPU.
Simulations with multiple GPUs typically rely on MPI data transport.
However, MPI operations do not inherently synchronize with GPU kernel execution, and explicit synchronization is required.
As most astrophysical simulation codes have historically been developed for CPU architectures, restructuring is required to enable efficient execution on GPUs.

Many astrophysical problems, such as star cluster formation and star formation, involve large dynamical ranges.
The adaptive mesh refinement (AMR) method enables the selective allocation of computational resources to regions requiring higher spatial and temporal resolution, and is an essential approach for performing high-resolution simulations with the grid-based methods \citep{1984JCoPh..53..484B, 1989JCoPh..82...64B}.
With AMR, the required spatial resolution can be achieved by introducing finer grids according to user-defined refinement criteria.
Furthermore, the adaptive time-step can be applied across different levels of refinement.
AMR-based simulation codes, such as \textsc{FLASH} \citep{2000ApJS..131..273F}, \textsc{RAMSES} \citep{2002A&A...385..337T}, \textsc{ENZO} \citep{2014ApJS..211...19B}, and \textsc{Athena++} \citep{2020ApJS..249....4S}, have been developed and applied to a broad range of astrophysical problems.

The development of GPU-accelerated AMR codes is being actively pursued in various projects, such as \textsc{GAMER-2} \citep{2018MNRAS.481.4815S}, \textsc{ARK-RT} \citep{2021A&A...646A.123B}, \textsc{CHOLLA} \citep{2015ApJS..217...24S}, \textsc{QUOKKA} \citep{2022MNRAS.512.1430W}, and \textsc{AthenaK} \citep{2026ApJS..283...27S}.
Nevertheless, applications of GPU-based simulations remain at a relatively early stage, particularly in the context of star and star cluster formation.
As high-performance computing continues to shift toward GPU architectures, further progress in GPU-enabled astrophysical simulations will be indispensable.

In this paper, we present \textsc{SFUMATO\#}, a new implementation of the AMR code incorporating the fluid dynamics, self-gravity, sink particle, non-equilibrium chemistry, thermal evolution, and radiation transfer solvers of the original \textsc{SFUMATO} code \citep{2007PASJ...59..905M} together with the radiation transfer module developed by \citet{2021MNRAS.506.5512F}. 
The code employs self-similar block-structured grids managed by an octree structure.
Self-gravity is solved using the multigrid method \citep[e.g.,][]{2007PASJ...59..905M, 2023ApJS..266....7T}.
This approach provides simple algorithms for the AMR code and enables uniform task distribution across GPU cores.
We adopt the CUDA and HIP programming frameworks, which allow low-level control over memory and task management.
MPI-based data transfer is also implemented, enabling simulations to be performed on multiple GPUs.

In the star formation, the thermal evolution determines the characteristic mass scales of fragmentation in collapsing clouds \citep{1985MNRAS.214..379L, 2005MNRAS.359..211L}.
Gas temperature is determined by various processes, including metal-line cooling, dust thermal emission, and chemical reactions \citep[e.g.,][]{2005ApJ...626..627O, 2010ApJ...722.1793O}. 
In the numerical simulations, the equations governing thermal and chemical evolution are typically solved using the implicit solvers based on the Newton-Raphson (NR) method due to the stiffness of the equations.
In such solvers, the number of iterations required for convergence can vary among computational processes.
On GPUs, this behavior can lead to warp divergence, which can significantly degrade computational performance.
In this study, we propose a new scheme for solving the non-equilibrium chemistry and thermal evolution based on a linearized implicit (LI) method, thereby avoiding the performance loss caused by warp divergence in iterative methods.

The remainder of this paper is organized as follows.
In Section \ref{numerical_method}, we describe the numerical methods implemented in our code, including the management of computational procedures on GPUs.
Section \ref{sec:test_problems} presents the results of a set of test problems.
Section \ref{section:performance} describes the performance tests of our code.
Finally, Section \ref{Section:discussion} is devoted to a summary of this work.

\section{NUMERICAL METHOD} \label{numerical_method}

\subsection{Basic equations} \label{section_basic_equation}

We solve the following RHD equations,
\begin{align}
  \frac{\partial \rho}{\partial t} + \nabla \cdot \left( \rho \bf{v} \right) = S_{\rm sink}, \label{eoc}
\end{align}
\begin{align}
  \frac{\partial \left(\rho \bf{v} \right)}{\partial t} + \nabla \cdot \left( \rho \bf{v} \bf{v} \right) + \nabla P  = \rho \left( \bf{g} + \bf{g}_{\rm sink,g} + \bf{f} \right), \label{eom}
\end{align}
\begin{align}
  	\frac{\partial \left( \rho E \right)}{\partial t} + \nabla \cdot \left[ \left( \rho E + P \right) \bf{v} \right] = \rho \left( {\bf{g} + \bf{g}_{\rm sink,g}  + \bf{f}} \right) \cdot {\bf{v}} + \Gamma - \Lambda, \label{energy_eq}
\end{align}
\begin{align}
  \frac{\partial E_{r}}{\partial t} + \nabla \cdot {\bf F}_{r} = S - \alpha_{\rm E} \tilde{c} E_{r}, \label{eq_radmoment1}
\end{align}
\begin{align}
  \frac{1}{\tilde{c}} \frac{ \partial {\bf F}_{r}}{\partial t} + \tilde{c} \nabla \cdot {\bf P}_{r} = - \alpha_{\rm F} {\bf F}_r, \label{eq_radmoment2}
\end{align}
together with Poisson's equations
\begin{align}
  \nabla^2 \Phi = 4 \pi G \rho, \label{eq_poission}
\end{align}
\begin{align}
  {\bf g} = - \nabla \Phi, \label{eq_gravity}
\end{align}
where $E$ is the total energy defined as,
\begin{align}
  E = \frac{|\bf v|^2}{2} + \left( \gamma_{\rm ad} - 1 \right)^{-1} \frac{P}{\rho}, \label{eq_total_energy}
\end{align}
$\rho$ is the gas density, $\bf v$ is the velocity, $P$ is pressure, $\bf g$ is gravity, $\bf{g}_{\rm sink,g}$ is gravity from sink particles, $\bf f$ is the radiation pressure force, $\gamma_{\rm ad}$ is the adiabatic exponent, $E_r$ is the radiation energy density, ${\bf F}_r$ is the radiation energy flux, ${\bf P}_r$ is the radiation pressure tensor, $S$ is the source term of radiation, $\alpha_{E}$ and $\alpha_{\rm F}$ are the energy density and flux weighted absorption coefficients, and $\Phi$ is the gravitational potential.
In Equation \eqref{eoc}, $S_{\rm sink}$ denotes the gas accretion term on sink particles and vanishes in cells outside the sink particle interior.
The details of $\bf{g}_{\rm sink,g}$ and $S_{\rm sink}$ are described in Section \ref{sink_particle}.
In Equation \ref{energy_eq}, $\Gamma$ and $\Lambda$ are the heating and cooling functions, and they are formulated in Table \ref{Tab:thermal_process} in Appendix \ref{apd:thermal_process}.

To close the radiation moment equations \eqref{eq_radmoment1} and \eqref{eq_radmoment2}, we employ the M1 closure \citep{1984JQSRT..31..149L}. 
The radiation pressure tensor is expressed in terms of the Eddington tensor $({\bf D}_r)$ and the radiation energy density $({E}_r)$ as
\begin{align}
  {\bf P}_r = {E}_r {\bf D}_r. \label{eq_et}
\end{align}
In the M1 closure scheme, the Eddington tensor $({\bf D}_r)$ is approximated as \citep{1984JQSRT..31..149L}
\begin{align}
  {\bf D}_{r} = \frac{1-\chi}{2} {\bf I} + \frac{3 \chi -1}{2} {\bf n}{\bf n}, \label{eq_M1closure}
\end{align}
where
\begin{align}
  {\bf n} = \frac{{\bf F}_r}{|{\bf F}_r|}, ~ \chi = \frac{3 + 4 f^2}{5 + 2 \sqrt{4-3f^2}}, ~ {\rm and} ~ f = \frac{|{\bf F}_r|}{\tilde{c} E_r}. \label{eq_M1c2}
\end{align}
The radiation transfer equations are solved explicitly, in the same manner as the hydrodynamics solver.
To prevent the timestep from becoming prohibitively small, we employ the reduced speed of light approximation, a standard approach in star and galaxy formation simulations \citep[e.g.,][]{2001NewA....6..437G, 2013MNRAS.436.2188R, 2013ApJS..206...21S}.
The reduced light speed is denoted as $\tilde{c}$ in Equation \eqref{eq_radmoment1} and \eqref{eq_radmoment2}, and we adopt $\tilde{c} = 10^{-4} c$ in this study.
In the update step, the fluxes at the cell interfaces are evaluated.
We prepare the two options for computing the numerical fluxes: global Lax-Friedrichs (GLF) and Harten-Lax-van Leer (HLL) flux functions \citep{harten1983upstream, 2013MNRAS.436.2188R}.
The GLF scheme produces more diffusive fluxes than the HLL scheme.
Although the HLL flux is accurate for propagation normal to the cell interface, it introduces additional diffusion for diagonal propagation.
We adopt three frequency bins: (1) extreme ultraviolet (EUV; $13.6~{\rm eV} < h\nu$), (2) far-ultraviolet (FUV: $6~{\rm eV} < h \nu < 13.6~{\rm eV}$), and (3) infrared (IR) photons. 
In addition, we estimate the energy density of Lyman-Werner photons, which dissociate $\rm H_2$, assuming a constant intensity ratio relative to the FUV photons.

We also incorporate the non-equilibrium chemistry equations, 
\begin{align}
  \frac{\partial \left( y_{i} n_{\rm H} \right)}{\partial t} + \nabla \cdot \left( y_{i} n_{\rm H} {\bf v} \right) = n_{\rm H} R_i, \label{eq_chemq}
\end{align}
where $n_{\rm H}$ is the number density of hydrogen nuclei, $y_i$ is the fractional abundance of the $i$-th chemical species, 
and $R_i$ is the reaction rate.
The details of the adopted chemical network and reaction rates are described in Section~\ref{sec_chemistry} and Appendix~\ref{apd:chemical_network}.

The dust temperature $(T_{\rm d})$ is also calculated to account for the thermal decoupling between gas and dust grains.
We assume that dust grains have a constant specific heat capacity $C_{\rm d}$ and no thermal diffusion.
Under these assumptions, the equation for dust temperature can be treated as a passive scalar:
\begin{align}
  \frac{\partial \left( \rho C_{\rm d} T_{\rm d} \right)}{\partial t} + \nabla \cdot \left( \rho C_{\rm d} T_{\rm d} {\bf v} \right) = \rho \mathcal{E}_{\rm d}, \label{eq_dust_temp}
\end{align}
where $\mathcal{E}_{\rm d}$ represents heating and cooling of dust grains due to interactions with the gas and radiation field.

\subsection{Simulation method} \label{sec_simulation_method}

The numerical algorithms to solve equations \eqref{eoc}-\eqref{eq_dust_temp} are based on the finite-volume approach, employing an operator-splitting approach to advance the system in time.
The algorithms implemented in the original \textsc{sfumato} code have been described in detail in previous works \citep{2007PASJ...59..905M, 2015ApJ...801...77M, 2021MNRAS.506.5512F}.
Therefore, we do not repeat them here, but instead focus on the additional features incorporated into \textsc{sfumato\#}.

\subsubsection{Grid structure} \label{sec:grid_structure}

In the \textsc{sfumato} AMR framework, self-similar block-structured grids with an octree-based structure are employed. 
The parent block is divided into 8 child grids when the refinement criterion is satisfied.
Each block has $N^3$ cells, where $N$ denotes the number of cells along each direction.
Although the number of cells per block is fixed, the cell width varies with the grid-level.
The grid at level $l$ achieves a spatial resolution $2^l$ times finer than that of the coarsest level.

\subsubsection{Data and task management on GPUs}

We assign one GPU to each MPI process, and the data are distributed to MPI processes in units of blocks.
Blocks are ordered using a Peano–Hilbert space-filling curve and then distributed among MPI processes.
Each block is assigned a unique ID, and lists describing the parent–child and neighboring relationships are created and managed on the hosts.

The two options for allocating memories of block data are provided.
The first is device memory allocation on GPUs using \texttt{cudaMalloc} / \texttt{hipMalloc}, and the second is unified memory allocation on the CPUs using \texttt{cudaMallocManaged} / \texttt{hipMallocManaged}.
Unified memory is particularly advantageous for large-scale problems that exceed the available device memory.
Moreover, since unified memory can also be accessed from the hosts, CPU cores can directly execute tasks.
The drawback is the increased communication cost associated with data transfers to GPUs.
However, with the recent emergence of architectures in which CPUs and GPUs share the same memory, the transfer cost in unified memory is expected to become negligible in such systems.

On GPUs, computational processes are parallelized at the cell level, except for grid-level reductions and numerical flux calculations.
In CUDA/HIP, the execution order of kernels on the GPU is controlled by streams and events.
We employ non-blocking streams and assign the functions without waiting for the completion of previous kernels.
Blocking with events is used when MPI communication is performed or when stream synchronization is required to monitor executed tasks.

\subsubsection{Time Marching} \label{section:time_marching}

We advance equations \eqref{eoc}-\eqref{eq_dust_temp} independently for each block.
As in the original \textsc{SFUMATO} code, two modes of time integration are implemented: synchronous and adaptive time stepping.
In the synchronous mode, all grid levels are updated sequentially using a common time step.
This mode is particularly suitable for self-gravitating gas, since changes in the density distribution and gravitational potential on fine grids immediately affect the coarser grids with second-order accuracy.
The adaptive time mode follows the scheme of \citet{1989JCoPh..82...64B}.
In this approach, coarser grids are updated with a longer time step, thereby reducing the total number of updates and lowering the computational cost.
We could calculate the self-gravitating gas with the first-order temporal accuracy in this mode.

In the adaptive timestep mode, the time step at each grid-level is determined following the procedure described in Section 5.2 of \citet{2007PASJ...59..905M}.
The time step on a finer grid, $\Delta t^h$, is constrained by both the time step of the coarser level, $\Delta t^{H}$, and the CFL-limited time step, $\Delta t^h_{\rm CFL}$, according to 
\begin{align}
    \Delta t^h = \Delta t ^{H} 2^{-n}, \label{eq:timestep}
\end{align}
where 
\begin{align}
    n = \min \{m \in \mathbb{N} \mid \Delta t ^{H} 2^{-m} \le \Delta t^h_{\rm CFL}\}. \label{eq:timestep2}
\end{align}
This ensures that the cumulative time on the finer grid remains synchronized with that of the coarser grid while satisfying the CFL condition.

If radiation transfer calculations are included, the timestep is further constrained by the CFL condition based on the reduced speed of light, $\tilde{c}$ appearing in equations \eqref{eq_radmoment1} and \eqref{eq_radmoment2}.
Since $\tilde{c}$ is generally larger than the gas velocity, the resulting timestep is smaller than that determined by the CFL condition for hydrodynamics alone.
The gravity and motions of sink particles depend only on the spatial distribution of the gas density.
Consequently, the timestep limited by the propagation speed of radiation is unnecessarily short for updating gravity and sink positions.
To address this, we provide an additional option in the adaptive timestep mode that updates these quantities only at intervals determined by the hydrodynamics CFL conditions on the finest grid level.

\subsubsection{Hydrodynamics} \label{sec_hydro}

\begin{figure*} 
\begin{center}
  \includegraphics[width=140mm]{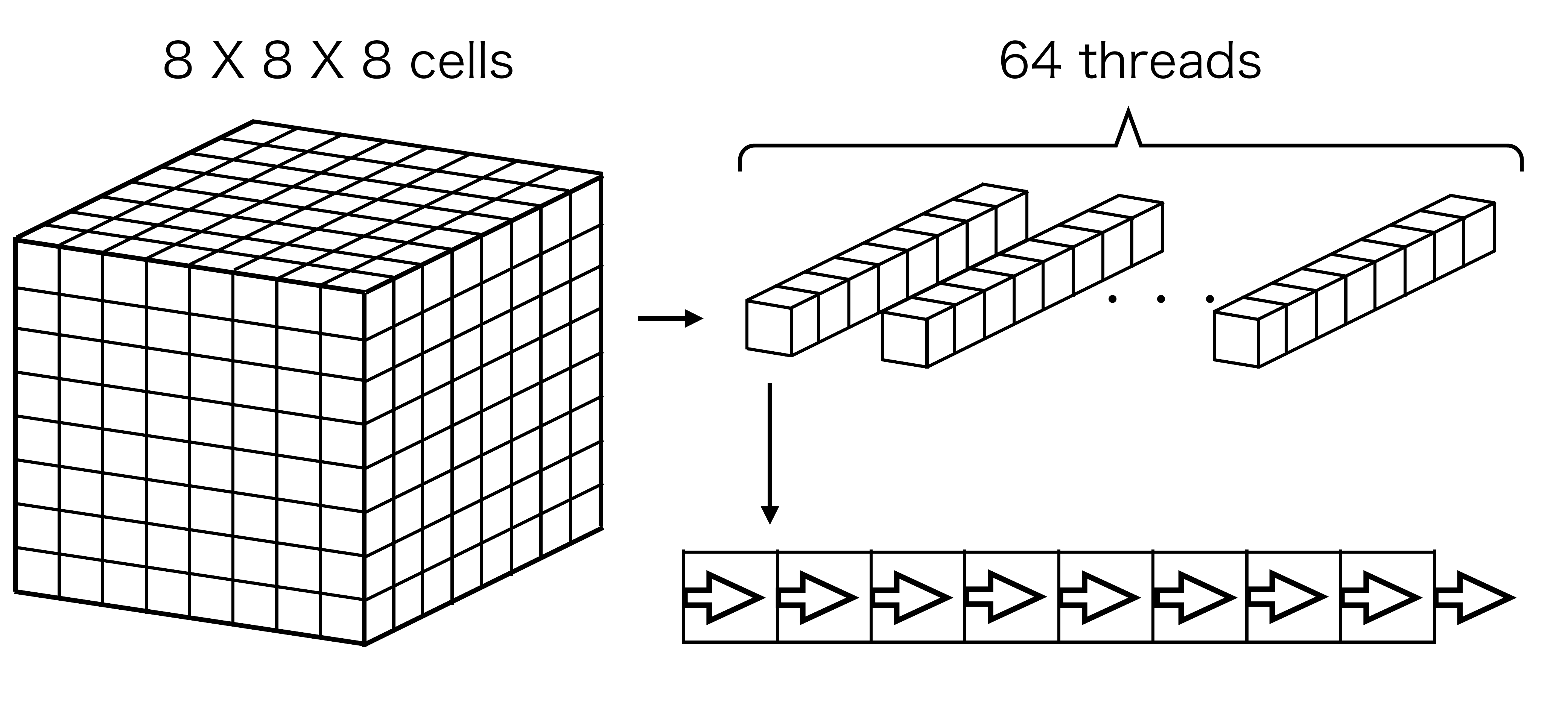}
\end{center}
 \caption{
Parallelization of numerical flux calculations at cell interfaces within a block.
Each block contains $8^3$ cells, and the computation is parallelized using 64 threads for each group of 8 cells aligned along the update direction.
The thick arrows indicate the numerical fluxes at the cell interfaces.
{Alt text: Schematic illustration showing the strategy for calculating numerical fluxes in our code. }
 }
 \label{fig:grid_flux}
\end{figure*}

The hydrodynamics equations \eqref{eoc}-\eqref{energy_eq} are solved on a block-structured grid, omitting the terms on the right-hand sides.
We employ a monotone upstream-centered scheme for conservation laws (MUSCL) in conjunction with a predictor-corrector method for time integration, thereby achieving second-order accuracy in both space and time \citep[e.g.,][]{1990nyjw.book.....H}.
In the hydrodynamics module, the conservative variables at cell centers are updated using numerical fluxes at cell interfaces.
These fluxes are computed with Riemann solvers based on either the HLLC scheme \citep{toro1994restoration} or the method of \citet{1981JCoPh..43..357R}.

In the update of hydrodynamics, interpolation of variables in ghost cells and reflux are also required.
Each block contains ghost cells that overlap with neighboring blocks.
The values in ghost cells are assigned using spatial and temporal interpolation as described in Section 5.3 of \citet{2007PASJ...59..905M}. 
After the update at each level, reflux corrections at interfaces between parent blocks are recalculated \citep{1989JCoPh..82...64B}.

For the hydrodynamics module, GPU parallelization is optimized using the following strategies. Most kernels employ cell-level parallelization. In the calculation of numerical fluxes, it is more efficient to parallelize over groups of cells aligned along the flux direction, as illustrated in Figure \ref{fig:grid_flux}, which allows reuse of the computed fluxes.
In the fiducial configuration, each grid contains $8^3$ cells, and $64$ threads are used to update the conservative variables.
This approach ensures a sufficient number of GPU threads, which is essential for achieving efficient computational performance.

\subsubsection{Self-gravity} \label{sec:self-gravity}

The gravitational potential $\Phi$ and the gravity $\bf g$ are computed using the multigrid method, which is widely used in AMR codes to solve Poisson's equation.
In the multigrid solver, three types of cycles are executed \citep{2007PASJ...59..905M}: (a) FMG-cycle on AMR hierarchy, (b) V-cycle on AMR hierarchy, and (c) FMG-cycle on the coarsened uniform base grid.
In \textsc{SFUMATO\#}, cycles (a) and (b) are executed by all processes, whereas cycle (c) is executed only by the primary-rank process.
Further details of the multigrid algorithm are described in Section 6 of \citet{2007PASJ...59..905M} \citep[see also,][]{2003ApJ...583..296M}.

In adaptive timestep mode, the elapsed time differs across grid levels.
Coarser grid levels use larger timesteps and therefore advance further in time.
To maintain consistency, we apply quadratic temporal interpolation of the density on the coarser grids in the same manner as in the hydrodynamics module.
Specifically, we use density values stored in the hydrodynamics update at the current, previous, and half timesteps, denoted as $\rho^{H,n}$, $\rho^{H,n+1/2}$, and $\rho^{H,n+1}$, where $n$ indicates the time level ($t^n$) and $H$ denotes the coarser grid level.
We note that $\rho^{H,n}$ and $\rho^{H,n+1}$ are second-order accurate in time, whereas $\rho^{H,n+1/2}$ is only first-order accurate.
The interpolated value $\rho^H(t)$ for $t^n \le t \le t^{n+1}$ is given by
\begin{align}
    \rho^H (t) &= \left( \rho^{H,n+1} - \rho^{(1), H, n+1} \right) \left( \frac{t-t^n}{t^{n+1}-t^{n}} \right)^2 \nonumber \\ 
    &+ \frac{(t^{n+1}-t) \rho^{H,n} + (t-t^n) \rho^{(1), H, n+1}}{t^{n+1}-t^n}, \label{eq:interp_rho}
\end{align}
where $\rho^{(1), H, n+1}$ is the first-order accuracy at the time $t^{n+1}$, defined as
\begin{align}
    \rho^{(1), H, n+1} = 2 \rho^{H, n+1/2}-\rho^{H, n}. \label{eq:inter_rho2}
\end{align}

\subsubsection{Sink particles} \label{sink_particle}

The sink particle method is also implemented in this code, following the approach described in \citet{2015ApJ...801...77M}.
Sink particles are introduced according to the creation condition based on \citet{2010ApJ...713..269F}.
The sink radius is defined as half of the Jeans length at the threshold density for sink creation, $\rho_{\rm thr}$.

The mass accretion procedure follows \citet{2010ApJ...724.1006M}.
For cells located within the sink radius, the density in excess of the threshold value $\rho_{\rm thr}$ is removed and added to the sink mass.
For the corresponding cells, we introduce $S_{\rm sink}$ in Equation \eqref{eoc} such that the gas density is reduced to the threshold value $\rho_{\rm thr}$.
Simultaneously, the sum of the removed momentum from cells is also added to the sink particle.

Sink particles move as Lagrangian particles under the influence of momentum gained through accretion and the gravitational forces from both the gas and other sink particles.
The gravitational force from the gas acting on a sink particle is evaluated by averaging the gravity over a cube of side length $r_{\rm sink}$ centered on the sink,
\begin{align}
    {\bf g}_{\rm gas} = \sum_n W_n {\bf g}({\bf x}_n), \label{eq:grav}
\end{align}
where ${\bf g}({\bf x}_n)$ is obtained from Poisson's equation \eqref{eq_poission} and \eqref{eq_gravity}.
The kernel function $W_n$ is defined as $W_n = V_{\rm overlap}/V_{\rm cube}$, where $V_{\rm overlap}$ is the overlapping volume between each cell and the cube volume $V_{\rm cube}$.
The gravitational forces of the $j$th sink from other sink particles are given by
\begin{align}
    {\bf g}_{\rm sink}({\bf r}_{j}) = \sum_{i \neq j} {\bf g}_{{\rm sink}, i}({\bf r}_{j}, {\bf r}_{i}). \label{eq:grav_sink}
\end{align}
The contribution from the $i$th sink is 
\begin{align}
    {\bf g}_{{\rm sink}, i}({\bf r}_{j}, {\bf r}_{i}) = 
    \begin{cases}
    \displaystyle
        - \frac{G m_i}{|{\bf r}_{j} - {\bf r}_i|^3} \left( {\bf r}_{j} - {\bf r}_i \right) \, ({\rm for} \, |{\bf r}_{j} - {\bf r}_i| \ge r_{\rm soft}) \\
    \displaystyle   - \frac{G m_i}{r_{\rm soft}^3} \left( {\bf r}_{j} - {\bf r}_i \right) \, ({\rm for} \, |{\bf r}_{j} - {\bf r}_i| < r_{\rm soft})
    \end{cases}, \label{eq_grav_sink}
\end{align}
with $m_i$ the mass of the $i$th sink and the softening length $r_{\rm soft}$ set to $r_{\rm soft} = r_{\rm sink}$.
The gravitational force exerted by sink particles on the gas, represented by $\bf{g}_{\rm sink,g}$ in Equations \eqref{eom} and \eqref{energy_eq}, is given by
\begin{align}
    {\bf g}_{\rm sink, g}({\bf r}_{g}) = \sum_{i} {\bf g}_{{\rm sink}, i}({\bf r}_{g}, {\bf r}_{i}), \label{eq:grav_flud_from_sink}
\end{align}
where ${\bf r}_{g}$ denotes the position of each gas cell.

The computational costs of Equations \eqref{eq_grav_sink} and \eqref{eq:grav_flud_from_sink} scales as $\mathcal{O}(n_{\rm sink}^2)$, where $n_{\rm sink}$ is the total number of sink particles.
For example, in massive star cluster formation, the number of stars can exceed $10^5$ \citep[e.g.,][]{2023MNRAS.524.1422F}, making direct evaluation prohibitively expensive.
To avoid this calculation becoming a bottleneck, we also implement the following option.
In Equations \eqref{eq_grav_sink} and \eqref{eq:grav_flud_from_sink}, only sink particles on the maximum grid level, which correspond to growing stars, are treated explicitly.
The contributions of the other sink particles are incorporated into the density distribution of equation \eqref{eq_poission},
 and their gravitational potential and forces are computed simultaneously with the gas using the multigrid method. 
The sink particle density field is constructed with the cloud-in-cell (CIC) interpolation scheme \citep{1981csup.book.....H}.
The sink mass is interpolated onto the cell density using the same kernel function $W_n$ as in equation \eqref{eq:grav}. In this way, the contributions from both gas and sink particles can be evaluated simultaneously through equation \eqref{eq:grav}.
This approach reduces the computational cost to $\mathcal{O}(n_{\rm sink}n_{\rm sink, max})$, where $n_{\rm sink, max}$ is the number of sink particles at the maximum grid level.
Since $n_{\rm sink, max} \ll n_{\rm sink}$ in general, this method significantly reduces the cost of sink dynamics calculation.  

\subsubsection{Radiation source} \label{sec:radiation_sorce}

We estimate the luminosity of sink particles using the pre-calculated stellar evolution tables.
We employ the stellar evolution models of \citet{2009ApJ...691..823H} \citep[see also,][]{2018MNRAS.473.4754F}, which provide 
the stellar radius $R_*$ and luminosity $L_*$ as functions of stellar masses, accretion rates, and stellar metallicity.
The emissivities of EUV, FUV, and IR photons, and their mean energy, are derived from the SED models of \citet{1997A&AS..125..229L} and \citet{2019A&A...621A..85H} for OB-stars.

Since the tabulated stellar models do not account for accretion history, a rapid variation in the accretion rate can result in spurious stellar evolution.
To mitigate this, we adopt the accretion rate averaged over the characteristic stellar evolutionary timescale $t_{\rm evol}$, defined as
\begin{align}
    t_{\rm evolv} = \min(t_{\rm KH}, t_{\rm acc}, t_{\rm life}), \label{eq:evolv_time}
\end{align}
where $t_{\rm KH} = (GM_*)/(R_*L_*)$ is the Kelvin-Helmholtz time-scale, $t_{\rm acc} = M_* /\dot M_*$ is the accretion time-scale, and $t_{\rm life}$ denotes the age of the sink particle \citep{2020MNRAS.497..829F}.
The mean accretion rate of each sink particle is updated as 
\begin{align}
    \dot M_{\rm mean}^{n+1} = \frac{\dot M_{\rm mean}^{n} f_* t_{\rm evolv} + \Delta m}{ f_* t_{\rm evolv} + \Delta t}, \label{eq:mdot_mean}
\end{align}
where $\dot M_{\rm mean}^{n+1}$ and $\dot M_{\rm mean}^{n}$ are the mean accretion at each time step, $\Delta m$ is the accreted mass, and $\Delta t$ is the timestep.
We adopt $f_* = 0.1$, which sets the effective averaging timescale for accretion rate updates.

\subsubsection{Thermochemical processes} \label{sec_chemistry}

We evaluate the right-hand side terms of equations \eqref{eom}-\eqref{eq_radmoment2}, \eqref{eq_chemq}, and \eqref{eq_dust_temp} after computing the flux terms.
In the previous code of \citet{2021MNRAS.506.5512F}, the equations for chemistry, dust temperature, and IR radiation were updated using the NR method.
However, this approach causes warp divergence on GPUs.
To avoid computational delay, we instead adopt the LI method without iterative calculations.
In the following, we describe the strategies implemented in our code.
Hereafter, we refer to the solver that simultaneously integrates the non-equilibrium chemistry and the energy equation as the thermochemical solver.

\textbf{Chemistry network}: We solve non-equilibrium reactions of 4 species: $\rm H$, $\rm H_2$, $\rm H^+$, $\rm e$.
This chemical network is based on \citet{2010ApJ...722.1793O}, where $\rm H^{-}$ is treated as an intermediate species.
We additionally include $\rm H$ photoionization, $\rm H_2$ photodissociation, and cosmic-ray ionization.
Details of the network are provided in Appendix \ref{apd:chemical_network}. 
The equations of the chemical species are stiff, and thus we update their abundances using the following scheme.
Discretization of equation \eqref{eq_chemq} is written as
\begin{align}
  y_{i}^{t + \Delta t} - y_{i}^{t} = R_{i}^{t+\Delta t} \Delta t, \label{eq_chem2}
\end{align}
where the reaction rates at the advanced timestep are estimated as
\begin{align}
  R_{i}^{t+\Delta t} = R_{i}^{t} + \sum_{j} \frac{\partial R_i}{\partial y_j}  \Delta y_j, \label{eq_chem3} 
\end{align}
with
\begin{align}
  \Delta y_j = y_j^{t +\Delta t} - y_j^{t}. \label{eq_y}
\end{align}
Assuming $\Delta y_j \ll y_j$, substitution of equation of \eqref{eq_y} into equation \eqref{eq_chem3} yields the equation 
\begin{align}
  \sum_j \left( \delta_{ij} - \frac{\partial R_i}{\partial y_j} \Delta t \right) \Delta y_j = R_i^{t} \Delta t, \label{eq_deq}
\end{align}
where $\delta_{ij}$ is the Kronecker delta function.
Solving equation \eqref{eq_deq} gives $\Delta y_i$ and hence $y_i^{t + \Delta t}$.
We update only the abundances of  $\rm H_2$ and $\rm e$, while the abundances of $\rm H$ and $\rm H^+$ are determined from conservation of hydrogen nuclei and charge neutrality.

\textbf{Dust temperature:} 
Dust temperature is typically determined by solving the energy balance equation with iterative methods.
However, such iterative methods are unsuitable for GPU computation.
Instead, we evaluate the dust temperature using a scheme by introducing a pseudo specific heat capacity, $C_{\rm d}$, as in equation \eqref{eq_dust_temp}.
The realistic value of $C_{\rm d}$ is of order $10^4~{\rm erg \, g^{-1} \, K^{-1}}$ at solar metallicity \citep{1985A&A...144..147L}, after accounting for the dust-to-gas mass ratio \citep[e.g.,][]{2011piim.book.....D}.
Because this value is extremely small, the timescale over which energy balance is established on dust grains is correspondingly short.
As a result, the dust temperature is usually derived by assuming an instantaneous energy balance.
In the LI method, however, the timestep for the chemistry and thermal processes is constrained by $C_{\rm d}$. 
Using the realistic value, therefore, requires a large number of timesteps to advance the calculation.
For this reason, it is necessary to adopt an appropriate value of $C_{\rm d}$ that avoids excessively small timesteps.
In Section \ref{sec:BE}-\ref{Sec:appropriate_value_dustcap}, we investigate choices of $C_{\rm d}$ using a series of test problems.
The source terms of the dust temperature, $\mathcal{E}_{\rm d}$ in equation \eqref{eq_dust_temp}, include: (1) dust thermal emission, (2) absorption of UV photons, (3) absorption of IR photons, (4) energy transfer between gas and dust grains.
The details of the method are provided in Appendix \ref{apd:dust_temp}.

\textbf{Radiation components:}
We account for the absorptions of EUV and FUV photons after updating the chemical abundances and the gas and dust temperatures. 
Details of them are given in Appendix C of \citet{2021MNRAS.506.5512F}.
In addition, we update the energy density of IR radiation ($E_{\rm IR}$).
In high-density gas, IR photons strongly couple with the thermal emission of dust grains and energy transfer between gas and dust grains.
Thus, we adopt the method described in Appendix \ref{apd:dust_temp} to update the IR radiation energy density.
We note that photon injection from radiation sources is incorporated at the beginning of each hydrodynamic time step.

\textbf{Time stepping:} We update the source terms of \eqref{eom}-\eqref{eq_radmoment2}, \eqref{eq_chemq}, and \eqref{eq_dust_temp} using a subcycling time step. 
In high-density gas, the equations governing dust temperature form a stiff system of equations.
To avoid prohibitively small time steps, we proceed as follows.
First, we estimate the timescales of $T_{\rm d}$, $T_{\rm g}$, and $E_{\rm IR}$ as
\begin{align}
  \Delta t_{\rm dust} = \min \left(\frac{T_{\rm d}}{\dot T_{\rm d}}, \frac{T_{\rm g}}{\dot T_{\rm g}}, \frac{E_{\rm IR}}{\dot E_{\rm IR}} \right), \label{eq:timescale_of_dust}
\end{align}
where we only consider the energy transport between gas and dust grains for gas temperature $T_{\rm g}$.
If the number of subcycles required is less than the subcycle limit $(N_{\rm cyc})$, we update all components using subcycling. 
Here, we adopt $N_{\rm cyc} = 300$.
In this case, the timescales of chemical reactions and the energy equation ($\Delta t_{\rm chem}$ and $\Delta t_{\rm energy}$) are estimated as
\begin{align}
    \Delta t_{\rm chem} = \min \left ( \frac{y_{\rm H_2}}{R_{\rm H_2}}, \frac{y_{\rm e}}{R_{\rm e}}  \right), \label{eq:timescale_chemi}
\end{align}
\begin{align}
    \Delta t_{\rm energy} = \frac{P}{(\gamma_{\rm ad -1})|\Gamma- \Lambda|}. \label{eq:timescale_energy} 
\end{align}
We then update each equation with a time step
\begin{align}
    \Delta t = f_{\rm chem} \min\left( \Delta t_{\rm dust}, \Delta t_{\rm chem}, \Delta t_{\rm energy} \right). \label{eq:timestep_subcycle}
\end{align}
Here, $f_{\rm chem}$ is a parameter that controls the time-step size of the thermochemical solver.
Unless otherwise stated, we adopt $f_{\rm chem} = 0.03$ throughout this paper.
The update schemes for $T_{\rm d}$ and $E_{\rm IR}$ are described in Appendix \ref{apd:dust_temp}. 
In this procedure, iterative calculations are required only when gas and dust grains are tightly coupled.
If cell data are allocated in host memory as unified memory, the CPU can also be employed to update these equations.
We have implemented a mode in which the time evolution of chemical and thermal processes is performed on the CPU as well.
In that case, we measure the average update time per cell on both CPU and GPU, and assign cells dynamically according to their relative performance.

To enable a direct comparison with the LI method, we also implemented a solver that integrates the non-equilibrium chemistry and energy equations using the NR method.
In the NR method, the iterations are divided into three groups to improve the convergence: (i) the gas temperature ($T_{\rm g}$), (ii) the dust temperature ($T_{\rm d}$) and the IR radiation energy density ($E_{\rm IR}$), and (iii) the chemical species.
The iteration loops for groups (ii) and (iii) are embedded within the iteration loop for the gas temperature.
In the NR update for the gas temperature, the numerical derivative is evaluated using a central difference scheme.
Consequently, the iteration loops for groups (ii) and (iii) are executed three times per gas-temperature iteration, corresponding to the evaluations required for the trial value of $T_{\rm g}$ and for the numerical derivative.
Convergence is achieved when the relative error of each physical quantity becomes smaller than $10^{-5}$.
For the time integration, the hydrodynamics timestep ($\Delta t_{\rm hydro}$) is normally adopted.
However, if convergence is not achieved within the maximum allowed number of iterations ($N_{\rm itr,limit}$), a substepping procedure is applied. 
In this case, the timestep is reduced as 
\begin{align}
    \Delta t_{\rm sub} = \frac{\Delta t_{\rm hydro} - t_{\rm sub}}{2^i}, \label{eq:substep_NRchem}
\end{align}
where $t_{\rm sub}$ denotes the elapsed time within the current substepping procedure.
The initial value of $i$ is set to 1.
If convergence is achieved in a given substep, the next substep adopts $i = i_{\rm prev}-1$, where $i_{\rm prev}$ is the value of $i$ in the previous trial iteration, and $\Delta t_{\rm sub}$ is recalculated using Equation \eqref{eq:substep_NRchem}.
If convergence is not achieved,  $i = i_{\rm prev}+1$ is used, and the timestep is further reduced accordingly.
Furthermore, if the number of gas-temperature iterations within a substep exceeds $N_{\rm itr,limit}/2$, the solution is considered difficult to converge. 
In such cases, the convergence criterion is relaxed to $10^{-3}$, and the calculation is continued.
In the present implementation of the NR method, we adopt $C_{\rm d} = 10^4~{\rm erg \, g^{-1} \, K^{-1}}$ and $N_{\rm itr,limit}=20$.
This solver can be executed on both GPUs and CPUs when unified memory is employed.

\section{TEST PROBLEMS} \label{sec:test_problems}

In this section, we present a series of numerical tests to validate the \textsc{SFUMATO\#} code.
The hydrodynamic solver is validated using the shock tube test (Section \ref{sec:shock_tube}) and the double Mach reflection test (Section \ref{sec:double_mach_refrection}).
The multigrid method for self-gravity is examined in Section \ref{sec:self_gravity}.
Radiation transport with M1-closure is validated in Section \ref{sec:Radiation_transport}.
In Section \ref{sec:test_problem_chem}, we assess the LI method implemented in the chemistry solver through time-evolution tests of chemical abundances.
In Section \ref{sec:BE} and \ref{sec:HII_region}, we further verify the accuracy of the LI method through simulations of Bonnor-Ebert sphere collapse and H{\sc ii} region formation.
We also evaluate the performance of the thermochemical solvers in these sections.
Performance measurements are carried out on an NVIDIA GH200 GPU and an Intel Xeon Platinum 8468 CPU (48 cores).
Throughout this section, we employ the HLLC Riemann solver for the hydrodynamics solver and use FP64 for all real-valued quantities.

\subsection{Shock tube} \label{sec:shock_tube}
\begin{figure} 
\begin{center}
  \includegraphics[width=\columnwidth]{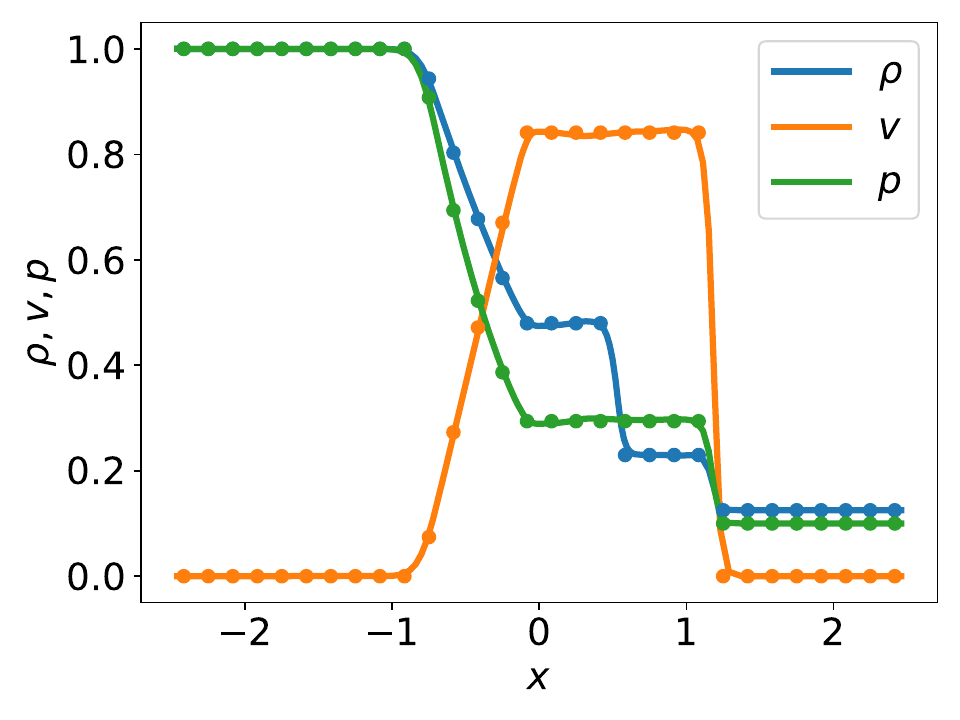}
\end{center}
 \caption{
 Density ($\rho$, blue), velocity ($v$, orange), and pressure ($p$, green) in the shock tube test. 
 Lines show the results of numerical simulations, and dots represent the exact solutions of the shock tube test.
 {Alt text: Three-line graph showing density, velocity, and pressure in the shock tube test.}
 }
 \label{fig_snap}
\end{figure}

We consider the shock tube problem as a test of hydrodynamics.
The initial condition consists of two stationary fluid states separated by a discontinuity at $x=0$.
On the left-hand side, we set $\rho_{\rm L} = 1$ and $p_{\rm L} = 1$, while on the right-hand side we adopt $\rho_{\rm R}=0.125$ and $p_{\rm R} = 0.1$.
The adiabatic index is $\gamma_{\rm ad} = 5/3$.
The problem is solved on a three-dimensional uniform grid with $32$ cells on each side.
Figure \ref{fig_snap} shows the numerical results for density, velocity, and pressure at $t = 0.63$.
The solid line represents the exact solution of this problem.
A comparison with the exact solutions demonstrates that the numerical results are in good agreement.

\subsection{Double Mach Reflection} \label{sec:double_mach_refrection}

We consider the double Mach reflection problem \citep{1984JCoPh..54..115W}, a standard test for hydrodynamics codes.
This test is also performed in Section 8.3 of \citet{2007PASJ...59..905M}.
A planar shock with Mach number $10$ propagates through a uniform medium with $\rho = 1.4$, $P = 1$, and $\gamma = 1.4$.
At the lower boundary of the computational box, a rigid wall exists for $1/6 \leq x \leq 4$, and the incident angle between the shock and the wall is $60^\circ$.
The computational domain spans $0 \leq x \leq 4$ and $0 \leq y \leq 1$, and is initially covered by $32 \times 8 \times 1$ blocks at level $l = 0$. 
We set the maximum refinement level at $l_{\rm max} = 4$, with each grid containing $8^3$ cells.
The minimum and maximum resolutions are $h = 1/64$ and $1/1024$, respectively.
Refinement is triggered according to the criterion
\begin{align}
    \max \left[ \mathcal{E}(\rho_{\rm i,j,k}),  \mathcal{E}(P_{\rm i,j,k})\right] \geq 10^{-2}, \label{eq:dmr_criterion},
\end{align}
where $\mathcal{E}$ is defined as 
\begin{align}
    \mathcal{E}(q_{\rm i,j,k}) \equiv \frac{|\partial_x^2 q_{i,j,k} + \partial_y^2 q_{i,j,k} + \partial_z^2 q_{i,j,k}|h^2}{q_{i,j,k}}. \label{eq:second_derivative}
\end{align}

\begin{figure} 
\begin{center}
  \includegraphics[width=\columnwidth]{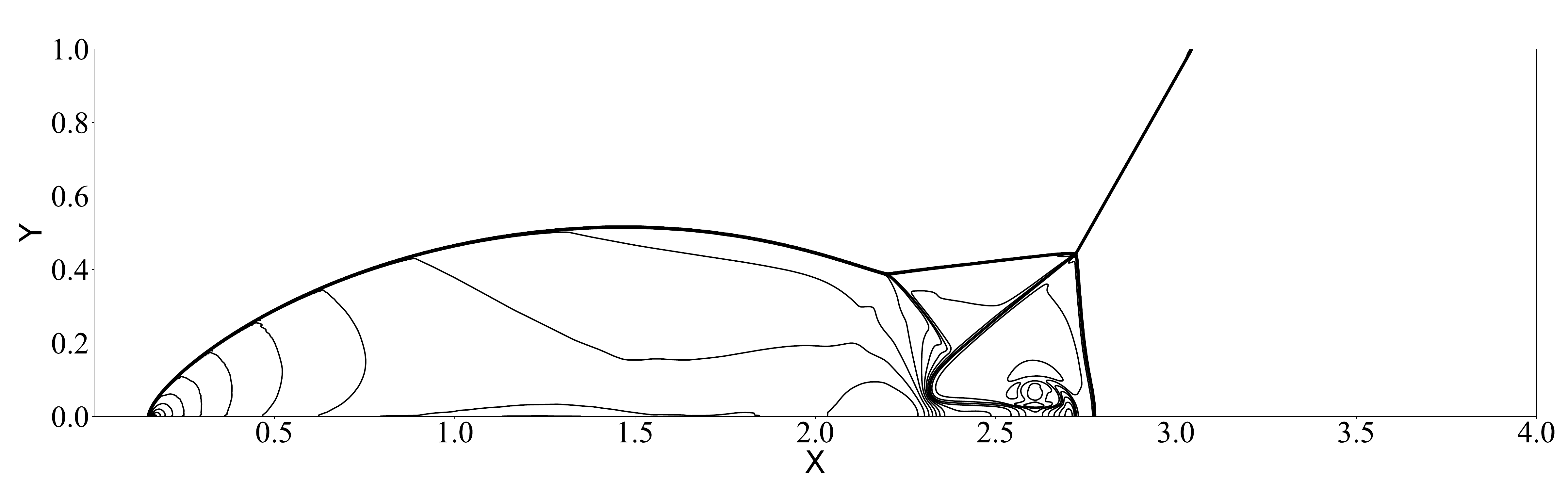}
\end{center}

\begin{center}
  \includegraphics[width=\columnwidth]{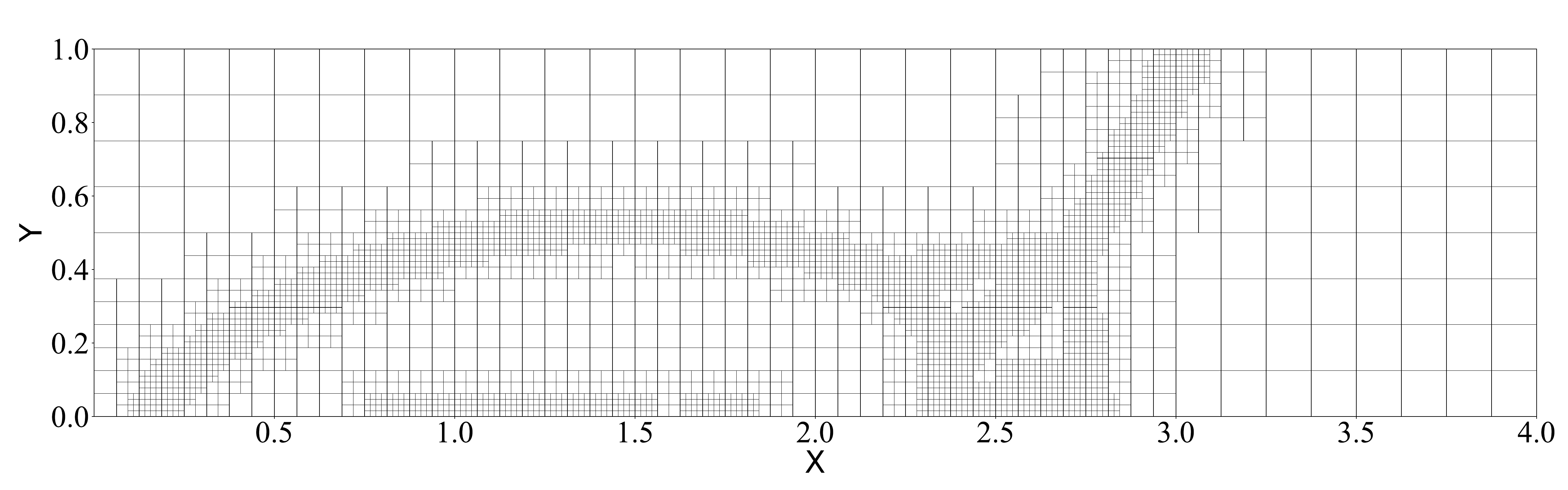}
\end{center}
  \caption{
  Density distribution for the double Mach reflection problem at $t = 0.2$ (top), and the corresponding grid structure (bottom).
  In the bottom panel, the black lines indicate the block boundaries.
  The top panel shows 30 contour lines in the range $2 < \rho < 24$.
  {Alt text: Two-panel figure.}
}
 \label{fig_dmf1}
\end{figure}

Figure \ref{fig_dmf1} shows the density and grid distributions at $t=0.2$.
Grid refinement occurs along the shocked region, but fine grids are effectively placed only in the limited regions.
The gas density smoothly connects across the interface between the coarser and finer blocks, and it indicates that refluxing is functioning correctly. 
An eddy structure appears behind the shock, and the shock structures also resolved well with the finest grid.

\subsection{Accuracy of Multigrid Method} \label{sec:self_gravity}

Next, we present a test of the multigrid method following the approach of \citet{2007PASJ...59..905M}.
Two uniform spheres with masses of $1$ and radii of $6/1024$ are placed at $(x,y,z) = (12/1024, 0, 0)$ and $ (-12/1024, 0, 0)$ within a computation domain $x,y,z \in [-0.5, 0.5]$.
The domain is covered by a nested grid with $8 \times 8 \times 8$ blocks at each level. 
We set the maximum refinement level at $l_{\rm max} = 4$, with each grid containing $16^3$ cells.

\begin{figure} 
\begin{center}
  \includegraphics[width=\columnwidth]{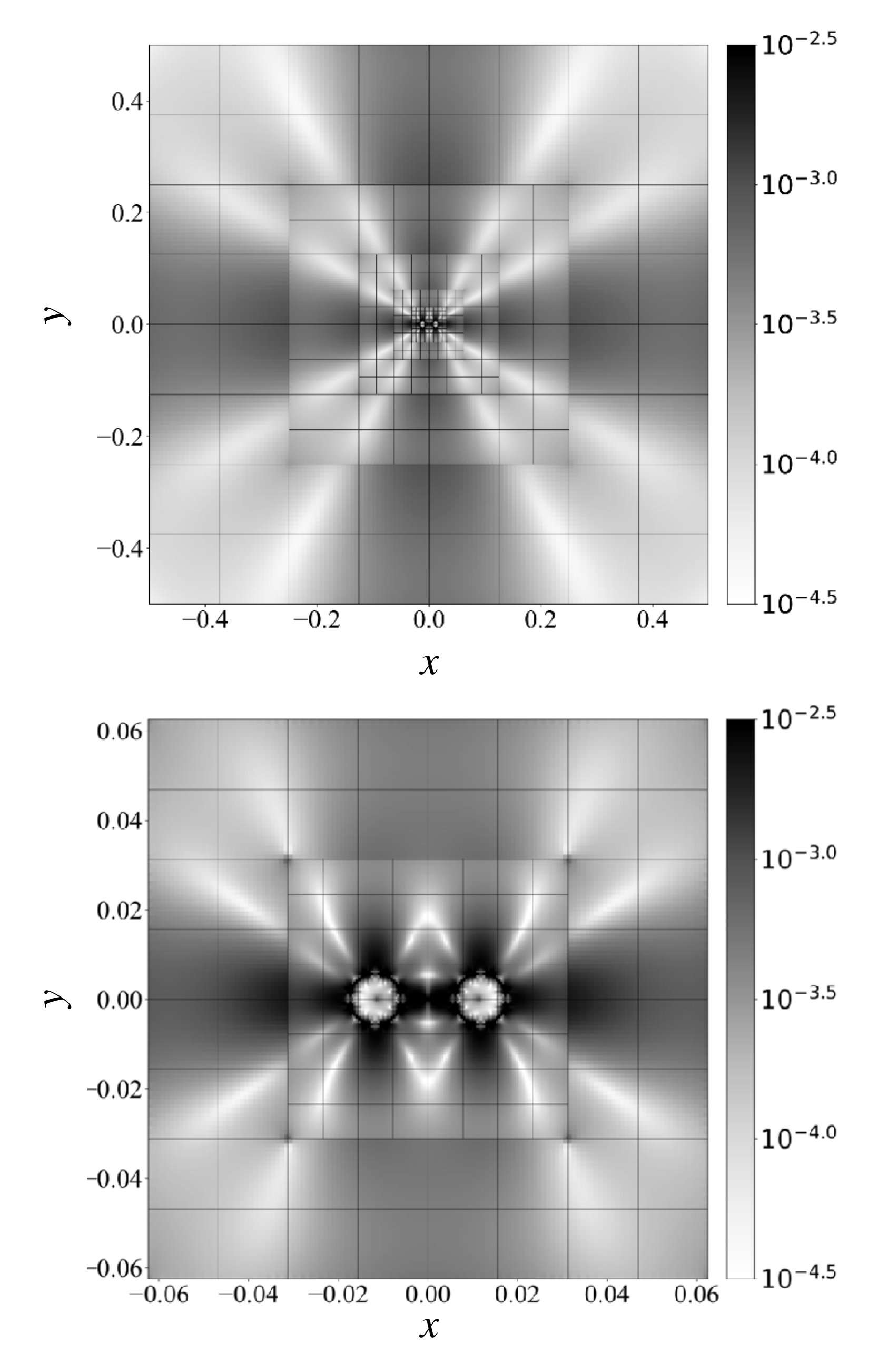}
\end{center}
 \caption{
 Distribution of the relative error, $\log (|{\bf g} - {\bf g_{\rm ex}}|/|{\bf g_{\rm ex}}|)$, of the numerically calculated gravity force in the plane of $z=0$ after $8$ iterations of the FMG-cycles. 
 The top panel shows the results across the entire computational domain at $z=0$, while the bottom panel shows a central region with refinement levels $3\leqq l$.
 The block lines indicate the block boundaries.
 {Alt text: Two-panel figure.}
 }
 \label{fig_grav_error}
\end{figure}

Figure \ref{fig_grav_error} shows the relative error of the numerically computed gravitational force, $|{\bf g} - {\bf g_{\rm ex}}|/|{\bf g_{\rm ex}}|$ where ${\bf g_{\rm ex}}$ denotes the exact gravity force calculated analytically.
The maximum error occurs near the spheres due to the sharp density gradient at their edge.
Elsewhere, the error remains below $10^{-3}$, and the error distribution is smoothly connected across the interfaces between coarser and finer grids.
This indicates that refluxing in the multigrid method is functioning correctly.

\subsection{Radiation transport} \label{sec:Radiation_transport}

Next, we test the radiation transfer of a beam, which is a standard test problem for moment-based approaches \citep{2013MNRAS.436.2188R}.
Photons are injected at a rate of $1.0$ per code unit time from the point $(x,y,z) = (0.1, 0.1, 0.5)$ in the computational domain $x,y,z \in [0, 1]$.
The speed of light is set to $1.2 \times 10^6$ in code units.
Here, we adopt the HLL flux function.
The refinement condition is defined as
\begin{align}
    \frac{|\partial_x^2 E_{r, {i,j,k}} + \partial_y^2 E_{r, {i,j,k}} + \partial_z^2 E_{r, {i,j,k}}|h^2}{E_{r, {i,j,k}}+E_{\rm r, 0}} \geq 10^{-2}, \label{eq:ray_criterion}
\end{align}
where $E_r$ and $h$ are the radiation energy density and cell width.
To ensure refinement occurs only along the ray path, we add $E_{\rm r,0} = 10^{-2}$ (in code units) to the denominator in Equation \eqref{eq:ray_criterion}.

\begin{figure} 
\begin{center}
  \includegraphics[width=\columnwidth]{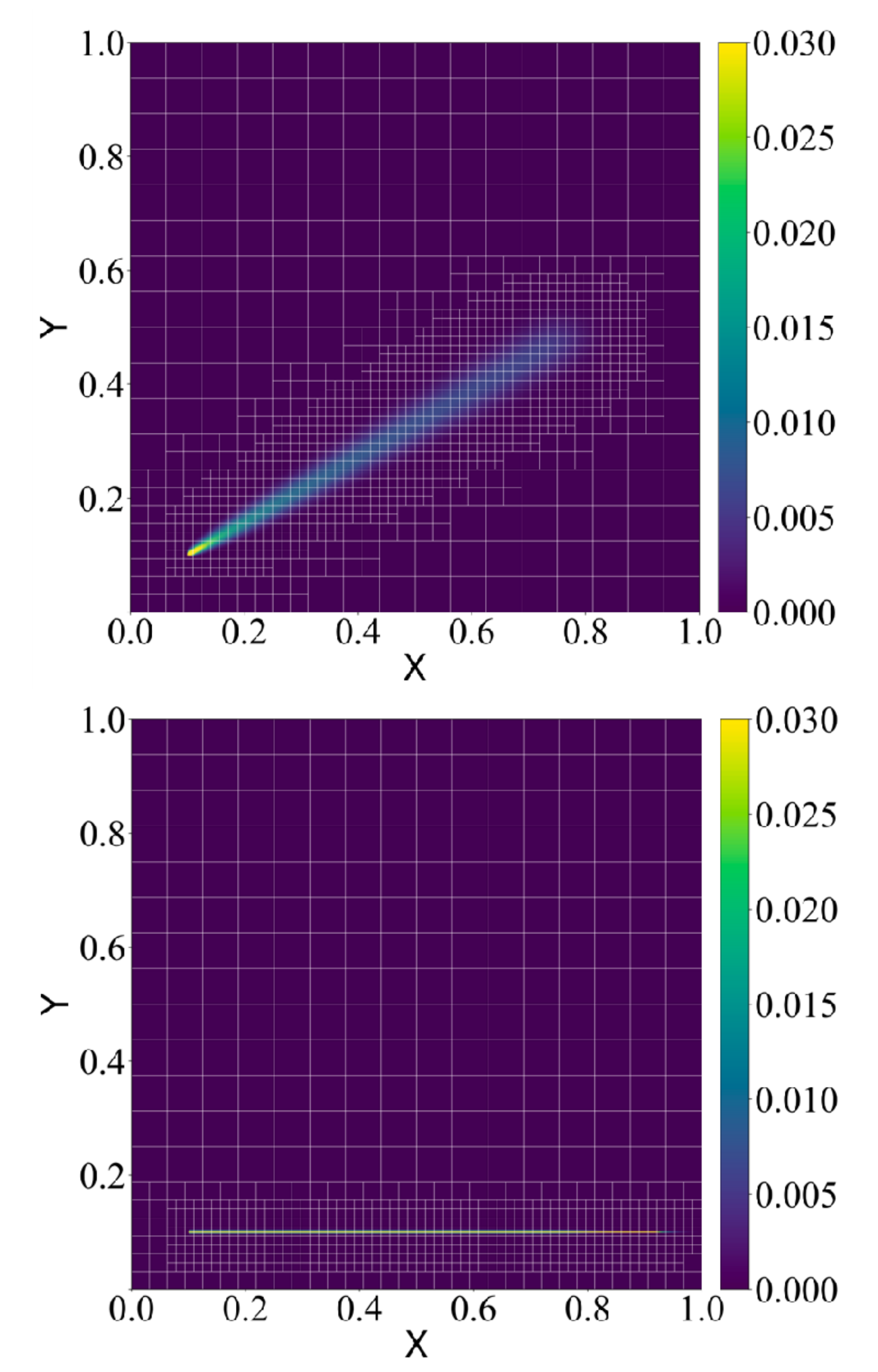}
\end{center}
 \caption{
    Photon densities in the beam experiments at $t=3.2 \times 10^{-7}$. 
    The top panel shows the result for an off-axis beam with an angle of $30^\circ$, while the bottom panel presents the result for a beam aligned with the $x$-axis.
    The white lines delineate the block boundaries.
     {Alt text: Two-panel figure.}
    }
 \label{fig_ray}
\end{figure}

Figure \ref{fig_ray} presents the results of the beam tests.
We consider both an off-axis beam at $30^\circ$ and a beam aligned with the $x$-axis.
In both cases, the beam propagates in a straight line, and the cell refinement is appropriately triggered along the ray.
The horizontal beam remains fully collimated, whereas artificial diffusion of radiation appears in the off-axis beams.
As described in Section \ref{section_basic_equation}, the dependence of accuracy on the propagation direction is a characteristic feature of the HLL flux function \citep[see also,][]{2013MNRAS.436.2188R}.

\subsection{Time evolution of chemical abundances} \label{sec:test_problem_chem}

As described in Section \ref{sec_chemistry}, we adopt a newly developed LI scheme to evolve the abundances of chemical species.
In this section, we assess the accuracy of this scheme by evolving the chemical network while keeping all other variables fixed.
Here, we consider two test problems: molecular hydrogen formation and hydrogen photoionization. 
The initial chemical abundances and other variables are listed in Table \ref{Tab:chemistryTest}, and the metallicity is fixed to $Z=Z_{\odot}$.

\subsubsection{Case 1: Molecular hydrogen formation} \label{sec:chemtest_H2form}

In collapsing star-forming clouds, ${\rm H_2}$ forms predominantly through reactions on dust grain surfaces
$2 {\rm H} + {\rm grain} \rightarrow {\rm H_2}$.
Assuming conservation of atomic and molecular hydrogen abundances, $y_{\rm H} + 2 y_{\rm H_2} =1$, the formation rate of molecular hydrogen is given as
\begin{align}
    \frac{d y_{\rm H_2}}{dt} =  \kappa_{\rm H8} n_{\rm H} y_{\rm H} =  \kappa_{\rm H8} n_{\rm H} \left( 1-2 y_{\rm H_2}\right), \label{eq:h2form}
\end{align}
where $\kappa_{\rm H8}$ is the reaction rate coefficient for $\rm H_{\rm 2}$ formation on grain surfaces (see also Table \ref{Tab:chemical_network}).
The analytical solution for the time evolution of $y_{\rm H_2}$ is
\begin{align}
    y_{\rm H_2} = \frac{1}{2} + \left( y_{\rm H_2, 0}- \frac{1}{2} \right) \exp(-2 \kappa_{\rm H8}n_{\rm H}t), \label{eq:H2form2}
\end{align}
where $y_{\rm H_2, 0}$ is the initial abundance at $t=0$.

\begin{table}
  \caption{Initial condition of the time evolution of chemical abundances.}
  \label{Tab:chemistryTest}
  \centering
  \begin{tabular}{l|cccccc}
    \hline
         case & $n_{\rm H} \, [ {\rm cm^{-3}} ]$ & $T_{\rm g}  \, [ {\rm K}]$ & $T_{\rm d}  \, [ {\rm K}]$ & $y_{\rm H}$ & $y_{\rm H_2}$ & $y_{\rm H^+}$ \\ \hline
       1 & $10^5$ & $10$ & $10$ & $1$ & $10^{-5}$ & $10^{-5}$ \\
       2 & $10^2$ & $12000$ & $75$ & $1$ & $10^{-5}$ & $10^{-5}$ \\
    \hline 
  \end{tabular}
  \begin{minipage}{1 \hsize}
  \vspace{3mm}
    Notes. Column 1: model name, Column 2: gas number density, Column 3: gas temperature, Column 4: dust temperature, Column 5-7: the abundances of $\rm H$, $\rm H_2$, and $\rm H^+$. 
  \end{minipage} 
\end{table}

\begin{figure} 
\begin{center}
  \includegraphics[width=\columnwidth]{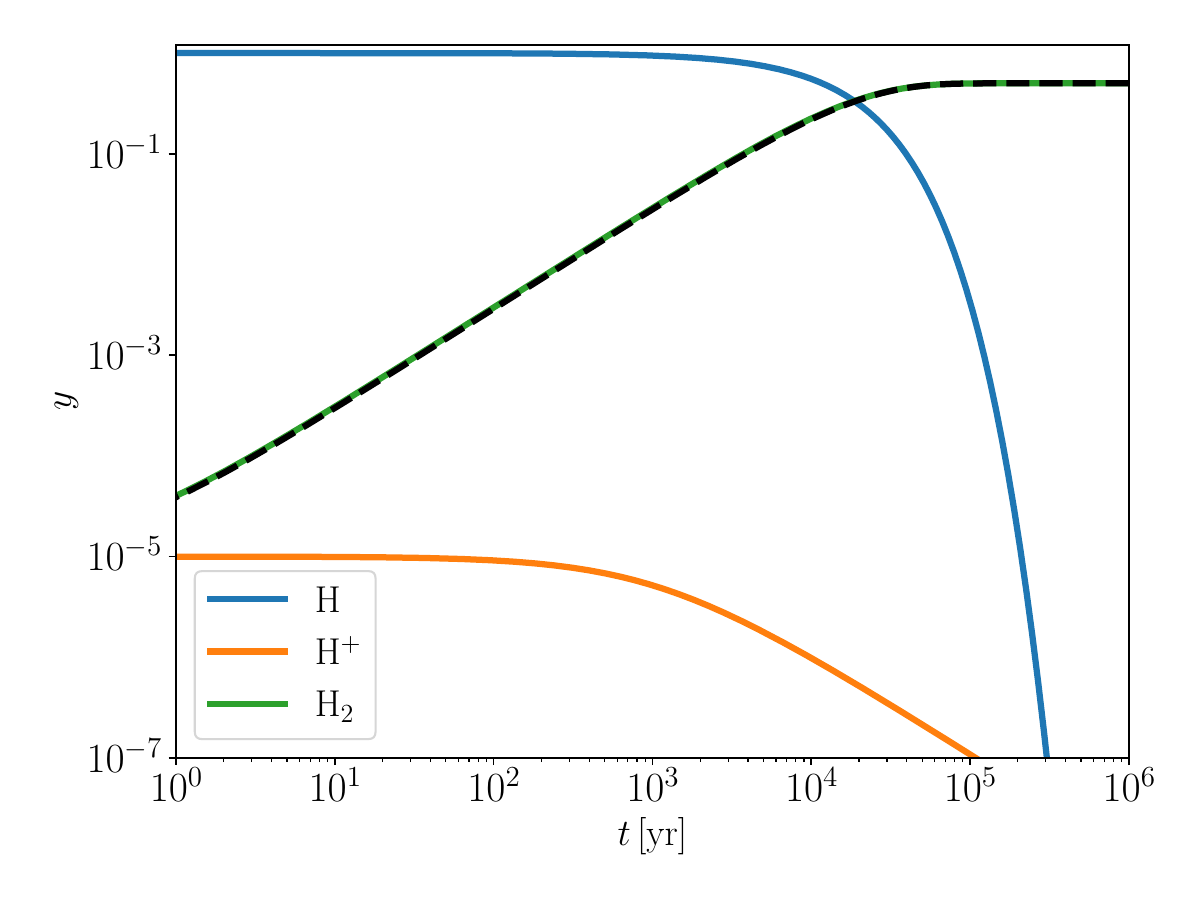}
\end{center}
 \caption{
 The time evolution of chemical abundances in a test problem of molecular hydrogen formation.
 The colored lines show the abundances of $\rm H$ (blue), $\rm H^+$ (orange), and $\rm H_2$ (green).
 The black dashed line shows the analytical solution given by Equation \eqref{eq:H2form2}.
 {Alt text: Time evolution of the abundances of $\rm H_2$, $\rm H$, and $\rm H^+$ shown as three colored lines with the analytical solution given by Equation \eqref{eq:H2form2} overplotted as a black dashed line.}
 }
 \label{fig_chem_H2form}
\end{figure}
\begin{figure} 
\begin{center}
  \includegraphics[width=\columnwidth]{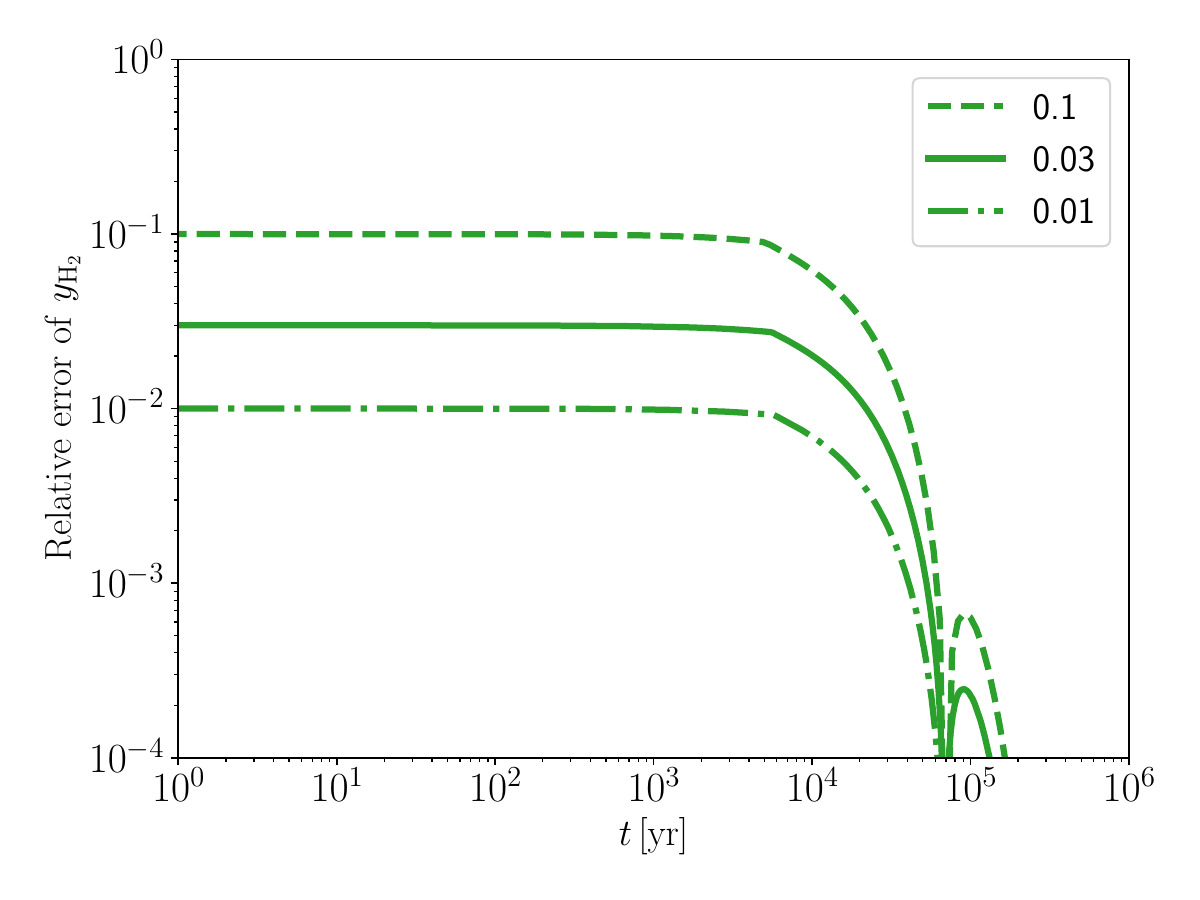}
\end{center}
 \caption{
 Time evolution of the relative error of $y_{\rm H_2}$  between the numerical results and the analytical solution given by Equation~\eqref{eq:H2form2}.
The dashed, solid, and dot-dashed lines show the cases with $f_{\rm chem} = 0.1$, $0.03$, and $0.01$, respectively.
 {Alt text: Graph showing the time evolution of the relative error in the molecular hydrogen abundance for different values of $f_{\rm chem}$.}
 }
 \label{fig_chem_H2form_error}
\end{figure}

As the initial condition, we adopt a fully atomic gas in a star-forming cloud.
We test the new solver by comparing the abundance of molecular hydrogen with the analytical solution given by Equation \eqref{eq:H2form2}.
Figure \ref{fig_chem_H2form} shows the time evolution of the abundances of $\rm H_2$, $\rm H$, and $\rm H^+$.
Here, we adopt the value of $f_{\rm chem} =0.03$, as defined in Equation \eqref{eq:timestep_subcycle}.
Atomic hydrogen is gradually converted into molecular hydrogen, and the gas becomes almost fully molecular at $t\sim 4\times 10^4~{\rm yr}$.
The abundance of $\rm H^+$ decreases due to recombination with electrons.

The black dashed line in Figure \ref{fig_chem_H2form} indicates the $\rm H_2$ abundance predicted by Equation~\eqref{eq:H2form2}.
Figure~\ref{fig_chem_H2form_error} shows the time evolution of the relative error in the molecular hydrogen abundance with respect to the analytical solution.
To examine the dependence of the solution accuracy on $f_{\rm chem}$, we also present results for $f_{\rm chem} = 0.1$ and $0.01$.
At early times, when the molecular hydrogen abundance is small, the relative error is approximately of order $f_{\rm chem}$.
In contrast, after $t \gtrsim 10^4~{\rm yr}$, when molecular hydrogen formation becomes significant, the errors decrease in all cases and fall below $10^{-3}$ for $t \gtrsim 10^5~{\rm yr}$.
From the perspective of computational cost, it is preferable to adopt as large a value of $f_{\rm chem}$ as possible. For $f_{\rm chem} = 0.03$, the relative error drops below $1$ per cent shortly after $t \sim 10^4~{\rm yr}$, indicating that this choice provides a reasonable balance between accuracy and computational efficiency.

\subsubsection{Case 2: Photoionization of atomic hydrogen }\label{sec:chemtest_HIIform}

Next, we test the photoionization of hydrogen.
Assuming conservation of neutral and ionized hydrogen abundances, $y_{\rm H} + y_{\rm H^+} = 1$, the photoionization rate of hydrogen is written as
\begin{align}
    \frac{dy_{\rm H^+}}{dt} = R_{\rm ph} y_{\rm H} = R_{\rm ph} \left( 1-y_{\rm H^+} \right), \label{eq:HIIform}
\end{align}
where $R_{\rm ph}$ is the photoionization rate of hydrogen.
The solution of Equation \eqref{eq:HIIform} is given as
\begin{align}
    y_{\rm H^+} = 1+(y_{\rm H^+, 0}-1)\exp(-R_{\rm ph} t), \label{eq::HIIform2}
\end{align}
where $y_{\rm H^+, 0}$ is the initial ionized hydrogen abundance at $t = 0$.

In an H{\sc ii} region, the equilibrium between photoionization and recombination determines the abundance of neutral hydrogen.
The equilibrium condition is given as 
\begin{align}
    R_{\rm ph}  y_{\rm H}  = \alpha_{\rm rec} y_{\rm H^+} y_{\rm e} n_{\rm H}, \label{eq:HIIeq}
\end{align}
where $\alpha_{\rm rec}$ is the recombination rate coefficient of ionized hydrogen.
In an H{\sc ii} region, $y_{\rm H^+} \sim 1$ and $ y_{\rm e} \sim 1$. 
Substituting these values into Equation \eqref{eq:HIIeq}, we obtain the neutral hydrogen abundance as
\begin{align}
    y_{\rm H} = \frac{\alpha_{\rm rec} }{R_{\rm ph}} n_{\rm H}. \label{eq:yelinHII}
\end{align}

\begin{figure} 
\begin{center}
  \includegraphics[width=\columnwidth]{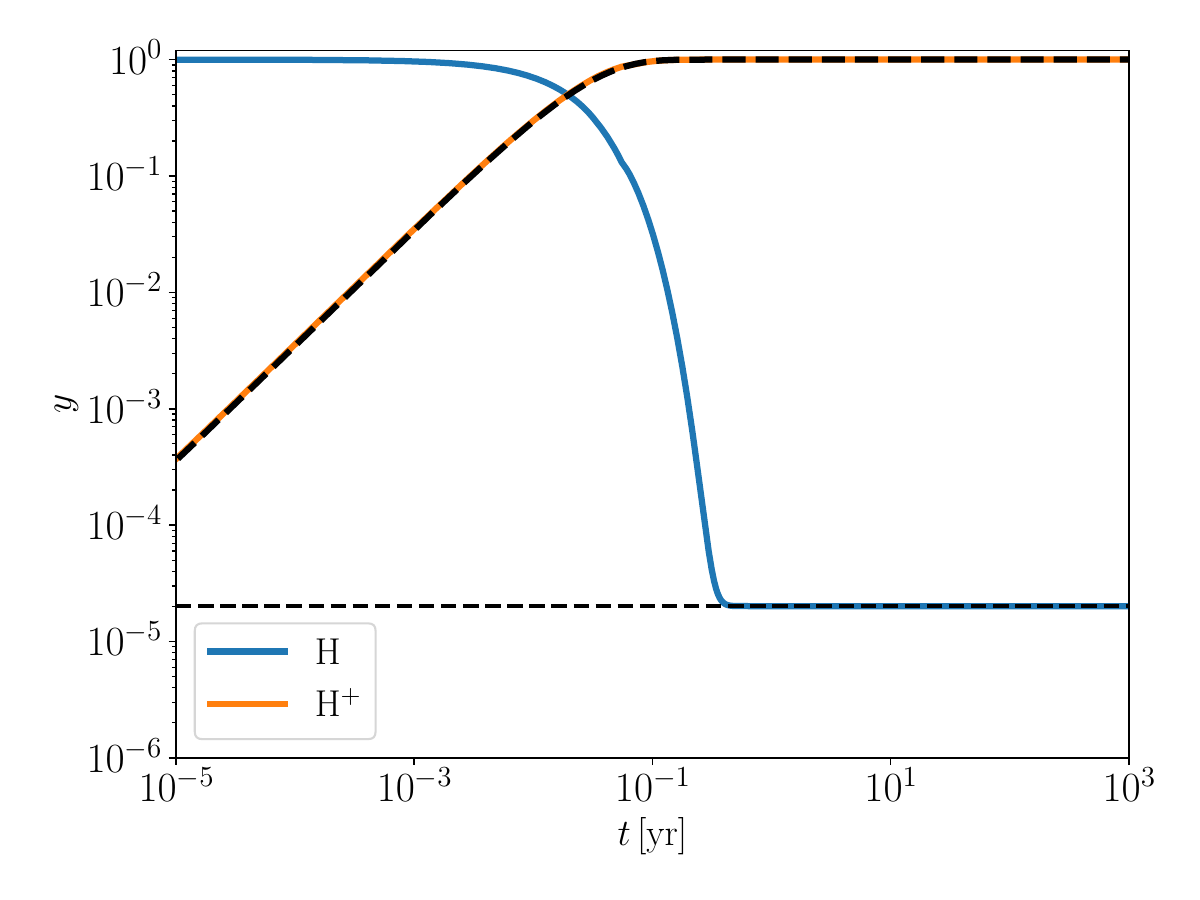}
\end{center}
 \caption{
 The time evolution of chemical abundances in a test problem of hydrogen photoionization.
 The colored lines show the abundances of $\rm H$ (blue) and $\rm H^+$ (orange).
 The black dashed line shows the analytical solution given by Equation \eqref{eq::HIIform2}. 
 The horizontal line shows the hydrogen abundance in an H{\sc ii} region given by Equation \eqref{eq:yelinHII}.
 {Alt text: Time evolution of the abundances of $\rm H$ and $\rm H^+$, shown as colored lines.
 The analytical solution from Equation \eqref{eq::HIIform2} and the hydrogen abundance from Equation \eqref{eq:yelinHII} are overplotted as black dashed and horizontal lines, respectively.}
 }
 \label{fig_chem_HIIform}
\end{figure}
\begin{figure} 
\begin{center}
  \includegraphics[width=\columnwidth]{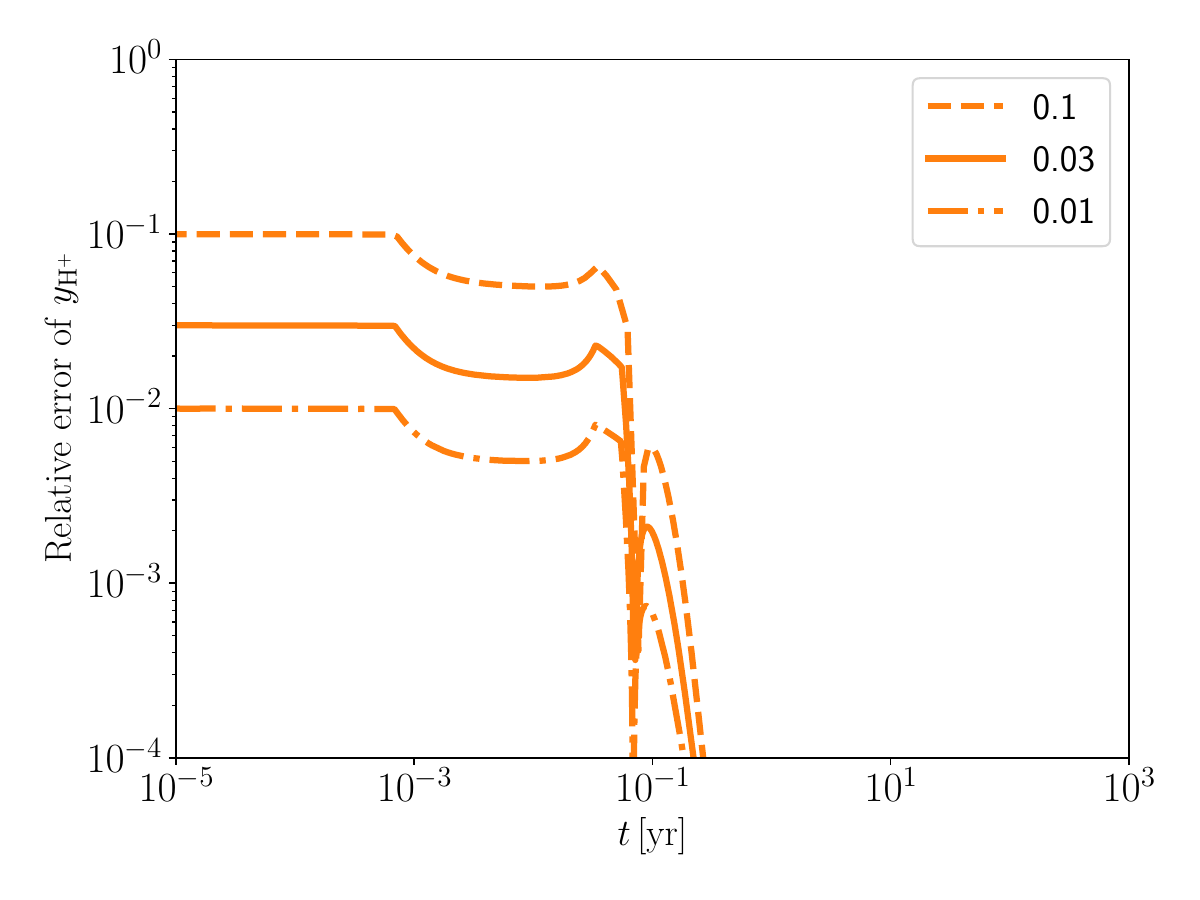}
\end{center}
 \caption{
 Time evolution of the relative error of $y_{\rm H^+}$ between the numerical results and the analytical solution given by Equation~\eqref{eq::HIIform2}.
The dashed, solid, and dot-dashed lines show the cases with $f_{\rm chem} = 0.1$, $0.03$, and $0.01$, respectively.
 {Alt text: Graph showing the time evolution of the relative error in the ionized hydrogen abundance for different values of $f_{\rm chem}$.}
 }
 \label{fig_chem_HII_error}
\end{figure}

In the numerical test, we consider hydrogen photoionization in gas with $n_{\rm H} = 10^2~{\rm cm^{-3}}$.
We adopt a photoionization rate of $R_{\rm ph} = 1.1\times 10^{-6} {\rm s^{-1}}$, which corresponds to the value at a distance of $0.5~{\rm pc}$ from a $40~M_{\odot}$ massive star.
As in Section \ref{sec:chemtest_H2form}, we adopt $f_{\rm chem}=0.03$.
Figure \ref{fig_chem_HIIform} shows the time evolution of the abundances of $\rm H$ and $\rm H^+$.
The abundance of $\rm H^+$ increases due to photoionization, and the gas becomes fully ionized at $t \sim 0.1~{\rm yr}$.
Meanwhile, the abundance of neutral hydrogen approaches a constant value of $2 \times 10^{-5}$, consistent with the value predicted by Equation~\eqref{eq:yelinHII}.

The black dashed line in Figure \ref{fig_chem_HIIform} represents the analytical solution for the ionized hydrogen abundance given by Equation~\eqref{eq::HIIform2}, which is accurately reproduced by the numerical solution.
As in Figure~\ref{fig_chem_H2form_error}, Figure~\ref{fig_chem_HII_error} shows the time evolution of the relative error in the ionized hydrogen abundance with respect to the analytical solution given by Equation~\eqref{eq::HIIform2}.
At early times, when the ionized hydrogen abundance is small, the relative error is of order $f_{\rm chem}$.
However, after $t \gtrsim 0.1~{\rm yr}$, when photoionization becomes significant, the errors decrease rapidly and become very small.
Since this timescale is much shorter than the typical hydrodynamical timescale in star cluster formation simulations, accurate calculations of ionized regions are expected for any choice of $f_{\rm chem}$ as shown in this figure.

\subsection{Collapse of Bonnor–Ebert Spheres} \label{sec:BE}

\begin{table}
  \caption{Initial condition of Bonnor-Ebert spheres.}
  \label{Tab:BEsphere}
  \centering
  \begin{tabular}{ccccc}
    \hline
      $Z \, [Z_{\odot}]$ & $T_{\rm g} \, [{\rm K}]$ & $y_{\rm H_2}$ & $y_{\rm H}$ & $y_{\rm H^{+}}$ \\ \hline
      $1$ & $5$ & $4.98 \times 10^{-1}$  & $4.40\times 10^{-3}$  & $1.27 \times 10^{-7}$ \\
      $10^{-2}$ & $21$ & $3.47 \times 10^{-2}$ &  $9.30 \times 10^{-1}$ & $1.99 \times 10^{-7}$ \\
      $10^{-4}$ & $190$ & $1.18 \times 10^{-3}$ & $9.98 \times 10^{-1}$  &  $8.29 \times 10^{-7}$ \\
    \hline 
  \end{tabular}
  \begin{minipage}{1 \hsize}
  \vspace{3mm}
    Notes. Column 1: metallicity, Column 2: gas temperature, Column 3-5: abundances of hydrogen molecules, hydrogen, and ionized hydrogen. 
  \end{minipage} 
\end{table}

In this section, we derive an appropriate value of the pseudo-specific heat capacity $C_{\rm d}$, that reproduces the solution obtained with the NR method through collapse simulations of critical Bonnor-Ebert (BE) spheres.
We then investigate the dependence of the computational cost of the thermochemical solver on the value of $C_{\rm d}$.
Simulations are performed, including hydrodynamics, self-gravity, thermochemical processes, and radiative transfer of IR photons.

The initial conditions are generated following \citet{2010ApJ...722.1793O}.
The gas temperature and chemical abundances are set to constant values, corresponding to those obtained from a pre-calculated one-zone model at a hydrogen number density of $n_{\rm H} = 10^4~{\rm cm^{-3}}$.
The details of the initial conditions are summarized in Table \ref{Tab:BEsphere}.
In the one-zone model, we adopt the same chemical network as in the numerical simulation, and the dust temperature and chemical abundances are obtained using the NR method.
The initial density distribution is constructed from an isothermal hydrostatic equilibrium sphere, which is then enhanced by a factor of $1.2$ to trigger gravitational collapse.
At the beginning of the simulation, we employ nested grids with a maximum refinement level of $l_{\rm max} = 2$, where each level contains $8^3$ blocks, each block consists of $8^3$ cells.
Blocks are further refined whenever the Jeans length is resolved by fewer than 5 cells.
The critical density for sink particle formation is set to $n_{\rm H} = 10^{11}~{\rm cm^{-3}}$, and the simulations are terminated at the epoch of sink formation.
We adopt the synchronous time-step mode, except in the Section \ref{sec:BE_adaptive_timestep}.

\subsubsection{Accuracy of the LI method in BE sphere tests} \label{sec:BE_LI_method}

\begin{figure} 
\begin{center}
  \includegraphics[width=\columnwidth]{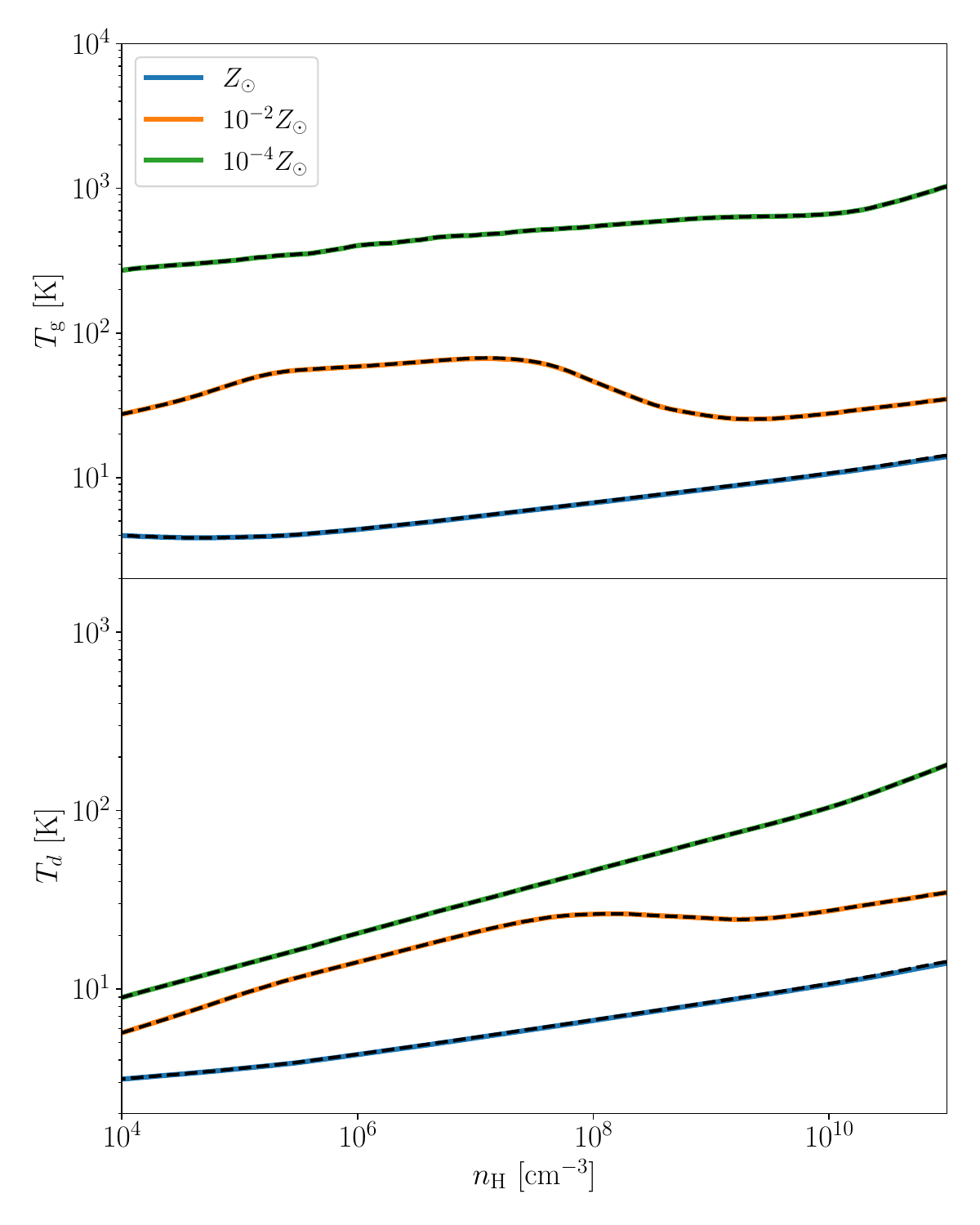}
\end{center}
 \caption{
  Gas temperature (top) and dust temperature (bottom) as a function of number density at the epoch of sink formation in the simulations of Bonnor-Ebert spheres.
  The colored lines represent the results obtained with the LI method using $C_{\rm d} = 10^7~{\rm erg \, g^{-1} \, K^{-1}}$.
  Each color corresponds to a different metallicity: $Z=1~Z_{\odot}$ (blue), $10^{-2}~Z_{\odot}$ (orange), and $10^{-4}~Z_{\odot}$ (green).
  The dashed lines show the results calculated with the NR method.
{Alt text: Graph showing the gas and dust temperature as a function of number density at the epoch of sink formation in collapse tests of Bonnor–Ebert spheres at different metallicities.}
 }
 \label{fig:BET_T}
\end{figure}

We perform simulations using both the LI and NR methods at metallicities of $Z=1~Z_{\odot}$, $10^{-2}~Z_{\odot}$, and $10^{-4}~Z_{\odot}$.
For $Z=1~Z_{\odot}$, we carry out the simulations with $C_{\rm d} = 10^4$, $10^5$, $10^6$, $10^7$, and $10^8~{\rm erg \, g^{-1} \, K^{-1}}$, whereas for the other metallicities we adopt $C_{\rm d} = 10^7 ~{\rm erg \, g^{-1} \, K^{-1}}$.
Figure \ref{fig:BET_T} shows the gas and dust temperature distributions, respectively, as functions of number density at the epoch of sink formation, with the results for $C_{\rm d} = 10^7 ~{\rm erg \, g^{-1} \, K^{-1}}$.
The corresponding results obtained with the NR method are also shown as black dashed lines in these figures.
In a collapsing cloud, the cooling and dynamical timescales are much shorter in the central dense regions.
As a result, the temperature distribution in the outer regions closely follows the temperature evolution of the central region at the epoch when their number densities are equal.
Hereafter, we compare these temperature distributions to evaluate the accuracy of the LI method.

\begin{figure} 
\begin{center}
  \includegraphics[width=\columnwidth]{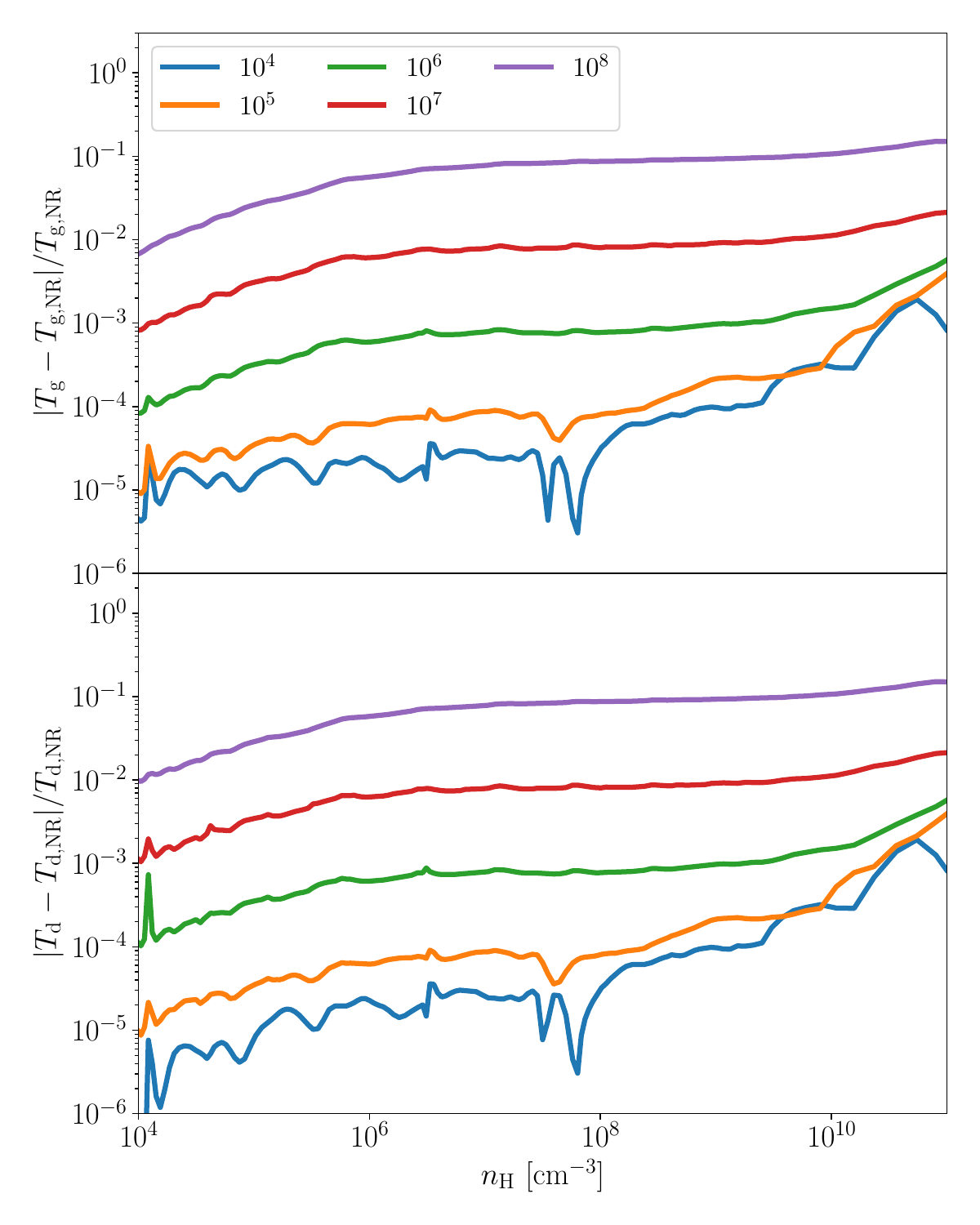}
\end{center}
 \caption{
 Relative errors of the gas temperature (top) and dust temperature (bottom) between the LI and NR methods in the BE sphere tests.
 Each color denotes a different pseudo-specific heat capacity: $C_{\rm d} = 10^4$ (blue), $10^5$ (orange), $10^6$ (green), $10^7$ (red), and $10^8~{\rm erg \, g^{-1} \, K^{-1}}$ (purple).
 {Alt text: Graph showing the relative error in gas temperature between the LI and NR methods for different values of the pseudo-specific heat capacity in the BE sphere tests.}
 }
 \label{fig:BE_T_err}
\end{figure}

Figure \ref{fig:BE_T_err} shows the relative errors in the gas and dust temperature distributions at the epoch of sink formation.
As discussed in Section \ref{sec_chemistry}, a realistic value of $C_{\rm d}$ is of order $10^4~{\rm erg \, g^{-1} \, K^{-1}}$ \citep{1985A&A...144..147L}.
Therefore, we adopt $C_{\rm d} = 10^4~{\rm erg \, g^{-1} \, K^{-1}}$ in the NR method.
Comparing the result of the LI method with $C_{\rm d} = 10^4~{\rm erg \, g^{-1} \, K^{-1}}$ to that of the NR method, the relative errors in both the gas and dust temperatures are of order $10^{-5}$ at $n_{\rm H} \lesssim 10^8~{\rm cm^{-3}}$, which corresponds to the convergence criterion of the NR method.
In this density range, the results obtained with the LI and NR methods can be regarded as consistent with each other.
At $n_{\rm H} > 10^{8} ~{\rm cm^{-3}}$, a discrepancy of order $10^{-3}$ appears.
However, since this difference remains sufficiently small, the LI method still accurately reproduces the results of the NR method for $C_{\rm d} = 10^4~{\rm erg \, g^{-1} \, K^{-1}}$.
When a larger value of $C_{\rm d}$ is adopted, a greater fraction of the energy generated by compressional heating in the collapsing cloud is stored in dust grains.
As a result, the gas temperature decreases, and the relative error increases for larger values of $C_{\rm d}$.
Nevertheless, the relative error remains below 1 percent even for $C_{\rm d} = 10^7~{\rm erg \, g^{-1} \, K^{-1}}$.
The parameter $C_{\rm d}$ determines the timestep of the dust temperature equation (see Appendix \ref{apd:dust_temp}).
Based on these results, when the dust temperature equation limits the global timestep in the LI method, we can increase $C_{\rm d}$ to $10^7~{\rm erg \, g^{-1} \, K^{-1}}$ to reduce the computation cost.
Figure \ref{fig:BE_T_err} also shows the relative errors in the dust temperature.
At $Z=Z_{\odot}$, dust grains are tightly coupled to the gas component, and the gas and dust temperatures are almost identical at $n_{\rm H} > 10^5~{\rm cm^{-3}}$.
Consequently, the relative errors between the LI and NR methods for the dust temperature are similar to those for the gas temperature.

\begin{figure} 
\begin{center}
  \includegraphics[width=\columnwidth]{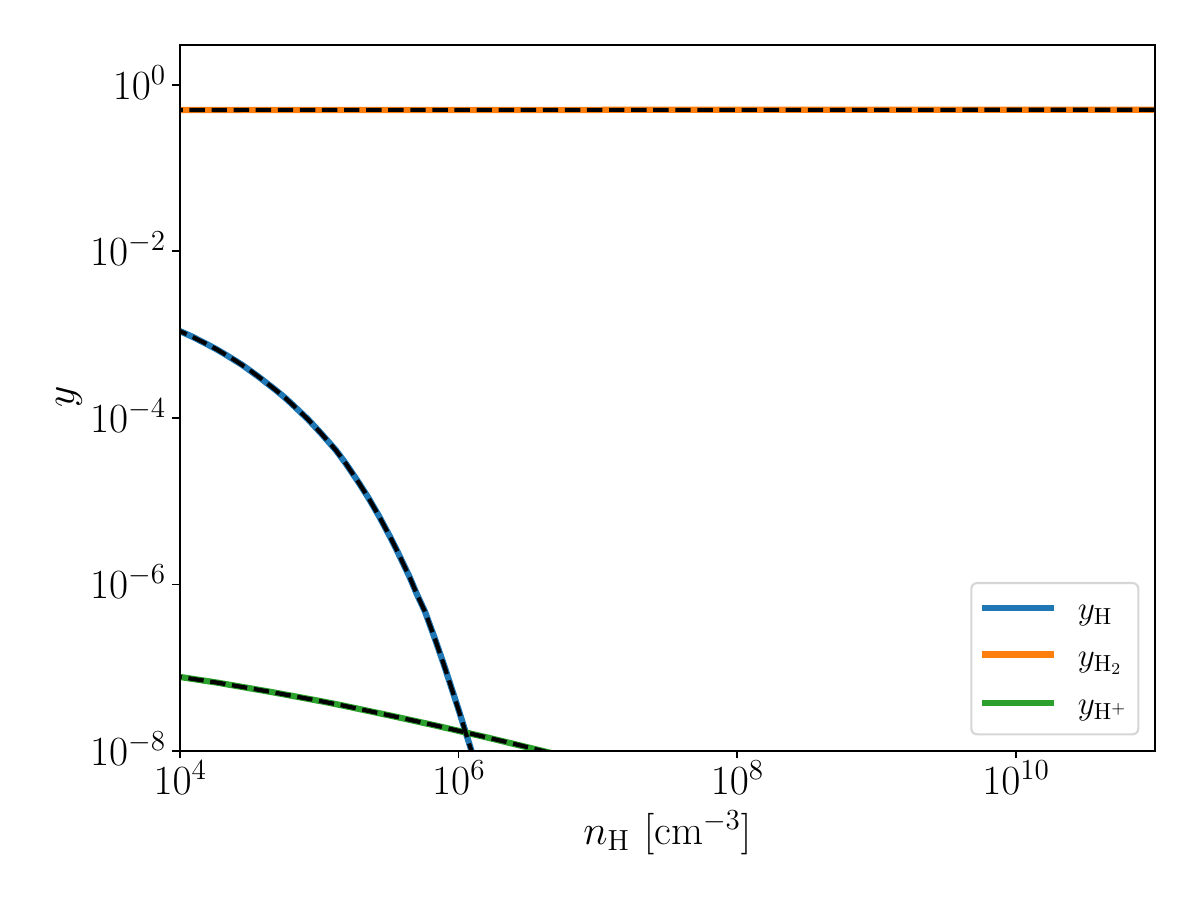}
\end{center}
 \caption{
 The abundance distributions of chemical species for the case with $Z=1~Z_{\odot}$. 
 The colored lines show the results obtained with the LI method using $C_{\rm d} = 10^7~{\rm erg \, g^{-1} \, K^{-1}}$.
 The blue, orange, and green lines correspond to  $\rm H$, $\rm H_2$, and $\rm H^+$, respectively.
 The black dashed lines show the results obtained with the NR method.
{Alt text: Graph showing the abundance distributions of chemical species obtained with the LI and NR methods at  $Z=Z_{\odot}$.}
 }
 \label{fig:chemical_abundance_Z0}
\end{figure}

Figure \ref{fig:chemical_abundance_Z0} shows the abundances of $\rm H$, $\rm H_2$, and $\rm H^+$ obtained using both of the LI and NR methods at $Z=1~Z_{\odot}$.
At this metallicity, gas is almost fully molecular owing to reactions on the surface of dust grains.
The new solver accurately reproduces the NR results, even for the low-abundance species $\rm H$ and $\rm H^+$.

\begin{figure} 
\begin{center}
  \includegraphics[width=\columnwidth]{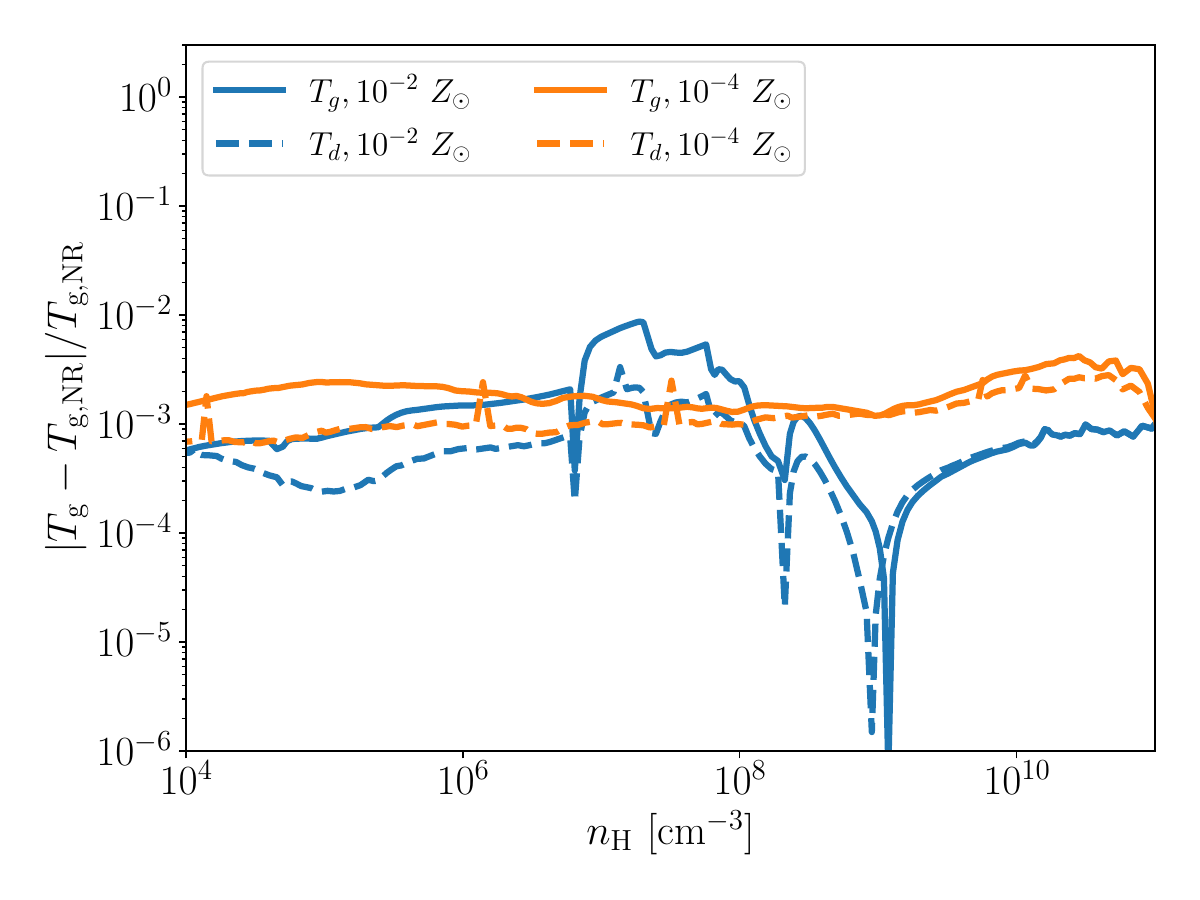}
\end{center}
 \caption{
 The relative errors in the gas and dust temperature between the LI and NR methods at $Z=10^{-2}~Z_{\odot}$ and $10^{-4}~Z_{\odot}$. 
 We adopt $C_{\rm d} = 10^7~{\rm erg \, g^{-1} \, K^{-1}}$ for the LI method. 
 The blue and orange lines correspond to the cases with $Z=10^{-2}~Z_{\odot}$ and $10^{-4}~Z_{\odot}$, respectively.
 The solid and dashed lines represent the relative errors in the gas and dust temperatures.
{Alt text: Graph showing the relative errors of gas and dust temperature at $Z=10^{-2}~Z_{\odot}$ and $10^{-4}~Z_{\odot}$.}
 }
 \label{fig:BE_Z-2_err}
\end{figure}

In low-metallicity environments, the reduced dust abundance decreases the total thermal energy reservoir of dust grains.
As a result, the LI method performs efficiently in high-density regions where dust grains are tightly coupled to the gas.
Figure \ref{fig:BE_Z-2_err} shows the relative errors in the gas and dust temperature for $Z=10^{-2}~Z_{\odot}$ and $Z=10^{-4}~Z_{\odot}$.
At $Z=10^{-2}~Z_{\odot}$, the relative errors remains below $10^{-2}$ over the entire density range and further decrease to $\leqq 10^{-3}$ at $n_{\rm H} \geqq 10^{8}~{\rm cm^{-3}}$, where gas-dust thermal coupling becomes effective at this metallicity.
At $Z=10^{-4}~Z_{\odot}$, the dust temperature remains lower than the gas temperature, and dust grains continue to be heated even at $n_{\rm H} =10^{11} ~{\rm cm^{-3}}$. 
Although the improvement relative to the $Z=10^{-2}~Z_{\odot}$ case is modest, the relative error remains below $10^{-2}$, indicating that the LI method successfully reproduces the results obtained with the NR method at $Z=10^{-4}~Z_{\odot}$.

\subsubsection{Performance of chemistry solvers in BE sphere tests} \label{section:performance_of_chemsolver_BE}

\begin{table*}
  \caption{Performance of chemistry solver in the BE sphere tests.}
  \label{Tab:performance_chemsolver_BE}
  \centering
  \begin{tabular}{l|cccc} \hline
        & average time $[\rm s]$ & performance & $N_{\rm sub, av}$ & $N_{\rm itr}$ \\ \hline 
    NR on CPU & $1.55 $ & $2.03 \times 10^6$                 & $1.00$ & $1.32$ ($3.96, 8.24$) \\
    NR on GPU & $1.14 \times 10^{-1}$&  $2.76 \times 10^7$   & $1.00$ & $1.32$ ($3.96, 8.24$) \\
    LI $10^4$ & $1.06 \times 10^{-1}$ & $2.98 \times 10^{7}$ & $1.13$ & $3.01$ ($1.20\times10^{-4}$) \\ 
    LI $10^5$ & $1.04 \times 10^{-1}$ & $3.02 \times 10^{7}$ & $1.04$ & $3.00$ ($1.53 \times 10^{-5}$)\\ 
    LI $10^6$ & $1.03 \times 10^{-1}$ & $3.06 \times 10^{7}$ & $1.01$ & $0$ ($0$)\\
    LI $10^7$ & $1.04 \times 10^{-1}$ & $3.01 \times 10^{7}$ & $1.00$ & $0$ ($0$) \\
    LI $10^8$ & $1.03 \times 10^{-1}$ & $3.06 \times 10^{7}$ & $1.00$ & $0$ ($0$) \\ \hline
  \end{tabular}
  \begin{minipage}{1 \hsize}
  \vspace{3mm}
  Notes. Column 1: average time per update step of the chemistry solver; Column 2: number of cell updates per device second; Column 3: average number of substeps ($N_{\rm sub, av}$); Column 4: iteration statistics ($N_{\rm itr}$). For the NR method, the value represents the average number of iterations required for the gas temperature. 
  The numbers in parentheses indicate the average number of iterations for the chemical abundances and for the dust temperature and IR radiation energy density per gas-temperature iteration loop. 
  For the LI method, the value represents the average number of iterations in cells where the $(T_{\rm d}, E_{\rm IR})$ iteration is performed, while the number in parentheses gives the fraction of cells that require iterations. 
  The GPU tests are performed using an NVIDIA GH200, and the CPU test uses an Intel Xeon Platinum 8468 (48 cores).
  The model labeled ``LI $10^4$'' corresponds to the LI method with $C_{\rm d} = 10^4~{\rm erg \, g^{-1} \, K^{-1}}$.
  \end{minipage} 
\end{table*}

Table \ref{Tab:performance_chemsolver_BE} summarizes the computational performance, as well as the average numbers of iterations and substeps, of the chemistry solvers at the epoch of sink formation.
We measure the performances of the LI method with $C_{\rm d} = 10^4-10^8~{\rm erg \, g^{-1} \, K^{-1}}$ and the NR method on both a GPU and CPUs.
In the BE tests, the performance of the LI method shows little dependence on $C_{\rm d}$.
The NR method on a GPU is slightly slower than the LI method, but its computational performance is sufficient.
On the other hand, the NR method on CPUs is about one-tenth as fast as on the GPU.

In the collapsing BE spheres, each timescale in Equation \eqref{eq:timestep_subcycle} is not smaller than the timesteps determined by the CFL condition for radiative transfer.
Even at the epoch of sink formation, the average number of subcycle steps remains close to unity regardless of $C_{\rm d}$. 
Thus, the performance of the LI method is independent of the values of $C_{\rm d}$ in the BE tests.
For the NR method, simulations of the BE sphere usually require only $1$-$2$ iterations for convergence of gas temperature, allowing the same timestep as that of the hydrodynamics to be used.
Consequently, the performance is comparable to that of the LI method.
On the other hand, CPU execution is approximately an order of magnitude slower, primarily reflecting the difference in computational throughput between CPUs and GPUs.

\subsubsection{Adaptive timestep mode} \label{sec:BE_adaptive_timestep}

\begin{figure} 
\begin{center}
  \includegraphics[width=\columnwidth]{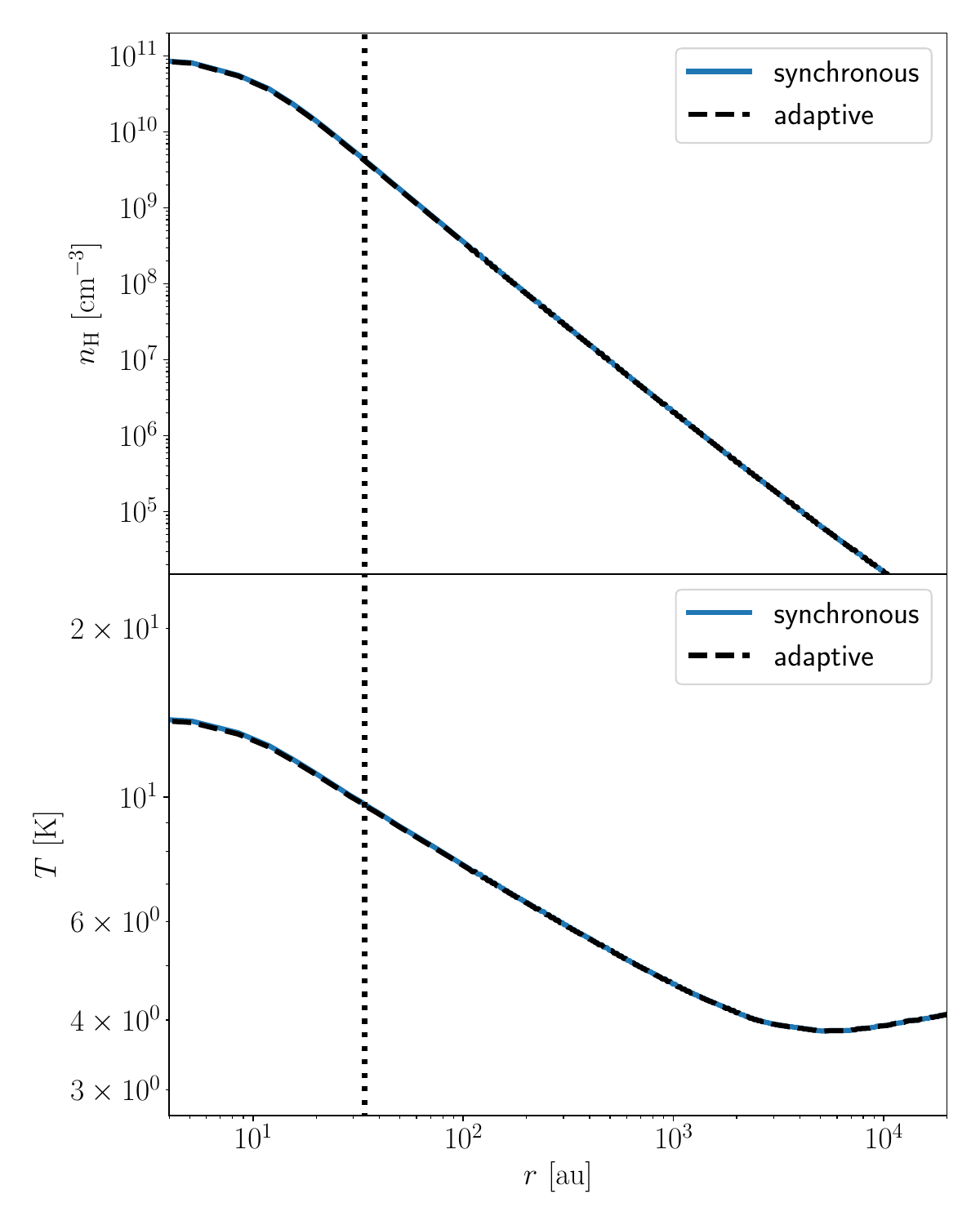}
\end{center}
 \caption{
    Radial profiles of gas density (top) and temperature (bottom) at the epoch of sink formation for $Z=1~Z_{\odot}$. 
    Each line corresponds to a simulation using either a synchronous (blue) or adaptive (black) timestep.
    The vertical dashed line indicates the sink radius, $r_{\rm sink} = 34~{\rm au}$.
    {Alt text: Radial profiles of gas density and temperature comparing simulations with synchronous and adaptive timestep.}
    }
 \label{fig:single_adaptive_nH}
\end{figure}

\begin{figure} 
\begin{center}
  \includegraphics[width=\columnwidth]{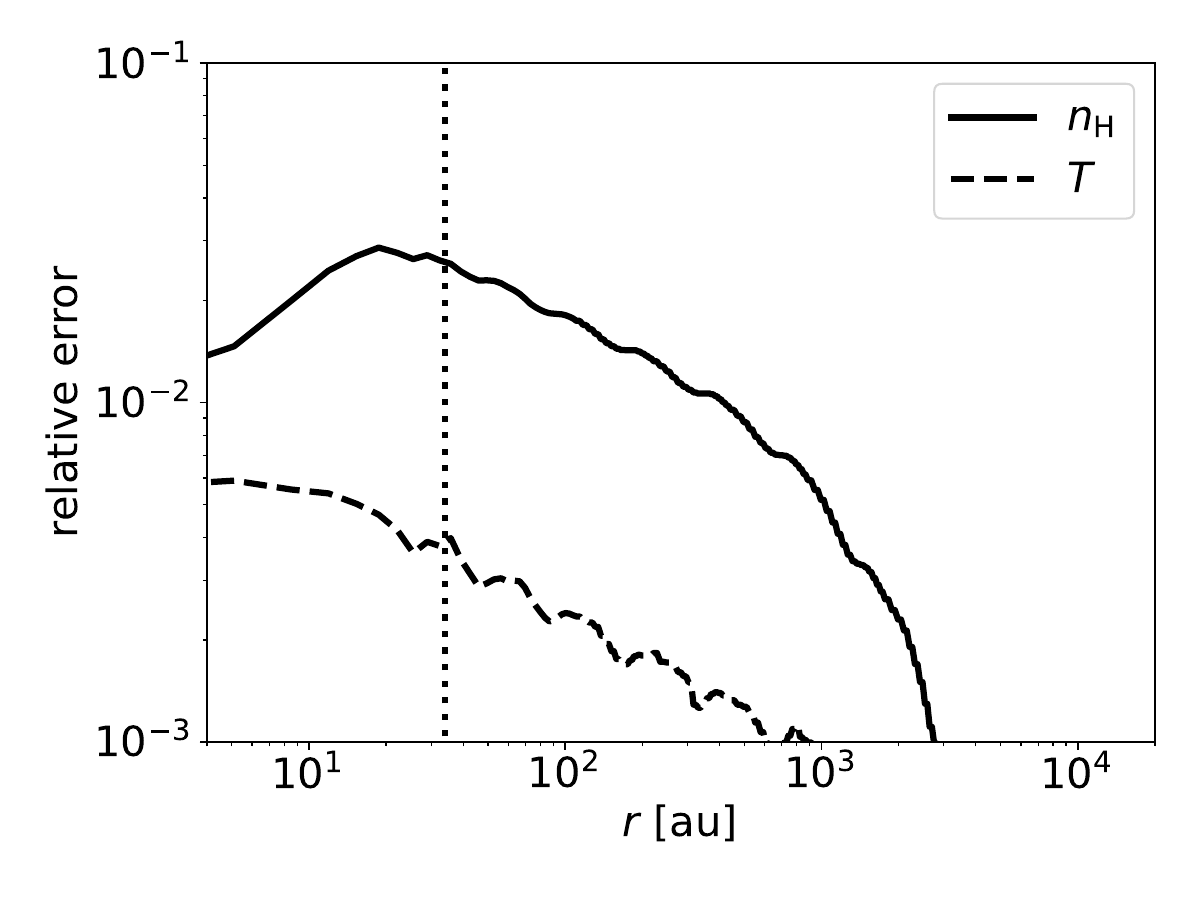}
\end{center}
 \caption{
 Radial distribution of the relative errors in number density ($n_{\rm H}$, dashed line) and gas temperature ($T$, solid line) between the synchronous and adaptive timestep models.
 As in Figure \ref{fig:single_adaptive_nH}, the vertical dashed line indicates the sink radius of $r_{\rm sink} = 34~{\rm au}$.
 {Alt text: Radial profiles of the relative errors comparing simulations with synchronous and adaptive timesteps.}
}
 \label{fig:BE_error}
\end{figure}

As described in Section \ref{sec:self-gravity}, we implement an adaptive timestep mode in our code.
To test the accuracy of this mode in the presence of self-gravity, we performed an additional run for the case with  $Z=Z_{\odot}$.
In this section, we adopt the LI method with $C_{\rm d} = 10^7~{\rm erg \, g^{-1} \, K^{-1}}$.
Figure \ref{fig:single_adaptive_nH} shows the radial distributions of number density and gas temperature at the epochs of sink formation.
The elapsed time to sink formation is identical in both simulations, $6.63\times 10^5~{\rm yr}$.
Figure \ref{fig:BE_error} shows the relative errors in the radial distributions of gas number density and temperature between the synchronous and adaptive timestep modes.
Although small discrepancies ($<3\times 10^{-2}$) are present, the overall evolution of them agrees well.
Moreover, the regions with relatively larger error are contained within the sink radius ($r_{\rm sink} = 34~{\rm au}$), and their subsequent evolution is therefore masked by the sink particle. 
We conclude that the adaptive timestep mode can be safely applied to simulations of star formation.

\subsection{H{\sc ii} region formation} \label{sec:HII_region}

In this section, we simulate the formation of an H{\sc ii} region in a static medium.
Following \citet{2021MNRAS.506.5512F}, we set up a uniform hydrogen number density of $n_{\rm H} = 10^2~{\rm cm^{-3}}$ and place a massive star with a mass of $M_* = 40~M_{\odot}$ at the center of the computational domain.
The domain size is $6~{\rm pc}$ on each side, and we employ a nested grid with a maximum refinement level of $l_{\rm max} = 3$.
Initially, all hydrogen is assumed to be in molecular form.
As in the BE sphere tests presented in Section \ref{sec:BE}, we run the simulations using both the LI and NR methods. 
We consider cases with $C_{\rm d} = 10^4$, $10^5$, $10^6$, $10^7$ and $10^8~{\rm erg \, g^{-1} \, K^{-1}}$ to verify the LI method and to evaluate the performance of the thermochemical solvers.

The ionization front is expected to reach the Str\"omgren radius, defined as
\begin{align}
  R_{\rm st} &= \left( \frac{3 f_{\rm ion} S_{\rm EUV}}{4 \pi \alpha_{\rm B} n_{\rm H}^2} \right)^{1/3} \nonumber \\ 
  &= 2.67 ~{\rm pc} \left( \frac{S_{\rm EUV}}{1.36 \times 10^{49} ~{\rm s^{-1}}} \right)^{1/3} \left( \frac{f_{\rm ion}}{0.45} \right)^{1/3} \left( \frac{n_{\rm H}}{10^2~{\rm cm^{-3}}} \right)^{-2/3}, \label{Rst}
\end{align}
where $S_{\rm EUV}$, $\alpha_{\rm B}$, and $f_{\rm ion}$ denote the emissivity of ionizing photons, the case-B recombination rate coefficient, and the fraction of ionizing photons absorbed by neutral hydrogen, respectively.
We adopt $\alpha_{\rm B} = 2.63 \times 10^{-13} ~{\rm cm^{-3} \, s^{-1}}$ \citep{1989agna.book.....O}, and take $f_{\rm ion}$ from the dusty medium model of \citet{1972ApJ...177L..69P}.

\subsubsection{Accuracy of the LI method in H{\sc ii} region formation tests}

\begin{figure} 
\begin{center}
  \includegraphics[width=\columnwidth]{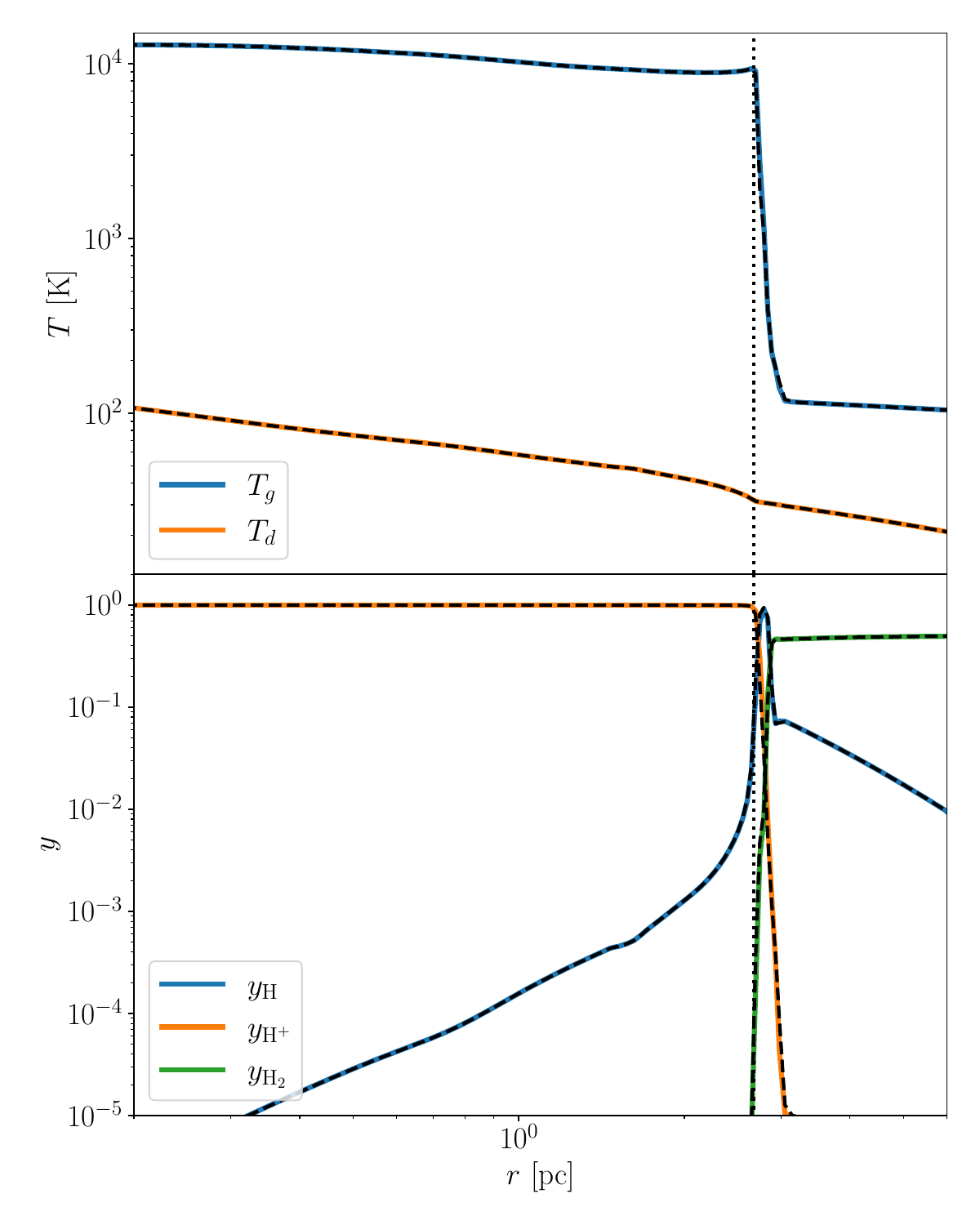}
\end{center}
 \caption{
 Radial profiles of temperatures and chemical abundances at $t = 2 \times 10^5 ~{\rm yr}$ in the H{\sc ii} region formation test. 
  The colored lines show the results obtained with the LI method using $C_{\rm d} = 10^7~{\rm erg \, g^{-1} \, K^{-1}}$, while the black-dashed lines show the results obtained with the NR method.
 Upper panel: Radial profiles of the gas and dust temperatures; the blue and orange lines represent the gas and dust temperatures, respectively.
 Lower panel: Radial profiles of the chemical abundances of $\rm H$ (blue), $\rm H^+$ (orange), and $\rm H_2$ (green).
 The vertical dashed line indicates the Str\"omgren radius given by Equation \eqref{Rst}.
 {Alt text: Radial profiles of gas and dust temperatures and chemical abundances obtained with the LI and NR methods in the H{\sc ii} region formation test.}
 }
 \label{fig_THII}
\end{figure}

Figure \ref{fig_THII} shows the radial profiles of the temperatures and chemical abundances at $t = 2 \times 10^5 ~{\rm yr}$.
The colored and black dashed lines show the results obtained with the LI and NR methods.
For the LI method, we adopt $C_{\rm d} = 10^7~{\rm erg \, g^{-1} \, K^{-1}}$.
The ionization front is located at $\sim 2.7~{\rm pc}$, which coincides with the Str\"omgren radius estimated from Equation \eqref{Rst}.
Within the H{\sc ii} regions, the gas is heated to $T_{\rm g} \sim 10^4~{\rm K}$ and is fully ionized by EUV photons from the massive star.
Immediately behind the ionization front, the gas drops steeply, reaching $\sim 100~{\rm K}$ in the outer region. 
The abundance of ionized hydrogen also decreases, and a photodissociation region with a thickness of approximately $0.3~{\rm pc}$ appears.
In regions beyond $3~{\rm pc}$, the gas remains molecular and is not affected by photodissociation. 
Dust grains are also heated to $100~{\rm K}$ by the absorption of UV photons, and their temperature gradually decreases toward the outer regions.

\begin{figure} 
\begin{center}
  \includegraphics[width=\columnwidth]{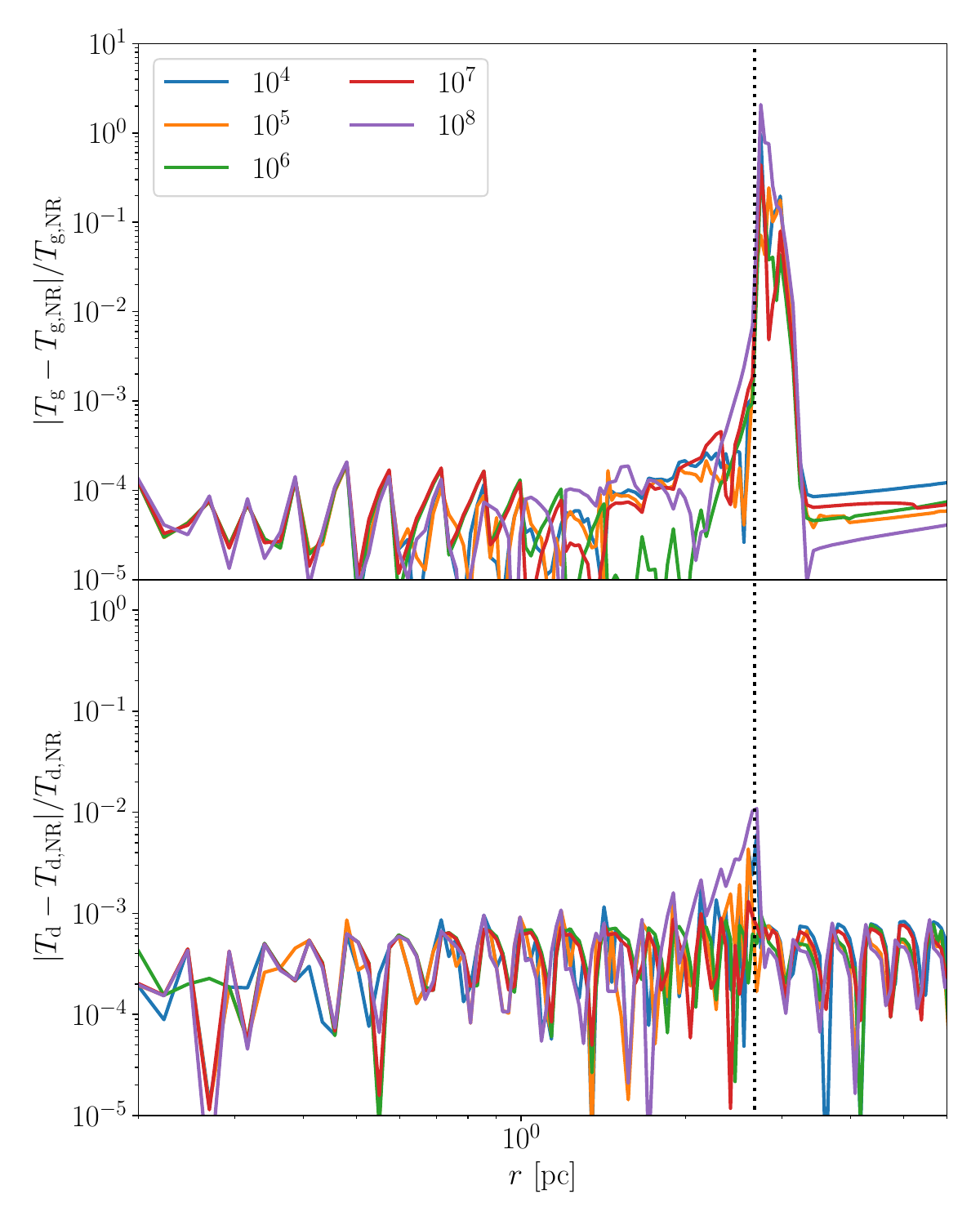}
\end{center}
 \caption{
  Relative errors in the gas temperature (top) and dust temperature (bottom) between the LI and NR methods in the H{\sc ii} region formation tests.
  Each colors denotes a different pseudo-specific capacity:$C_{\rm d} = 10^4$ (blue), $10^5$ (orange), $10^6$ (green), $10^7$ (red), and $10^8~{\rm erg \, g^{-1} \, K^{-1}}$ (purple).
 {Alt text: Graph showing the relative error in gas and dust temperatures between the LI and NR methods for different values of the pseudo-specific heat capacity in the H{\sc ii} region formation tests.}
 }
 \label{fig:HII_Terr}
\end{figure}

The top panel of Figure \ref{fig:HII_Terr} shows the relative errors in the gas and dust temperatures between the LI and NR methods.
In the H{\sc ii} region, the gas temperature is determined by the balance between radiative heating and metal-line cooling processes \citep[e.g.,][]{2011piim.book.....D}.
Consequently, the relative errors in the gas temperature are almost independent of $C_{\rm d}$.
In the H{\sc ii} and the outer neutral regions, the relative errors are less than $10^{-3}$, and the LI method accurately reproduces the results of the NR method.
In contrast, in the region immediately behind the ionization front, where the gas temperature decreases sharply, the relative error increases to values of $\sim 10^{-1}-1$.
In this region, the gas temperature drops rapidly from $10^4~{\rm K}$ to $10^{2}~{\rm K}$, while this transition is resolved by only $\leqq10$ computational cells.
As a result, discrepancies arise between the different solvers.
Star formation is suppressed in photodissociation regions (PDRs) owing to the high gas temperature \citep[e.g.,][]{1998ApJ...501..192D, 2015A&A...580A..49I, 2020MNRAS.497.5061I, 2022MNRAS.511.3346F}.
Therefore, uncertainties in the gas temperature in the PDRs are expected to have only a minor impact on studies of star formation. 
However, particular care is required when modeling gas temperature within the PDRs located just behind the ionization front.
The relative errors in the dust temperature are smaller than $\sim 2\times 10^{-2}$ and are also independent of $C_{\rm d}$.
In these tests, dust grains are heated by UV photons.
The dust temperature is determined by the balance between UV radiative heating and thermal emission, and heat storage by dust grains is almost negligible.

\subsubsection{Performance of chemistry solvers in H{\sc ii} region formation tests}

\begin{table*}
  \caption{Performance of chemistry solver in the H{\sc ii} region formation tests.}
  \label{Tab:performance_chemsolver_HII}
  \centering
  \begin{tabular}{l|cccc} \hline
        & average time $[\rm s]$ & performance  & $N_{\rm sub, av}$ & $N_{\rm itr}$ \\ \hline 
    NR on CPU & $45.5 $ & $ 1.84 \times 10^5$ & $1.25$ & $15.4$ ($3.16, 6.16$) \\
    NR on GPU & $2.71$  & $ 3.09 \times 10^6$ & $1.25$ & $15.4$ ($3.16, 6.16$) \\
    LI $10^4$ & $3.45$  & $ 2.43 \times 10^6$ & $42.6$ & $2.08$ ($2.22 \times 10^{-1}$)  \\ 
    LI $10^5$ & $1.88$  & $ 4.45 \times 10^6$ & $12.1$ & $2.06$ ($1.79 \times 10^{-1}$) \\ 
    LI $10^6$ & $9.42 \times 10^{-1}$ & $8.90 \times 10^{6}$ & $13.3$ & $2.04$ ($1.95 \times 10^{-3}$) \\
    LI $10^7$ & $5.21 \times 10^{-1}$ & $1.61 \times 10^{7}$ & $4.71$ & $2.48$ ($1.09 \times 10^{-4}$)  \\
    LI $10^8$ & $3.58 \times 10^{-1}$ & $2.34 \times 10^{7}$ & $1.33$ & $3.11$ ($1.29 \times 10^{-5}$)  \\ \hline
  \end{tabular}
  \begin{minipage}{1 \hsize}
  \vspace{3mm}
  Notes. Labels are the same as in Table \ref{Tab:performance_chemsolver_BE}.
  \end{minipage} 
\end{table*}

As in Section \ref{section:performance_of_chemsolver_BE}, we measure the performance of the chemistry solvers, summarized in Table \ref{Tab:performance_chemsolver_HII}.
In simulations of H{\sc ii} region formation, unlike the BE sphere tests, the governing equations become highly stiff owing to UV heating of gas and dust grains and hydrogen photoionization. 
When solving the chemical network and the energy equations of gas and dust with the NR method, an average of $15.4$ iterations is required for the convergence of the gas temperature. 
In some cells, particularly around the ionization front, the NR method fails to converge when the hydrodynamical timestep is adopted, and a substepping procedure is therefore applied. 
As a result, the performance becomes roughly one-tenth of that in the BE sphere tests.
The GPU execution still achieves more than an order of magnitude higher than the CPU execution. 
Therefore, the use of GPUs is also effective for executing the NR method.
For the LI method, the timestep is determined by the characteristic timescale of the dust energy equation.
When adopting the realistic heat capacity of dust grains, $C_{\rm d} = 10^4~{\rm erg \, g^{-1} \, K^{-1}}$, the average number of substeps becomes as large as $\sim 40$, and iterations of the dust temperature and IR radiation energy density are required in about $20$ percent of the cells.
Consequently, the performance becomes slightly lower than that of the NR method.
As the value of $C_{\rm d}$ increases, however, the number of substeps and cells requiring iterations decreases, leading to improved performance.
In particular, adopting $C_{\rm d} = 10^7~{\rm erg \, g^{-1} \, K^{-1}}$, we achieve a speedup of about a factor of five compared to the NR method.

\subsection{Appropriate values of dust pseudo heat capacity} \label{Sec:appropriate_value_dustcap}

In numerical simulations, it is ideal to adopt either the NR method or the LI method with a realistic dust heat capacity of $C_{\rm d} = 10^4~{\rm erg \, g^{-1} \, K^{-1}}$.
For problems in which the timestep of the thermochemistry solver is comparable to or longer than that of the radiation-hydrodynamics solver throughout the computational domain, such as the BE collapse discussed in Section \ref{sec:BE}, the use of these methods is recommended.
However, in problems where the timestep becomes extremely short, such as inside H{\sc ii} regions, applying these approaches in GPU-based calculations results in prohibitively high computational costs.
In such cases, it is necessary to reduce the computational cost by adopting a larger pseudo heat capacity value than $10^4~{\rm erg \, g^{-1} \, K^{-1}}$.
This strategy is particularly effective for star cluster formation simulations, in which molecular clouds are disrupted by the formation of H{\sc ii} regions.
As demonstrated in Section \ref{sec:HII_region}, increasing $C_{\rm d}$ significantly reduces the computational cost, while the simulation results remain essentially insensitive to the choice of $C_{\rm d}$. 
On the other hand, the BE tests indicate that errors associated with an increased pseudo heat capacity become non-negligible when $C_{\rm d}$ is too large.
When the specific internal energy is taken into account, the specific heat capacity of the gas itself is approximately $k_{\rm B}/m_{\rm H} \sim 8 \times 10^7~{\rm erg \, g^{-1} \, K^{-1}}$.
If $C_{\rm d}$ exceeds this value, the energy stored in dust grains would exceed that stored in the gas.
Considering this constraint together with the results of the BE tests, the value of $C_{\rm d}$ should be limited to $\lesssim 10^{7}~{\rm erg \, g^{-1} \, K^{-1}}$ to ensure accurate calculations.
Considering both numerical accuracy and computational efficiency, we conclude that adopting $C_{\rm d} = 10^7~{\rm erg \, g^{-1} \, K^{-1}}$ is an appropriate and practical choice for star cluster formation simulations.

\section{Performance} \label{section:performance}
\begin{table*}[t]

\caption{
Computational performance of the hydrodynamics and self-gravity solvers on a uniform grid.
}
\centering
\resizebox{\textwidth}{!}{
\begin{tabular}{l|ccc|ccc|ccc|ccc}
\hline
GPUs
& \multicolumn{3}{c|}{ Hydro $8^3$}
& \multicolumn{3}{c|}{ Hydro $16^3$}
& \multicolumn{3}{c|}{ Self-gravity $8^3$}
& \multicolumn{3}{c}{  Self-gravity $16^3$} \\
\cline{2-13}
 & Time & Perf. & Scaling
 & Time & Perf. & Scaling
 & Time & Perf. & Scaling
 & Time & Perf. & Scaling \\
\hline
1  & $6.33$ & $265$ & $1$     & $4.47$  & $375$ & $1$    & $9.73$ & $172$ & $1$      & $8.48$ & $198$  & $1$ \\
2  & $3.41$ & $246$ & $0.929$ & $2.87$  & $293$ & $0.780$& $6.76$ & $124$ & $0.720$  & $6.76$ & $124$  & $0.627$  \\
4  & $1.86$ & $225$ & $0.849$ & $1.32$  & $318$ & $0.848$& $6.26$ & $67.0$ & $0.389$ & $7.44$ & $56.4$ & $0.285$ \\
8  & $1.02$ & $205$ & $0.775$ & $0.740$ & $283$ & $0.755$& $4.69$ & $44.7$ & $0.260$ & $5.79$ & $36.2$ & $0.183$ \\
\hline
\end{tabular}
}
\begin{tablenotes}[para,flushleft]
\item \parbox{\textwidth}{
  \begin{minipage}{1 \hsize}
  \vspace{3mm}
    Notes. Time denotes the execution time per step in units of $10^{-2}\, \rm{s}$. Perf. represents the computational performance in units of $10^6$ cell updates per device second.
    Scaling corresponds to strong-scaling efficiencies with respect to the single-device case.
    The labels $8^3$ and $16^3$ denote simulations using $8^3$ and $16^3$ blocks, respectively.
  \end{minipage} 
}
\end{tablenotes}
\label{tab:performance_scaling_uniform}
\end{table*}
\begin{table*}[t]

\caption{
Computational performance of the hydrodynamics and self-gravity solvers on an AMR grid.
}
\centering
\resizebox{\textwidth}{!}{
\begin{tabular}{l|ccc|ccc|ccc|ccc}
\hline
GPUs
& \multicolumn{3}{c|}{ Hydro $8^3$}
& \multicolumn{3}{c|}{ Hydro $16^3$}
& \multicolumn{3}{c|}{ Self-gravity $8^3$}
& \multicolumn{3}{c}{  Self-gravity $16^3$} \\
\cline{2-13}
 & Time & Perf. & Scaling
 & Time & Perf. & Scaling
 & Time & Perf. & Scaling
 & Time & Perf. & Scaling \\
\hline
1  & $37.8$  & $178$ & $1$     & $21.1$  & $319$ & $1$    & $42.0$ & $160$  & $1$     & $37.2$ & $180$  & $1$ \\
2  & $21.0$  & $160$ & $0.899$ & $11.7$  & $288$ & $0.903$& $29.2$ & $115$  & $0.719$ & $31.5$ & $107$  & $0.592$  \\
4  & $12.2$  & $138$ & $0.775$ & $6.53$  & $257$ & $0.806$& $23.6$ & $71.2$ & $0.446$ & $29.9$ & $56.2$ & $0.312$ \\
8  & $7.40$  & $113$ & $0.638$ & $3.87$  & $217$ & $0.680$& $21.6$ & $38.8$ & $0.243$ & $29.7$ & $28.2$ & $0.157$ \\
\hline
\end{tabular}
}
\begin{tablenotes}[para,flushleft]
\item \parbox{\textwidth}{
  \begin{minipage}{1 \hsize}
  \vspace{3mm}
    Notes. Labels are the same as in Table \ref{tab:performance_scaling_uniform}.
  \end{minipage} 
}
\end{tablenotes}
\label{tab:performance_scaling_AMR}
\end{table*}
\begin{table}[t]
\caption{
Performance of the hydrodynamics and self-gravity solvers on a CPU. 
}
\centering
\begin{tabular}{lcccc}
\hline
Grid & Physics & Block  & Time & Perf. \\
\hline
\multirow{4}{*}{Uniform}
  & \multirow{2}{*}{Hydro}
 &    $8^3$  & $62.8$ & $26.7$ \\
 &  & $16^3$ & $45.2$ & $37.1$ \\
 & \multirow{2}{*}{Self-Gravity}
 &    $8^3$  & $148$ & $11.3$  \\
 &  & $16^3$ & $50.0$& $33.6$ \\
\hline
\multirow{4}{*}{AMR}
 & \multirow{2}{*}{Hydro}
 &    $8^3$  & $855$ & $7.85$ \\
 &  & $16^3$ & $659$ & $10.2$ \\
 & \multirow{2}{*}{Self-Gravity}
 &    $8^3$  & $949$ & $7.08$  \\
 &  & $16^3$ & $264$ & $25.4$ \\
\hline
\end{tabular}
  \begin{minipage}{1 \hsize}
  \vspace{3mm}
    Notes. As in Table \ref{tab:performance_scaling_uniform}, Time denotes the execution time per step in units of $10^{-2}\, \rm{s}$, and Perf. represents the computational performance in units of $10^{6}$ cell updates per device second. Block denotes the block size.
  \end{minipage} 
\label{tab:cpu_single}
\end{table}


A primary goal of \textsc{SFUMATO\#} is to enable high-performance numerical simulations on GPUs.
In this section, we assess whether this objective has been achieved by measuring the computational performance of the major components of the code.
In Section \ref{section:performance_hydro_uniform}-\ref{section:performance_selfgrav_AMR}, we evaluate the strong-scaling performance of the hydrodynamics and self-gravity solvers, both of which require MPI communication in multi-process executions.
We assess performance on both uniform and AMR grids to investigate the impact of grid structures.
In these sections, we also measure and compare the performance obtained on a CPU using the original \textsc{SFUMATO} code.
We do not present a separate performance analysis of the radiative transfer solver, as its time-integration scheme is similar to that of the hydrodynamic solver.
In Section \ref{section:Star_forming_clouds}, we measure the performance of a full star cluster formation simulation in a giant molecular cloud (GMC) and quantify the computational costs of each solver when the radiative transfer and thermochemical processes are included.
In octree-based AMR codes, computational performance also depends on the number of cells contained in each block.
We therefore examine two commonly adopted configurations, in which each block contains either $8^3$ or $16^3$ cells, hereafter referred to as $8^3$ and $16^3$ blocks.
Following Section \ref{sec:test_problems}, all GPU-based performance measurements are carried out on NVIDIA GH200 GPUs, while CPU-based measurements are performed on an Intel Xeon Platinum 8468 (48 cores).
The GPU performance results are summarized in Tables \ref{tab:performance_scaling_uniform}, \ref{tab:performance_scaling_AMR}, \ref{tab:performance_scaling_GMC_888}, and \ref{tab:performance_scaling_GMC_16}.
Table \ref{tab:cpu_single} presents the corresponding results obtained on the CPU.
Throughout this section, we employ the HLLC Riemann solver for the hydrodynamics solver and use FP64 for all real-valued quantities.

\subsection{Hydrodynamics on uniform grids} \label{section:performance_hydro_uniform}

\begin{figure} 
\begin{center}
  \includegraphics[width=\columnwidth]{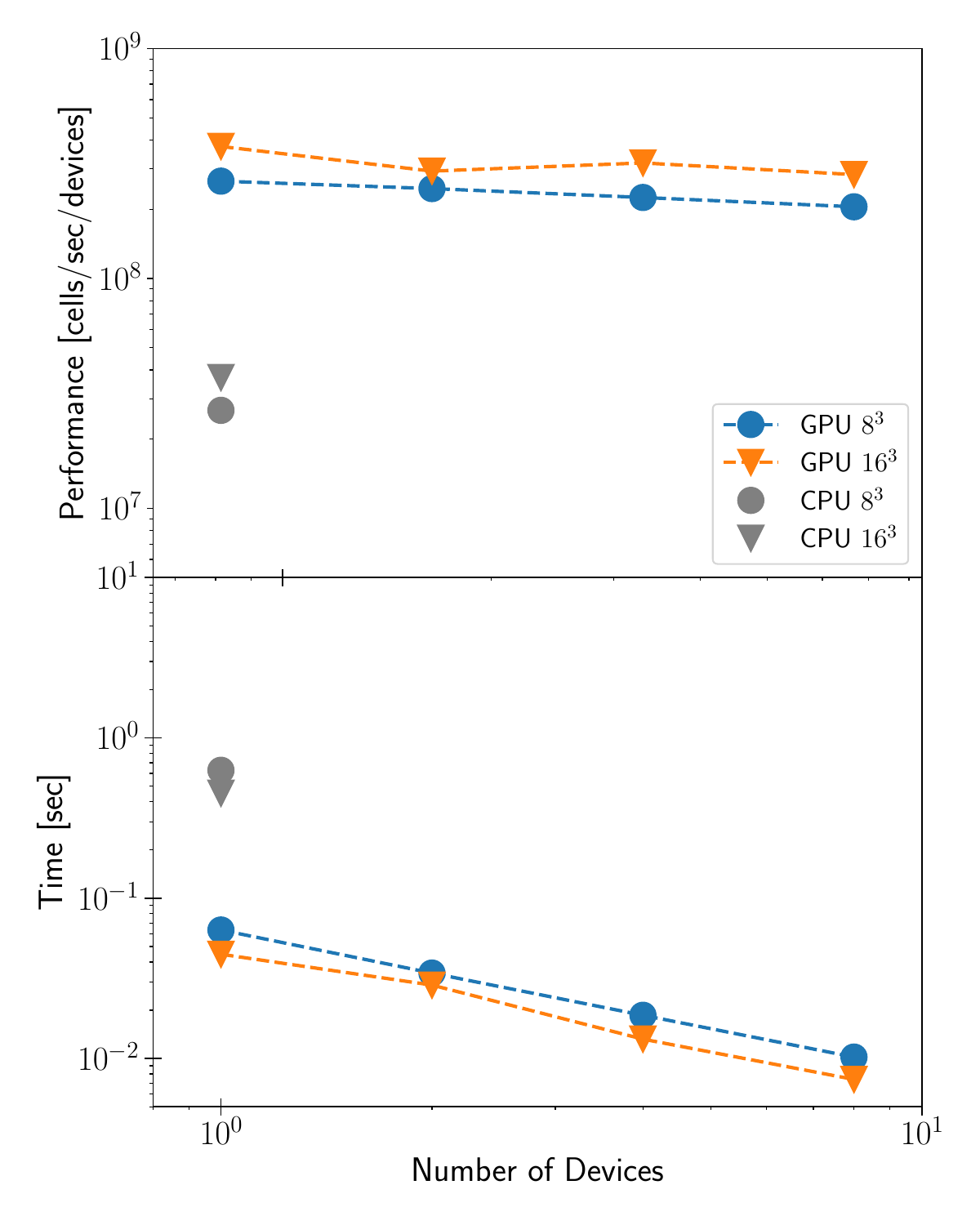}
\end{center}
 \caption{
  Strong-scaling test of the hydrodynamics solver on a uniform grid with $256^3$ cells and block sizes of $8^3$ and $16^3$.
  Each panel shows the performance in units of cell updates per device second (top) and the execution time per step (bottom).
  The colored lines show the GPU results, and the gray symbols represent the CPU cases.
 {Alt text: Performance test of the hydrodynamics solvers on a uniform grid with $256^3$ cells using GPUs and a CPU.}
 }
 \label{fig:uniform_hydro}
\end{figure}

To investigate the performance of the hydrodynamics solver on a uniform grid, we perform a strong-scaling test of the shock-tube problem using the same setup as described in Section \ref{sec:shock_tube}. 
In this test, we adopt a fixed uniform grid with $256^3$ cells.

Figure \ref{fig:uniform_hydro} shows the performance for the $8^3$ and $16^3$ blocks (see also Table \ref{tab:performance_scaling_uniform}).
With a single device, the hydrodynamics solver achieves $2.65\times 10^8$ and $3.75 \times10^8$ cell updates per device second for the $8^3$ and $16^3$ blocks, respectively.
In hydrodynamic calculations, MPI communication is required to exchange ghost-cell values across block boundaries, whereas the remaining computations can be carried out independently by each process.
As the number of devices increases, the strong-scaling efficiency gradually decreases.
However, it remains above $0.7$ with $8$ devices in both cases, demonstrating good scalability.

We also perform the same simulation on a CPU using the original \textsc{SFUMATO} code, as indicated by the gray symbols in Figure \ref{tab:performance_scaling_uniform} (see also Table \ref{tab:cpu_single}).
In both cases, the computational performance per device is approximately an order of magnitude higher on a GPU than on a CPU.
Considering the theoretical peak performance, the Intel Xeon Platinum 8468 and the GPU component of the NVIDIA GH200 differ by roughly a factor of ten.
Therefore, in terms of device capability, \textsc{SFUMATO\#} achieves performance comparable to that of the original \textsc{SFUMATO} code.

When the number of cells per block increases, communication overhead is reduced, and the number of cells processed by a single GPU kernel thread in numerical flux calculations increases.
As a result, the performance with $16^3$ blocks is improved by a factor of approximately $1.4$ compared to that with $8^3$ blocks.
Similar trends have also been reported for other AMR codes. 
For example, in \textsc{AthenaK} \citep{2026ApJS..283...27S} running on an NVIDIA A100 GPU, exhibits a performance improvement by a factor of $4.9$ when the block size is increased from a value of $16^3$ to $128^3$.
In addition, \textsc{AthenaK} achieves a performance of $10^9$ cell updates per device second for hydrodynamics simulations on a GH200 when using $128^3$ blocks.
By extrapolating from the A100 results, the performance for block size of $16^3$ is estimated to be approximately $2\times 10^8$ cell updates per device second on a GH200.
\textsc{SFUMATO\#} is capable of achieving a comparable level of performance with $16^3$ blocks.
On the other hand, as shown in Figure \ref{fig:grid_flux}, the update of numerical fluxes is computed by a single thread in each spatial direction, and further parallelization in this direction has not yet been implemented.
Moreover, memory allocation for blocks exceeds the maximum size allowed by \texttt{cudaMalloc}/\texttt{hipMalloc} for block sizes of $64^3$ or larger.
As a result, \textsc{SFUMATO\#} cannot be executed with block sizes larger than $32^3$, and thus cannot fully exploit the potential performance gains offered by larger blocks.
Nevertheless, since block sizes of $8^3$ and $16^3$ are commonly used in AMR simulations, the current implementation is already sufficiently practical. 
To support large-scale, uniform-grid simulations in the future, we plan to revise the code to enable the use of large block sizes, thereby achieving more efficient performance on uniform grids.

\subsection{Self-gravity on uniform grids} \label{section:performance_selfgrav_uniform}

Next, we measure the performance of the multigrid solver for self-gravity on a uniform grid with $256^3$ cells.
As the initial condition, we adopt a uniform-density sphere with a radius of $r_{\rm cl} = 1~{\rm pc}$ and a density of $\rho = 10^{-22}~{\rm g \, cm^{-3}}$.
The computational domain has a size of $3~{\rm pc}$ on each side.
The convergence criterion of the multigrid solver is defined such that the absolute value of the residual of the Poisson equation \eqref{eq_poission} is less than $10^{-3}$.

\begin{figure} 
\begin{center}
  \includegraphics[width=\columnwidth]{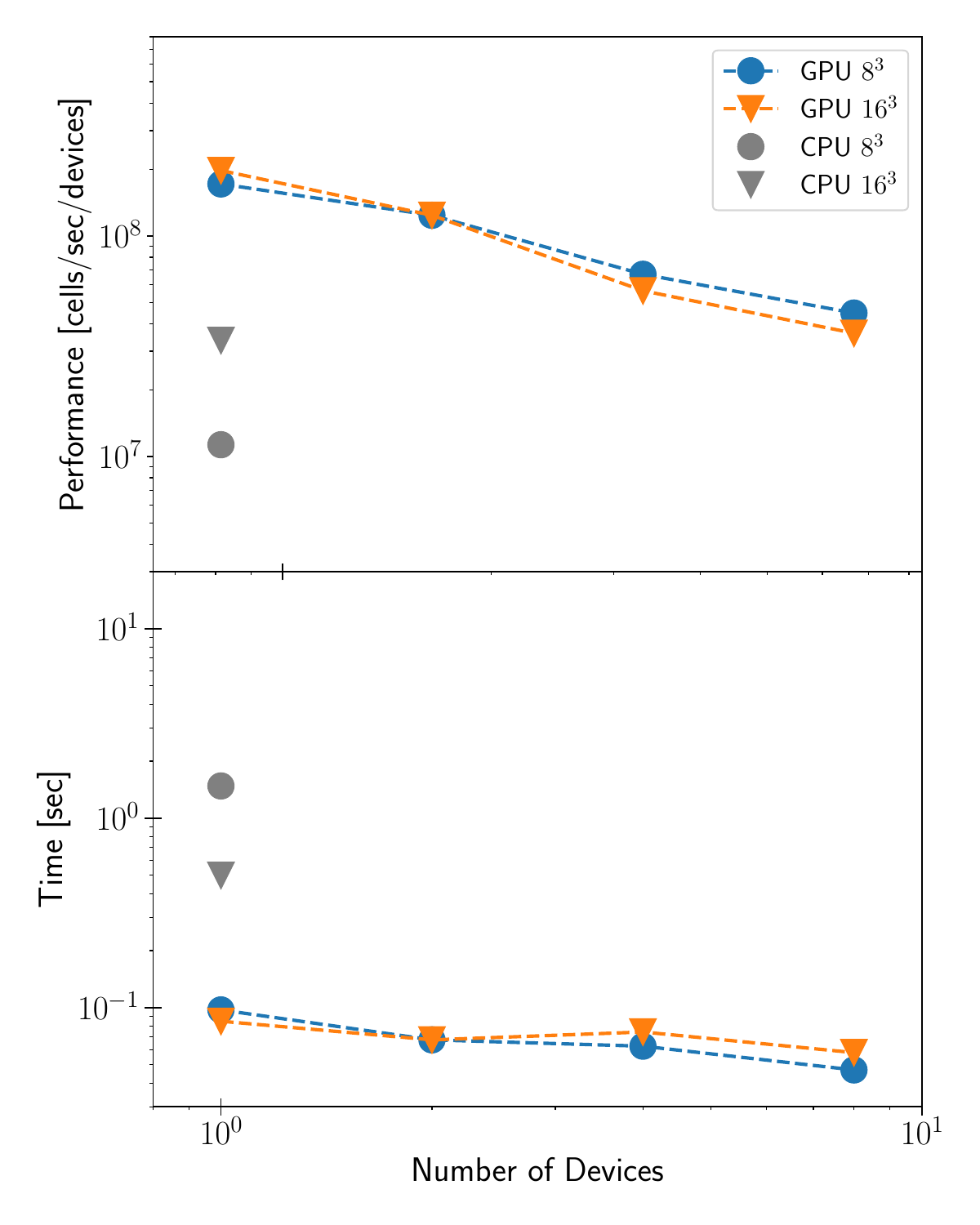}
\end{center}
 \caption{
 Same as Figure \ref{fig:uniform_hydro}, but for the self-gravity solver on a uniform grid with $256^3$ cells.
 {Alt text: Performance test of the self-gravity solver on a uniform $256^3$ grid using GPUs and a CPU.}
 }
 \label{fig:uniform_grav}
\end{figure}

Figure \ref{fig:uniform_grav} shows the performance of the self-gravity solver on the uniform grid.
With a single GPU, the self-gravity solver achieves performance of $1.72 \times 10^8$ and $1.98 \times 10^8$ cells per device second for the $8^3$ and $16^3$ blocks, which are comparable to those of the hydrodynamics solver.
As in the hydrodynamic solver, the $16^3$ blocks provide better single-device performance.
On the other hand, as the number of devices increases, the execution time decreases only marginally, and the performance per device correspondingly degrades.
In particular, when more than $4$ devices are used, the strong-scaling efficiency drops below $0.5$, and the execution performance is no longer advantageous.
As described in Section \ref{sec:self-gravity}, the multigrid solver is divided into three types of cycles.
During the FMG-cycle on the AMR hierarchy, the cells within each grid are progressively coarsened.
As the solver proceeds through the V-cycle on the AMR hierarchy and the FMG-cycle on the coarsened uniform base grid, the number of active cells continues to decrease.
Consequently, compared to the hydrodynamics solver, the number of active cells becomes extremely small for most of the multigrid cycles, leading to reduced parallel efficiency.
Moreover, as the number of processes increases, the cost of MPI communication becomes a performance bottleneck.
For the $16^3$ blocks, the number of coarsening steps in cycle (a) is larger than in the $8^3$ block case, resulting in a greater number of MPI communications.
As a result, the parallel efficiency for the $16^3$ blocks degrades more rapidly than that for the $8^3$ blocks.

The gray symbols in Figure \ref{fig:uniform_grav} and Table \ref{tab:cpu_single} indicate the results obtained on a CPU.
Compared to the CPU performance, the GPU execution performance is improved by factors of $15.2$ and $5.89$ for the $8^3$ and $16^3$ blocks, respectively.
Thus, the transition from CPU to GPU enables the self-gravity solver to achieve acceleration comparable to that of the hydrodynamics solver.

\subsection{Hydrodynamics on AMR grids}\label{section:performance_hydro_AMR}

In this section, we measure the performance of the hydrodynamics solver on an AMR grid.
We carried out the simulations using a nested-grid structure in which each level consists of $128^3$ cells, with levels prepared from level $0$ to level $3$.
The shock-tube test is adopted, using the same setups as described in Section \ref{section:performance_hydro_uniform}.

\begin{figure} 
\begin{center}
  \includegraphics[width=\columnwidth]{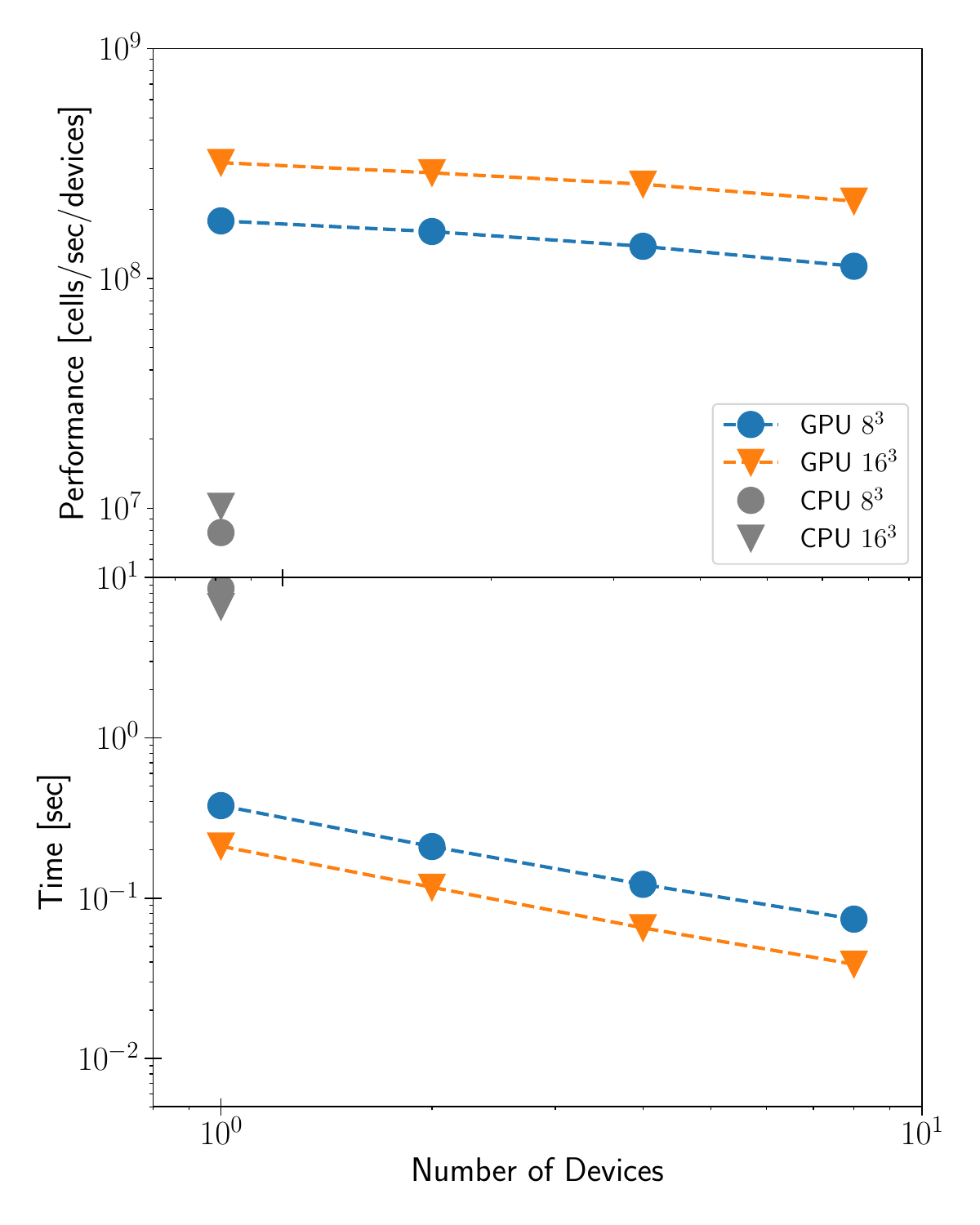}
\end{center}
 \caption{
 Same as Figure \ref{fig:uniform_hydro}, but for the hydrodynamics solver on the AMR grid.
 {Alt text: Performance test of the hydrodynamics solver on the AMR grid using GPUs and a CPU.}
 }
 \label{fig:nested_hydro}
\end{figure}

Figure \ref{fig:nested_hydro} shows the performance of the hydrodynamics solver with the $8^3$ and $16^3$ blocks (see also Table \ref{tab:performance_scaling_AMR}).
Compared with the uniform-grid case presented in Section \ref{section:performance_hydro_uniform}, both the performance and the strong-scaling efficiency are slightly degraded.
This degradation arises from the additional computational overhead associated with reflux operations at the level interfaces and correction of physical variables between coarse and fine cells.
Nevertheless, the strong-scaling efficiency remains above $0.6$ up to $8$ devices, indicating reasonably good scalability on the AMR grids.

The gray symbols in Figure \ref{fig:nested_hydro} also show the results obtained on a CPU (see also Table \ref{tab:cpu_single}).
On the CPU, the performance degradation caused by the use of the AMR grids is more pronounced than on the GPU.
In the GPU implementation, inter-level operations such as reflux corrections are managed using CUDA/HIP streams and are executed during periods when other kernels are idle, for example, during MPI communications for updating the next refinement level.
As a result, the performance ratio between the GPU and the CPU per device increases compared to the uniform-grid case, reaching factors of $22$ for $8^3$ blocks and $31$ for $16^3$ blocks.

\subsection{Self-gravity on AMR grids} \label{section:performance_selfgrav_AMR}

\begin{figure} 
\begin{center}
  \includegraphics[width=\columnwidth]{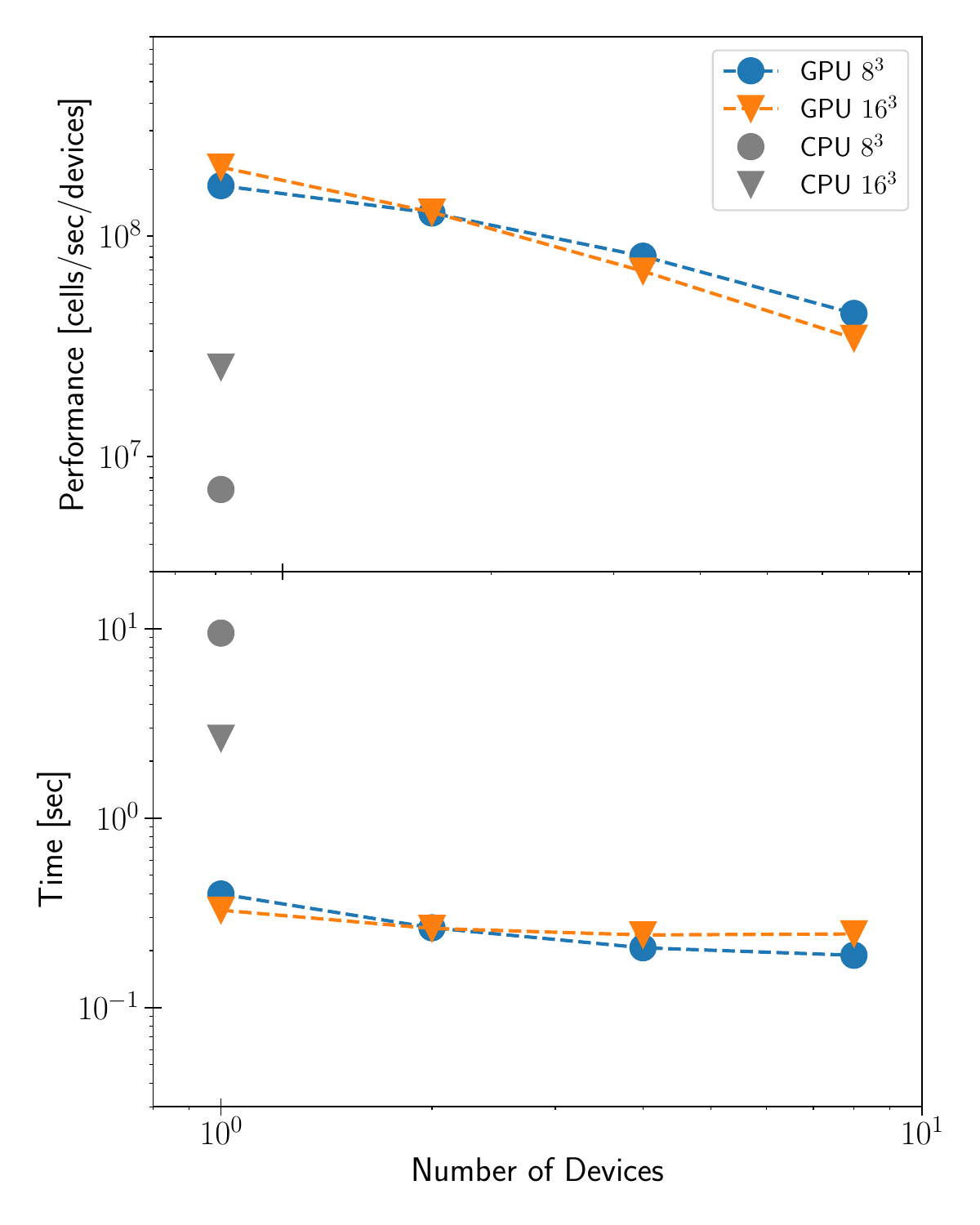}
\end{center}
 \caption{
  Same as Figure \ref{fig:uniform_hydro}, but for the self-gravity solver on the AMR grid.
 {Alt text: Performance test of the self-gravity solver on the AMR grid using GPUs and a CPU.}
 }
 \label{fig:nested_grav}
\end{figure}

Next, we measure the performance of the multigrid solver for self-gravity on an AMR grid.
We adopt the same uniform-density sphere setup as described in Section \ref{section:performance_selfgrav_uniform}.
The grid configuration follows that in Section \ref{section:performance_hydro_AMR}, 
where a nested grid is constructed from level $0$ to level $3$, each containing $128^3$ cells.
Figure \ref{fig:nested_grav} shows the performance in this test (see also Table \ref{tab:cpu_single}).
On a single device, the performance with $8^3$ blocks is comparable to that of the hydrodynamics solver.
For $16^3$ blocks, although the performance gain is smaller, the achieved performance is not inferior to that of the hydrodynamics solver.
In AMR grids, interpolation operations are required at the interfaces between different refinement levels \citep{2007PASJ...59..905M}.
However, their impact is less significant than in the hydrodynamics solver, and the overall performance remains comparable to that of the uniform-grid case presented in Section 
\ref{section:performance_selfgrav_uniform}.
Moreover, as in the uniform-grid case, the strong-scaling efficiency in parallel executions degrades with increasing numbers of devices due to increased MPI communication costs.

Compared to the execution performance on a CPU, the GPU achieves speedup factors of $23$ and $7$ for $8^3$ and $16^3$ blocks, respectively.
Since these speedups are comparable to those obtained in the uniform-grid case, this result indicates that a similar level of acceleration in self-gravity calculations can also be achieved on GPUs for AMR grids.

\subsection{Giant Molecular Cloud Simulations} \label{section:Star_forming_clouds}

\begin{figure} 
\begin{center}
  \includegraphics[width=\columnwidth]{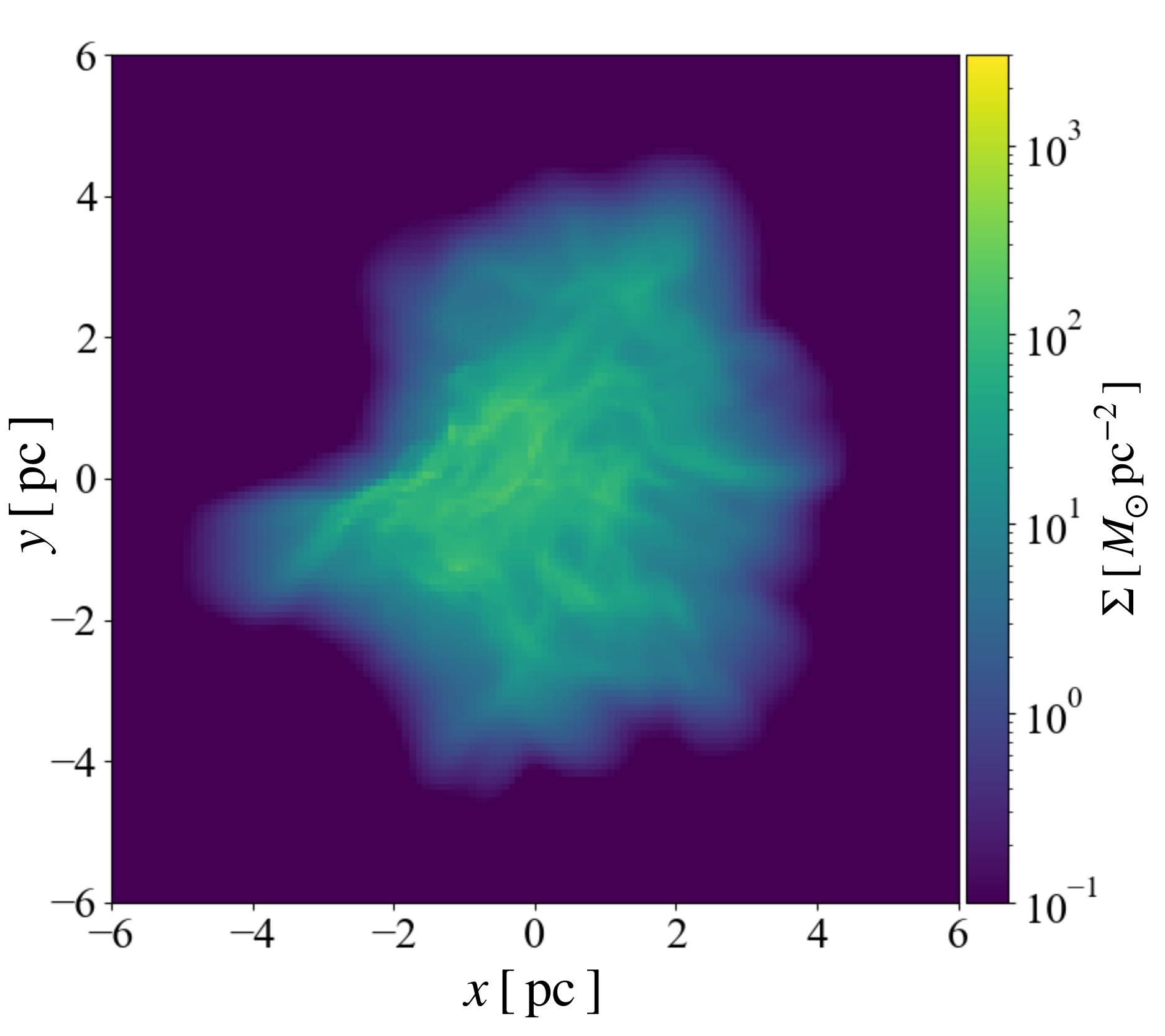}
\end{center}
 \caption{
 Density structure in the GMC simulation at $t=1.73~{\rm Myr}$ with the $8^3$ blocks.
 {Alt text: Density structure in the GMC simulation at $t=1.73~{\rm Myr}$.}
 }
 \label{fig:GMC2}
\end{figure}

\begin{table}
  \caption{Number of cells at each AMR level in the GMC simulations with block sizes of $8^3$ and $16^3$.}
  \label{Tab:cellnumber_inGMC_simulation}
  \centering
  \begin{tabular}{l|cc} \hline
              & $8^3$ Block & $16^3$ Block \\ \hline 
    Level $0$ & $2,097,152 \, (4,096)$ & $2,097,152 \, (512)$  \\
    Level $1$ & $2,195,456 \, (4,288)$ & $4,194,304 \, (1,024)$  \\
    Level $2$ & $1,343,488 \, (2,624)$ & $6,553,600 \, (1,600)$   \\ 
    Level $3$ & $258,048 \, (504)$     & $2,555,904 \, (624)$   \\ 
    Total     & $5,894,144 \, (11,512)$ & $15,400,960 \, (3,760)$  \\ \hline
  \end{tabular}
  \begin{minipage}{1 \hsize}
  \vspace{3mm}
  Notes. The numbers in parentheses indicate the corresponding number of blocks at each AMR level.
  \end{minipage} 
\end{table}

In this section, we measure the performance of \textsc{SFUMATO\#} in more realistic star formation simulations.
We consider a uniform-density sphere with mass $M_{\rm cl} = 10^3~M_{\odot}$ and radius $R_{\rm cl} = 3~{\rm pc}$, and impose a turbulent velocity field such that the virial parameter becomes $\alpha_{\rm vir} = 1$, following \citet{2020MNRAS.497..829F}.
The computational box size and metallicity are set to $L_{\rm box} = 6~{\rm pc}$ and $Z=Z_{\odot}$.
Cells are refined such that the Jeans length is resolved by at least $8$ cells.
In these simulations, we include the hydrodynamics, the radiative transfer of EUV, FUV, and IR photons, self-gravity, and the thermochemistry.
We adopt the LI method with $C_{\rm d} = 10^7~{\rm erg \, g^{-1} \, K^{-1}}$ for the chemistry solver.
To investigate the effects of the grid size on computational performance, we perform the simulations with $8^3$ and $16^3$ blocks, and measure the performance at a simulation time of $t=1.73~{\rm Myr}$. 
Figure \ref{fig:GMC2} shows the gas density distribution at this epoch.
At this stage, the refinement progresses up to level $3$.
Table \ref{Tab:cellnumber_inGMC_simulation} shows the cell and block number at $t=1.73~{\rm Myr}$.
The total cell count for the $16^3$ blocks exceeds that for the $8^3$ blocks by more than a factor of three.
We adopt the AMR with a self-similar block-structured grid using an octree structure.
In this approach, for the same cell size, a larger block size corresponds to a larger spatial region per block. 
When refinement occurs, the block is subdivided into eight child blocks, each containing the same number of finer cells.
Consequently, compared with the $8^3$ block case, the $16^3$ block configuration tends to generate additional cells in regions surrounding those that satisfy the refinement criteria.
As a result, the total number of cells in the computational domain becomes larger in the case of the $16^3$ blocks.

These differences in the spatial distributions of cells lead to non-identical numerical solutions, such as the turbulent velocity field, between the two setups.
The epoch shown in Figure \ref{fig:GMC2} corresponds to the onset of refinement at level $3$.
At this stage, transient density peaks produced by turbulent motions can satisfy the refinement criterion, triggering the creation of level-$3$ blocks.
In general, the total number of blocks is expected to be smaller in the $16^3$ block case, as each block covers a larger spatial volume.
At levels below $3$, gravitational collapse has already proceeded, and the density distributions are consistent between the two setups. 
Accordingly, the expected relationship in the number of blocks is reproduced, as shown in Table \ref{Tab:cellnumber_inGMC_simulation}.
In contrast, this relationship is reversed at level $3$.
This inversion arises from differences in the spatial distributions of density peaks generated by turbulence when different block sizes are employed.
Such discrepancies in the number of blocks are particularly pronounced immediately after the onset of refinement.
As the system evolves further, gravitational collapse is expected to proceed at level $3$ as well, and the block numbers will follow the same trend as in the other levels.

\begin{table*}[t]

\caption{
Computational performance of individual code components in the GMC simulations with $8^3$ blocks.
}
\centering
\resizebox{\textwidth}{!}{
\begin{tabular}{l|ccc|ccc|ccc|ccc}
\hline
GPUs
& \multicolumn{3}{c|}{ Hydro+RT}
& \multicolumn{3}{c|}{ Self-gravity}
& \multicolumn{3}{c|}{ Thermochemistry}
& \multicolumn{3}{c}{Total} \\
\cline{2-13}
 & Time & Perf. & Scaling
 & Time & Perf. & Scaling
 & Time & Perf. & Scaling
 & Time & Perf. & Scaling \\
\hline
1  & $26.1$  & $22.6$ & $1$     & $13.9$  & $42.5$ & $1$     & $22.5$ & $26.2$  & $1$     & $62.7$ & $9.39$  & $1$ \\
2  & $14.8$  & $19.9$ & $0.879$ & $19.1$  & $15.4$ & $0.363$ & $11.3$ & $26.0$  & $0.992$ & $45.5$ & $6.48$  & $0.689$  \\
4  & $8.72$  & $16.9$ & $0.749$ & $22.2$  & $6.62$ & $0.156$ & $5.75$ & $25.6$  & $0.977$ & $36.9$ & $3.99$  & $0.425$ \\
8  & $5.32$  & $13.9$ & $0.612$ & $24.9$  & $2.96$ & $0.0696$& $2.90$ & $25.4$  & $0.968$ & $33.3$ & $2.21$  & $0.236$ \\
\hline
\end{tabular}
}
\begin{tablenotes}[para,flushleft]
\item \parbox{\textwidth}{
  \begin{minipage}{1 \hsize}
  \vspace{3mm}
    Notes. ``Hydro + RT'' denotes the combined hydrodynamics and radiative transfer solvers. 
    ``Thermochemistry'' refers to the solver for the non-equilibrium chemistry and the energy equations.
    All other labels are defined in Table \ref{tab:performance_scaling_uniform}.
  \end{minipage} 
}
\end{tablenotes}
\label{tab:performance_scaling_GMC_888}
\end{table*}

\begin{table*}[t]

\caption{
Computational performance of individual code components in the GMC simulations with $16^3$ blocks.
}
\centering
\resizebox{\textwidth}{!}{
\begin{tabular}{l|ccc|ccc|ccc|ccc}
\hline
GPUs
& \multicolumn{3}{c|}{ Hydro+RT}
& \multicolumn{3}{c|}{ Self-gravity}
& \multicolumn{3}{c|}{ Thermochemistry}
& \multicolumn{3}{c}{Total} \\
\cline{2-13}
 & Time & Perf. & Scaling
 & Time & Perf. & Scaling
 & Time & Perf. & Scaling
 & Time & Perf. & Scaling \\
\hline
1  & $42.5$  & $36.2$ & $1$     & $22.8$  & $67.6$ & $1$     & $18.8$ & $81.5$  & $1$     & $85.0$ & $18.1$  & $1$ \\
2  & $23.8$  & $32.3$ & $0.891$ & $31.1$  & $24.8$ & $0.366$ & $10.0$ & $76.7$  & $0.936$ & $65.5$ & $11.7$  & $0.648$  \\
4  & $14.3$  & $27.0$ & $0.745$ & $36.7$  & $10.5$ & $0.155$ & $4.96$ & $77.6$  & $0.947$ & $56.5$ & $6.82$  & $0.376$ \\
8  & $8.69$  & $22.2$ & $0.611$ & $40.0$  & $4.81$ & $0.0711$& $2.63$ & $73.3$  & $0.895$ & $51.6$ & $3.73$  & $0.260$ \\
\hline
\end{tabular}
}
\begin{tablenotes}[para,flushleft]
\item \parbox{\textwidth}{
  \begin{minipage}{1 \hsize}
  \vspace{3mm}
    Labels are the same as in Table \ref{tab:performance_scaling_GMC_888}.
  \end{minipage} 
}
\end{tablenotes}
\label{tab:performance_scaling_GMC_16}
\end{table*}

\begin{figure} 
\begin{center}
  \includegraphics[width=\columnwidth]{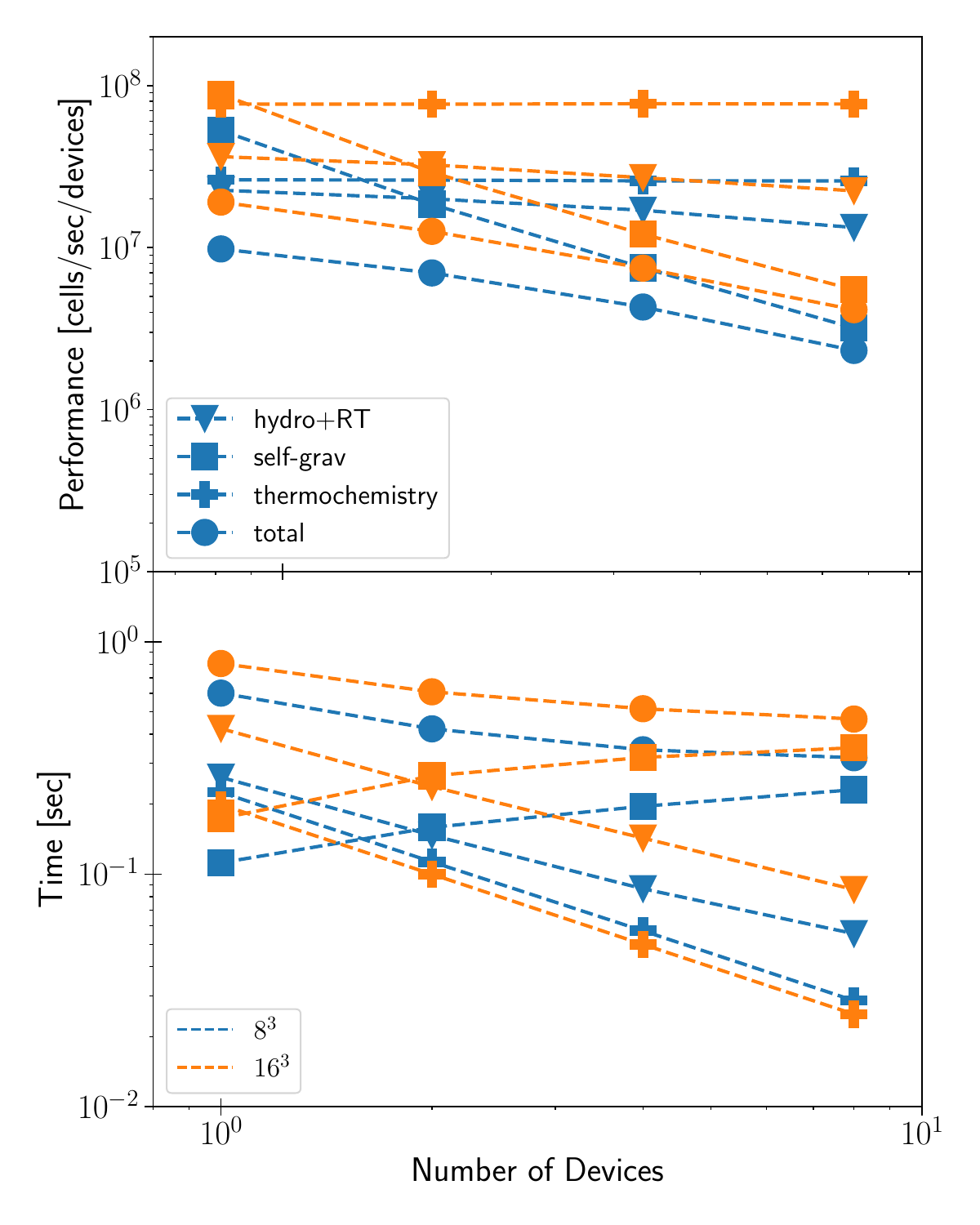}
\end{center}
 \caption{
 Same as Figure \ref{fig:uniform_hydro}, but for the GMC simulations.
 Each symbol represents the performance of the hydrodynamics and radiative transfer solver (triangles; Hydro+RT), self-gravity solver (squares; self-gravity), thermochemical solver (crosses; thermochemistry), and the entire code (circles; total).
 Blue and orange symbols correspond to the cases with $8^3$ and $16^3$ blocks, respectively.
 {Alt text: Performance of individual solvers and the total code in GMC simulations with $8^3$ and $16^3$ blocks.}
 }
 \label{fig:GMC}
\end{figure}

Figure \ref{fig:GMC} and Table \ref{tab:performance_scaling_GMC_888} and \ref{tab:performance_scaling_GMC_16} show the performance of the GMC simulations.
Here, the computational cost associated with the hydrodynamics, radiative transfer, and transports of chemical species and dust-deposited energy is collectively referred to as Hydro+RT.
For the $8^3$ blocks on a single device, the computational costs of the numerical flux calculation and the chemistry solver are comparable.
In contrast, the self-gravity solver is approximately a factor of two faster than the other major components of the code. 
Compared with the AMR grid tests presented in Section \ref{section:performance_hydro_AMR} and \ref{section:performance_selfgrav_AMR}, the GMC simulations exhibit a more complex distribution of finer cells concentrated in high-density regions.
As a result, the number of interfaces between finer and coarser cells increases.
This leads to higher computational costs associated with numerical-flux corrections and the interpolation of physical quantities on ghost cells at the block boundaries.
Consequently, the overall performance of both the hydrodynamics and self-gravity solvers is lower than that in the idealized AMR grid tests.
Moreover, the numerical flux calculation includes not only hydrodynamics but also the transport of radiation and chemical species (denoted as Hydro+RT in Figure \ref{fig:GMC}), resulting in a higher total computational cost than that of the self-gravity solver. 
In contrast, the thermochemical solver maintains the performance level expected from the results in Section \ref{sec:BE} and \ref{sec:HII_region}.

As the number of devices increases, the strong-scaling efficiency of Hydro+RT remains comparable to that obtained in the AMR test presented in Section \ref{section:performance_hydro_AMR}, staying above $0.6$ even with $8$ devices.
The thermochemical solver is executed independently on each device and achieves a strong-scaling efficiency exceeding $0.9$.
In contrast, the performance of the self-gravity solver degrades with increasing device count.
In particular, for GMC simulations, the strong-scaling efficiency tends to be lower because the complex grid configuration increases communication between different processes compared to the AMR grid test presented in Section \ref{section:performance_selfgrav_AMR}. 
In our implementation, blocks are evenly distributed among processes at each level.
When the number of grids at the finest level is extremely small, as in the present test, the workload at that level also becomes small, leading to degraded performance.
As a result, while the overall code performance retains a strong-scaling efficiency of approximately $0.6$ with two devices, further increases in the number of devices lead to degraded performance due to the self-gravity solver.
This indicates that the number of devices should be chosen according to the problem so that the performance of the self-gravity solver remains comparable to or better than that of Hydro+RT.

For the $16^3$ blocks, the number of cells processed per device second increases relative to the $8^3$ blocks.
In addition, the strong-scaling efficiencies of the individual solvers are similar to those obtained with the $8^3$ blocks.
However, as shown in Table \ref{Tab:cellnumber_inGMC_simulation}, the total number of cells generated by refinement increases by more than a factor of three compared to the $8^3$ case.
As a consequence, the time required to advance a single time step increases by more than a factor of $1.3$.
Although blocks with a larger number of cells generally achieve higher per-device performance, they also induce the creation of more cells in AMR simulations.
In problems that involve extensive refinement, this increase in the total cell count becomes a dominant factor and ultimately degrades the overall performance.
Moreover, in GPU-based computations, memory capacity is more constrained than on CPUs, making an increase in the total number of cells unfavorable from a memory-usage perspective.
Therefore, for star cluster formation simulations, the use of $8^3$ blocks is preferable to $16^3$ blocks.

\section{Summary} \label{Section:discussion}

In this work, we have developed \textsc{SFUMATO\#}, a new implementation of a self-gravitational radiation hydrodynamics code with adaptive mesh refinement (AMR).
\textsc{SFUMATO\#} is designed to perform radiation hydrodynamics simulations of star cluster formation in environments with various metallicities.
\textsc{SFUMATO\#} is built with the CUDA/HIP programming frameworks and supports execution on multiple GPUs.
The code incorporates modules for hydrodynamics, self-gravity, radiative transfer, non-equilibrium chemistry, and sink particle treatment.
Both synchronous and adaptive timestep modes are implemented.
The accuracy of our code in the hydrodynamics, self-gravity, and radiative transfer was tested by 
shock tube, double Mach reflection, gravitational potential around two spherical clouds, and radiation transfer of a beam.

We have developed new chemistry and thermal solvers based on a linearized implicit (LI) method, which avoids iterative calculations based on the Newton-Raphson (NR) method.
In the LI method, chemical abundances and the gas internal energy are integrated using a sub-stepping scheme applied independently in each cell.
By incorporating the heat capacity of dust grains, the dust temperature can also be evolved without resorting to iterative methods.
The accuracy and performance of the LI method are validated through tests of time evolution of chemical abundances, collapsing Bonnor-Ebert sphere, and H{\sc ii} region formation.
We show that the LI method reproduces the results obtained with the NR method with good accuracy.
From the perspective of computational cost, we find that using an increased pseudo dust heat capacity is preferable to adopting the realistic value.
We further demonstrate that the pseudo heat capacity must not exceed the specific heat capacity of the gas, which is on the order of $k_{\rm B}/m_{\rm H} \sim 8 \times 10^7~{\rm erg \, g^{-1} \, K^{-1}}$.
We also find that employing a pseudo heat capacity does not affect the simulation results for values of $C_{\rm d} \leqq 10^7~{\rm erg \, g^{-1} \, K^{-1}}$.
In this paper, we confirm that the LI method is valid for gas densities up to $n_{\rm H} \simeq 10^{11}~{\rm cm^{-3}}$.
Its applicability to higher-density regions, where stronger coupling between gas and dust grains occurs, will be examined in future work.

We measured the performance of the hydrodynamics and self-gravity solvers on both uniform and AMR grids.
The hydrodynamics solver exhibits good strong scaling, with efficiencies exceeding $0.6$ on both grid configurations, even with $8$ devices.
In contrast, the self-gravity solver shows degraded strong-scaling efficiency as the number of devices increases, due to the increasing cost of MPI communication.
Simulations of a realistic giant molecular cloud, additionally including radiative transfer and thermochemical processes, show that the self-gravity solver runs approximately twice as fast as each of the other solvers.
However, as the degree of parallelism increases, the relative cost of the self-gravity solver grows and gradually becomes the dominant bottleneck.
These results indicate that star cluster formation simulations should be performed with a number of GPU devices chosen such that the computational cost of the self-gravity remains comparable to that of the other solvers.

In the current formalism of \textsc{SFUMATO\#}, we neglect terms of order $\mathcal{O} (v/c)$ in radiation transfer, which arise from Lorentz transformations of radiation quantities and are essential for accurately treating the tight coupling between radiation and matter \citep[e.g.,][]{1984frh..book.....M}.
Such effects have been incorporated into radiation-hydrodynamics codes designed for optically thick regions \citep[e.g.,][]{2007ApJ...667..626K, 2011ApJS..196...20Z, 2013ApJS..206...21S, 2015MNRAS.449.4380R, 2024MNRAS.531.1228H, 2026ApJ...999..257K}.
In addition, the reduced speed of light approximation itself introduces energy non-conservation and can lead to different temperature distributions in optically thick media, although it accurately captures the radiation field in optically thin regions \citep{2026ApJ...999..257K}.
Recently, \citet{2026ApJ...999..257K} addressed these issues by developing a radiative transfer solver that combines radiative transfer and diffusion equations with the inclusion of $\mathcal{O} (v/c)$ terms, thereby overcoming the limitations of the reduced speed of light approximation.
In star cluster formation, regions that are optically thick to dust thermal emission can develop.
This effect becomes important in high-density clouds with solar metallicity and surface densities exceeding $\Sigma_{\rm cl} \geqq 10^5~M_{\odot} {\rm pc^{-2}}$ \citep{2022MNRAS.517.1313M}.
To properly capture the effect of radiation pressure due to dust thermal emission, we plan to implement the scheme of \citet{2026ApJ...999..257K} in a future study.
In contrast, the current implementation remains adequate for cases where star-forming clouds are disrupted by H{\sc ii} region formation or in low-metallicity environments.

An AMR hydrodynamics code with self-gravity can be applied to a wide range of astrophysical problems.
In particular, by utilizing modern supercomputers equipped with GPUs, which have been increasingly adopted in recent years, our code can tackle problems that were previously inaccessible due to resolution limitations.
In future work, we will apply \textsc{SFUMATO\#} to studies of gas accretion and disk formation during star cluster formation.

\section{ACKNOWLEDGEMENTS}
The authors would like to thank Kohji Yoshikawa, Hidenobu Yajima, and Kentaro Nagamine for useful discussions and comments. The numerical simulations were performed on the Miyabi-G and Pegasus provided by the Multidisciplinary Cooperative Research Program in the Center for Computational Sciences, University of Tsukuba, and the GPU cluster provided by the Center for Computational Astrophysics of the National Astronomical Observatory of Japan. This work was supported by JSPS KAKENHI Grant Numbers JP23K13139, JP26K00735 (HF) and JP23K03464 (TM).

\bibliographystyle{apj}
\bibliography{article} 

\appendix
\section{Chemical network} \label{apd:chemical_network}

\begin{table*}
  \caption{Chemical reactions.}
  \label{Tab:chemical_network}
  \centering
  \begin{tabular}{lllc}
    \hline
        Number & Reaction & Rate Coefficient & Reference \\ \hline
        H0  &  ${\rm H} + {\rm e} \rightarrow {\rm H^+} + 2{\rm e}$ & $k_{\rm H0} = \exp \lbrack  - 32.71396786 + 13.536556 \times  \left( \ln T_{\rm g} \left({\rm ev} \right) \right) -5.73932875 \left( \ln T_{\rm g} \left({\rm ev} \right) \right)^{2}$ & 1 \\
      & & \hspace{1.4cm} $+1.563154998 \left( \ln T_{\rm g} \left({\rm ev} \right) \right)^{3} -0.2877056   \left( \ln T_{\rm g} \left({\rm ev} \right) \right)^{4}  $ & \\
      & & \hspace{1.4cm} $ + 3.48255977 \times 10^{-2}  \left(\ln T_{\rm g} \left({\rm ev} \right) \right)^{5} -2.63197617 \times 10^{-3}  \left( \ln T_{\rm g} \left({\rm ev} \right) \right)^{6} $ & \\
      & & \hspace{1.4cm} $ + 1.11954395 \times 10^{-4}  \left( \ln T_{\rm g} \left({\rm ev} \right) \right)^{7} - 2.03914985 \times 10^{-6}  \left( \ln T_{\rm g} \left({\rm ev} \right) \right)^{8} \rbrack  $ & \\
        H1  &  ${\rm H^+} + {\rm e} \rightarrow {\rm H} + \gamma$ & $k_{\rm H1} = 2.753 \times 10^{-14}(315614/T_{\rm g})^{1.5} [1+(115188/T_{\rm g})^{0.407}]^{-2.242}$  & 2,3 \\ 
        H2  &  ${\rm H}   + {\rm e} \rightarrow {\rm H^-} + \gamma$ & $k_{\rm H2} = 1.4 \times 10^{-18} T_{\rm g}^{0.928} \exp(-T_{\rm g}/16200)$  &4 \\
        H3  &  ${\rm H^-} + {\rm H} \rightarrow {\rm H_2} + {\rm e}$ &  $k_{\rm H3} = 1.35 \times 10^{-9} \left[ T_{\rm g}^{0.098493} + 0.32852 T_{\rm g}^{0.5561} + 2.771 \times 10^{-7} T_{\rm g}^{2.1826} \right]$  & 5 \\  
            & & \hspace{6mm} $/ \left[ 1 + 6.191 \times 10^{-3} T_{\rm g}^{1.0461} + 8.9712 \times 10^{-11} T_{\rm g}^{3.0424} + 3.2576 \times 10^{-14} T_{\rm g}^{3.7741}  \right]$ & \\
        H4  &  ${\rm H_2} + {\rm H} \rightarrow 3 {\rm H}$ & see the reference & 6 \\
        H5  &  ${3 \rm H} + \rightarrow {\rm H_2} + {\rm H}$ & $k_{\rm H5} = 6 \times 10^{-32} T_{\rm g}^{-0.25} + 2 \times 10^{-31} T_{\rm g}^{-0.5}$  & 7 \\
        H6  &  $2 {\rm H}  + {\rm H_2} \rightarrow 2 {\rm H_2} $ &  $k_{\rm H6} = k_{\rm H5}/8$ & 8 \\
        H7  &  $2 {\rm H_2} \rightarrow 2 {\rm H} + {\rm H_2}$ & $k_{\rm H7} = k_{\rm high}^{1 - a} k_{\rm low}^{a}$ & 8 \\ 
      & & \hspace{2mm} $k_{\rm low} = 1.18 \times 10^{-10} \exp(-6.95 \times 10^{4} / T_{\rm g})$ &  \\
      & & \hspace{2mm} $k_{\rm high} = 8.125 \times 10^{-8} T_{\rm g}^{-1/2} \exp(-5.2 \times 10^{4} / T_{\rm g}) \left[ 1 - \exp(-6000/T_{\rm g}) \right]$ & \\
      & & \hspace{2mm} $a = \left( 1 + n / n_{\rm cr} \right)^{-1}$, $\log_{10} (n_{\rm cr}) = 4.845 - 1.3 \log_{10} \left(T_{\rm g} / 10^{4} \right) + 1.62 \left[ \log_{10} \left( T_{\rm g} / 10^{4}  \right) \right]^2$ & \\
        H8  &  $ 2 {\rm H} + {\rm grain} \rightarrow {\rm H_2}$ & $k_{\rm H8} = 6.0 \times 10^{-17} \sqrt{T_{\rm g}/300} f_{a} \left(Z/Z_{\odot} \right) $  & 9 \\
       & & \hspace{1.4cm}  $ \times \lbrack 1.0 + 4.0 \times 10^{-2} \sqrt{T_{\rm g}+T_{\rm d}} + 2.0 \times 10^{-3} T_{\rm g} + 8.0 \times 10^{-6} T_{\rm g}^{2} \rbrack^{-1} $ & \\
       & & \hspace{2mm} $f_{a} = [1.0 + \exp (7.5 \times 10^{2} (1/75 - T_{\rm d}^{-1}))]^{-1}$ & \\
        H9  &  $ {\rm H} + h \nu \rightarrow {\rm H^+} + {\rm e}$ & see the reference  & 10 \\
        H10 &  $ {\rm H_2} + h \nu \rightarrow 2{\rm H} $ & see the reference  & 10 \\ 
        H11 &  $ {\rm H} + {\rm cr} \rightarrow {\rm H^+} + {\rm e}$ & see the reference  & 10, 11, 12 \\
    \hline 
  \end{tabular}
  \begin{minipage}{1 \hsize}
  \vspace{3mm}
    Reference. (1) \citet{1997NewA....2..181A}, (2) \citet{1992ApJ...387...95F}, (3) \citet{2007ApJ...666....1G}, (4) \citet{1998A&A...335..403G}, (5) \citet{2010Sci...329...69K}, (6) \citet{1996ApJ...461..265M}, (7) \citet{2013ApJ...773L..25F}, (8) \citet{1983ApJ...271..632P}, (9) \citet{1985ApJ...291..722T}, (10) \citet{2021MNRAS.506.5512F}, (11) \citet{2017ApJ...843...38G}, (12) \citet{1974ApJ...193...73G}.
  \end{minipage} 
\end{table*}

In Table \ref{Tab:chemical_network}, we summarize the chemical network consisting of $\rm H$, $\rm H_2$, $\rm H^+$, and $\rm e^-$.
The abundance of $\rm H^-$ is obtained as an intermediate species, following the treatment described in Appendix A of \citet{2010ApJ...722.1793O}.
The photoionization rate of hydrogen and the photodissociation rate of $\rm H_2$ are calculated from the energy densities of EUV and FUV photons, respectively.
Details of these calculations are provided in Appendix C of \citet{2021MNRAS.506.5512F}.

\section{Thermal processes}\label{apd:thermal_process}

\begin{table*}
  \caption{Thermal processes.}
  \label{Tab:thermal_process}
  \centering
  \begin{tabular}{lllc}
     \hline 
     Number & Process & Rate ($\rm erg \, cm^{-3} \, s^{-1}$) & Reference \\ \hline
             \multicolumn{4}{c}{Heating}\\ 
     \TPHtwoform & $\rm H_2$ formation & $\Gamma_{\TPHtwoform} = [ \, 3.73 (1+ n_{\rm cr} / n_{\rm H})^{-1} k_{\rm H3} n({\rm H}) n({\rm H^{-}}) + 4.48 (1+ n_{\rm cr} / n_{\rm H})^{-1} k_{\rm H5} n^{3}({\rm H})$ & \TPPHM, \OMKI \\ 
        & & \hspace{7mm} $+ \, \left( 0.2 + 4.2 (1+ n_{\rm cr} / n_{\rm H})^{-1} \right) k_{\rm H7} n^{2}({\rm H_2})] ~ \rm{eV}  $ &  \\
        & & $n_{\rm cr} = 10^{6}T_{\rm g}^{-1/2}$ & \\
        & &  \hspace{5mm} $/ \left\{ 1.6 n({\rm H}) / n_{\rm H} \exp [-(400/T_{\rm g})^{2}] + 1.4 n({\rm H_2})/n_{\rm H} \exp[-12000/(T_{\rm g}+1200)]  \right\} ~ {\rm cm^{-3}} $ & \\
     \TPHphion & $\rm H$ photoionization & see the reference& \TPHFHYONE \\
     \TPHdisso & $\rm H_2$ photodissociation & see the reference& \TPPHM \\
     \TPPhotoele & Photoelectric heating &  see the reference & \TPPBakesTielens, \TPPFKSM \\
             \multicolumn{4}{c}{Cooling}\\ 
     \TPPHion & $\rm H$ ionization & $\Lambda_{\TPPHion} = 13.6 k_{\rm H0} n({\rm H}) n({\rm e}) {\rm eV}$ & \TPPHM, \OMKI \\
     \TPHtwodiso & $\rm H_2$ dissociation & $\Lambda_{\TPHtwodiso} = 4.48 \, [ \, k_{\rm H4} n({\rm H_2}) n({\rm H}) + k_{\rm H7} n^{2}({\rm H_2})  \, ] \, {\rm eV}$ & \TPPHM, \OMKI \\
     \TPHrecom & $\rm H$ recombination &  $\Lambda_{\TPHrecom} = \exp \lbrack \ln 10 \times (-25.87 + 0.4958 \log_{10} T_{\rm g} - 0.1052 (\log_{10} T_{\rm g})^2 + 0.04264 (\log_{10} T_{\rm g})^3$ & \TPFERLAND, \TPPSUGI \\ 
     & & \hspace{7mm} $ - 9.165 \times 10^{-3} (\log_{10} T_{\rm g})^4 + 5.491 \times 10^{-4} (\log_{10} T_{\rm g})^5 \rbrack n({\rm H^+}) n({\rm e}) $ &  \\
     \TPHminus & $\rm H^-$ free-bound & $\Lambda_{\TPHminus} = k_{\rm B} T_{\rm g} k_{\rm H2} n({\rm H}) n({\rm e})$ & \TPPHOSOKW \\
     \TPHexcit & $\rm H$ excitation & $\Lambda_{\TPHexcit} = 7.50 \times 10^{-19} [1 + (T_{\rm g}/10^5)^{1/2}]^{-1}\exp[-118348/T_{\rm g}]n({\rm e})n({\rm H^{+}})$ & \TPPCEN \\
     \TPHeplus & $\rm He^+$ excitation &  $\Lambda_{\TPHeplus} = 5.54 \times 10^{-17} T_{\rm g}^{-0.397} [1 + (T_{\rm g}/10^5)^{1/2}]^{-1}\exp[-473638/T_{\rm g}] n({\rm e})n({\rm He^{+}}) $  & \TPPCEN \\
     \TPFreeFree & Free-free & $\Lambda_{\TPFreeFree} = 1.426 \times 10^{-27} T_{\rm g}^{1/2} g_{\rm ff} (T_{\rm g}) n(\rm H^+) n(\rm e) $ & \TPPSKang \\
     & & $g_{\rm ff} (T_{\rm g}) = 0.79464 + 0.1243 \log_{10} T_{\rm g}$ \hspace{5mm} $T_{\rm g} < 3.2 \times 10^5~{\rm K}$ & \\
        & & \hspace{9.5mm} $= 2.13164 - 0.1240 \log_{10} T_{\rm g}$ \hspace{5mm} $T_{\rm g} > 3.2 \times 10^5~{\rm K}$ & \\ 
     \TPCOMPT & Compton & $\Lambda_{\TPCOMPT} = 1.017 \times 10^{-37} T_{\rm CMB}^4 (T_{\rm g} - T_{\rm CMB}) n({\rm e})$ & \TPPCEN \\
     \TPLINE & Line cooling & $\Lambda_{\TPLINE} = \Lambda_{\rm H_2} + \Lambda_{\rm CII} + \Lambda_{\rm OI}  + \Lambda_{\rm OII}  + \Lambda_{\rm OIII}$, see the reference & \TPHFHYONE\\
     \TPCOOL & Gas-grain heat transfer & $\Lambda_{\TPCOOL} =  5.83 \times 10^{-8} n_{\rm H} \rho ( T_{\rm g} / 10^{3})^{1/2} [ 1 - 0.8 \exp(- 75/ T_{\rm g})] ( T_{\rm g} - T_{\rm d} ) ( Z / Z_{\odot} )  $& \TPPHM, \OMKI\\
     \hline
  \end{tabular}
  \begin{minipage}{1 \hsize}
  \vspace{3mm}
    Reference. 
    (\TPPHM) \citet{1979ApJS...41..555H}, 
    (\OMKI) \citet{2000ApJ...534..809O}, 
    (\TPHFHYONE) \citet{2021MNRAS.506.5512F}, 
    (\TPPBakesTielens) \citet{1994ApJ...427..822B},
    (\TPPFKSM) \citet{2020MNRAS.497..829F},
    (\TPFERLAND) \citet{1992ApJ...387...95F}, 
    (\TPPSUGI) \citet{2017MNRAS.469...62S}, 
    (\TPPHOSOKW) \citet{2016ApJ...824..119H}, 
    (\TPPCEN) \citet{1992ApJS...78..341C},
    (\TPPSKang) \citet{1987ApJ...318...32S}. 
    Notes. $n({\rm H})$, $n({\rm H_2})$, $n({\rm H^+})$, $n({\rm H^-})$, and $n({\rm e})$ denote the number densities of ${\rm H}$, ${\rm H_2}$, ${\rm H^+}$, ${\rm H^-}$, and electrons, respectively.
    For the abundance of ionized helium, $n({\rm He^+})$, we assume that all helium atoms are fully ionized.
    $T_{\rm CMB}$ denotes the cosmic microwave background (CMB) temperature.
  \end{minipage}
\end{table*}

In Table \ref{Tab:thermal_process}, we summarize the heating and cooling processes implemented in our code.
Line cooling from $\rm H_2$, C{\sc ii}, O{\sc i}, O{\sc ii}, and O{\sc iii} is included, and these processes are denoted by $\Lambda_{\rm H_2}$, $\Lambda_{\rm CII}$, $\Lambda_{\rm OI}$, $\Lambda_{\rm OII}$, and $\Lambda_{\rm OIII}$, respectively.
\citet{2005ApJ...626..627O} demonstrated that the temperature evolution of collapsing clouds can be well reproduced by assuming that all carbon and oxygen atoms are in the form of C{\sc ii} and O{\sc i}, respectively.
In addition, O{\sc ii} and O{\sc iii} act as the primary coolants in H{\sc ii} regions. 
In this work, we estimate the abundances of O{\sc i}, O{\sc ii}, and O{\sc iii} under the assumption of ionization equilibrium, following the method described in  Appendix B of \citet{2021MNRAS.506.5512F}.
We also adopt an $\mathrm{H_2}$ self-shielding prescription for FUV photons, and the photoelectric heating rates include the effect of recombination between dust grains and electrons, as described in \citet{2022MNRAS.511.3346F}.

\section{Computational procedure for dust temperature} \label{apd:dust_temp}

We simultaneously update the source terms of energy density of IR radiation, dust temperature, and energy transfer between gas and dust grains.
Retaining only the relevant terms, equations \eqref{energy_eq}, \eqref{eq_radmoment1}, and \eqref{eq_dust_temp} can be written as
\begin{align}
  C_{\rm d} \frac{\partial T_{\rm d}}{\partial t} &= \kappa_{\rm d}^0(T_{\rm IR}) \tilde{c} E_{\rm IR} + \mathcal{E}_{\rm UV} -\kappa_{\rm d}^0(T_{\rm d}) a c T_{\rm d}^4 \nonumber \\
  & - R_gn_{\rm H}\sqrt{\frac{T_{\rm g}}{T_{\rm high}}} \left[ 1 - 0.8 \exp(-T_{\rm low}/T_{\rm g})\right] \left( T_{\rm d} - T_{\rm g} \right), \label{ap_eq:dust_temp}
\end{align}
\begin{align}
\frac{\partial E_{\rm IR}}{\partial t} &= \rho f_{\rm d} \left[ \kappa_{\rm d}^0(T_{\rm d}) c a T_{\rm d}^4 - \kappa_{\rm d}^0(T_{\rm IR}) \tilde{c} E_{\rm IR} \right], \label{ap_eq:Eir}
\end{align}
\begin{align}
\frac{\partial e_{\rm th}}{\partial t} &=  R_g n_{\rm H} f_{\rm d}\sqrt{\frac{T_{\rm g}}{T_{\rm high}}} \left[ 1 - 0.8 \exp(-T_{\rm low}/T_{\rm g})\right] \left( T_{\rm d} - T_{\rm g} \right), \label{ap_eq:Tg}
\end{align}
where $\kappa_{\rm d}^0$ is dust opacity at the solar metallicity, $\mathcal{E}_{\rm UV}$ is the dust absorption rates of EUV and FUV photons, $f_{\rm d}$ is the dust abundance ratio relative to that of the solar metallicity, $e_{\rm th}$ is specific internal energy, $T_{\rm high} = 1000~{\rm K}$, and $T_{\rm low} = 75~{\rm K}$.
The IR radiation temperature $T_{\rm IR}$ is defined as $aT_{\rm IR}^4 = (\tilde{c}/c) E_{\rm IR}$, where the factor $(\tilde{c}/c)$ represents a correction introduced by adopting the reduced speed of light approximation.
The third terms on the right-hand side of Equations \eqref{ap_eq:dust_temp} and \eqref{ap_eq:Tg} represent the energy coupling between gas and dust grains denoted as $\Lambda_{\TPCOOL}$ in Table \ref{Tab:thermal_process}, with $R_{g} = 5.83 \times 10^{-8}$.
\citep{1979ApJS...41..555H, 2005ApJ...626..627O}.
We use the opacity table driven by \citet{1993ApJ...402..441L}.
As in the chemical network solver, equations \eqref{ap_eq:dust_temp}-\eqref{ap_eq:Tg} are updated using the following method.
The right-hand side terms are evaluated at the advanced step $t + \Delta t$.
The differences in dust temperature $T_{\rm d}$, IR radiation energy density $E_{\rm IR}$, and gas temperature $T_{\rm g}$ are expressed as follows,
\begin{align}
  \Delta T_{\rm d} = T_{\rm d}^{t+\Delta t} - T_{\rm d}^{t}, \label{ap_eq:Td}
\end{align}
\begin{align}
  \Delta E_{\rm IR} = E_{\rm IR}^{t+\Delta t} - E_{\rm IR}^{t}, \label{ap_eq:Eir2}
\end{align}
\begin{align}
  \Delta T_{\rm g} = T_{\rm g}^{t+\Delta t} - T_{\rm g}^{t}. \label{ap_eq:Tg_3}
\end{align}
This leads to the following difference equations:
\begin{align}
  &\left( C_{\rm d} + R_{\rm g} n_{\rm H} \sqrt{\frac{T_{\rm g}^{t}}{T_{\rm high}}} \gamma_{\rm d} \Delta t + ac \Delta t \left(T_{\rm d}^t\right)^3 \beta_{\rm d} \right) \Delta T_{\rm d} - \tilde{c} \Delta t \beta_{\rm IR} \Delta E_{\rm IR} \nonumber \\
& + R_g n_{\rm H}\sqrt{\frac{T_{\rm g}^t}{T_{\rm high}}} \Delta t  \zeta \Delta T_{\rm g} \nonumber \\
& = \Delta t \big[ \kappa_{\rm d}^0(T_{\rm IR}^t) \tilde{c} E_{\rm IR}^t + \mathcal{E}_{\rm UV} -\kappa_{\rm d}^0(T_{\rm d}^t) a c \left(T_{\rm d}^t\right)^4 \nonumber \\
& \hspace{4cm}- R_gn_{\rm H}\sqrt{\frac{T_{\rm g}^t}{T_{\rm high}}} \gamma_{\rm d} \left( T_{\rm d}^t - T_{\rm g}^t \right)  \big], \label{ap_eq:Td2}
\end{align}
\begin{align}
  & - \rho \Delta t c a \left(T_{\rm d}^t\right)^3 f_{\rm d}\beta_{\rm d} \Delta T_{\rm d} + \left( 1 + \rho \Delta t \tilde{c} f_{\rm d} \beta_{\rm IR} \right) \Delta E_{\rm IR} \nonumber \\
&= \rho \Delta t f_{\rm d} \left[ \kappa_{\rm d}^0(T_{\rm d}^t) c a \left(T_{\rm d}^t\right)^4 - \kappa_{\rm d}^0(T_{\rm IR}^t) \tilde{c} E_{\rm IR}^t \right]
\end{align}
\begin{align}
  & - \Delta t f_{\rm d} R_{\rm g} n_{\rm H} \sqrt{\frac{T_{\rm g}^t}{T_{\rm high}}} \gamma_{\rm d} \Delta T_{\rm d} + \left( \alpha_{\rm g} - \Delta t  f_{\rm d} R_{\rm g} n_{\rm H} \zeta \sqrt{\frac{T_{\rm g}^t}{T_{\rm high}}} \right) \Delta T_{\rm g} \nonumber \\
& = \Delta t f_{\rm d} R_{\rm g} n_{\rm H} \gamma_{\rm d} \sqrt{\frac{T_{\rm g}^t}{T_{\rm high}}} \left( T_{\rm d}^t - T_{\rm g}^t \right), \label{ap_eq:Tg_2}
\end{align}
where
\begin{align}
  \alpha_{\rm g} = \frac{1}{\gamma_{\rm ad}-1} \frac{k}{\mu m_{\rm H}}, \label{ap_eq:alpha_g} 
\end{align}
\begin{align}
   \beta_{\rm d} = 4 \kappa_{\rm d}^0\left( T_{\rm d}^t \right) + \frac{d \kappa_{\rm d}^0 \left( T_{\rm d}^t \right)}{d  T_{\rm d}^t } T_{\rm d}^t, \label{ap_eq:beta_d}
\end{align}
\begin{align}
   \beta_{\rm IR} = \kappa_{\rm d}^0\left( T_{\rm IR}^t\right) + \frac{d \kappa_{\rm d}^0 \left( T_{\rm IR}^t \right)}{d  T_{\rm IR}^t } \frac{T_{\rm IR}^t}{4}, \label{ap_eq:beta_IR}
\end{align}
\begin{align}
  \gamma_{\rm d} = 1-0.8 \exp(-T_{\rm low}/T_{\rm g}^t), \label{ap_eq:gamma_d}
\end{align}
\begin{align}
  \zeta = \left[ \frac{\gamma_{\rm d}}{2} \left( \frac{T_{\rm d}^t}{T_{\rm g}^t} -1 \right) + (\gamma_{\rm d} -1) \frac{T_{\rm low}}{T_{\rm g}^t}\left( \frac{T_{\rm d}^t}{T_{\rm g}^t} -1 \right) - \gamma_{\rm d} \right], \label{ap_eq:zeta}
\end{align}
$\mu$ is the mean molecular weight, and $m_{\rm H}$ is the mass of hydrogen nucleus. 

\end{document}